\documentclass[10pt,letterpaper,fignum,man,floatsintext]{apa7}
\usepackage{epsfig}
\usepackage{lineno}
\usepackage{graphics}
\usepackage{times}
\usepackage{epsfig}
\usepackage{graphicx}
\usepackage{amsmath}
\usepackage{array}
\usepackage{textcomp}
\usepackage{setspace}
\usepackage{afterpage}
\usepackage{url}
\usepackage[natbibapa]{apacite}

\newcommand{\humanmodelsystem}{1}
\newcommand{\modeldemo}{2}
\newcommand{\perculturescores}{3}
\newcommand{\commonasterismmt}{1}
\newcommand{\commonasterism}{S2}
\newcommand{\modelasterism}{S3}
\newcommand{\perculturescoresvsothers}{S13}
\newcommand{\systemcnt}{27}

\newcommand{\captionletter}[1]{{#1}}

\title{Perceptual grouping explains similarities in constellations across cultures} 
\shorttitle{Perceptual grouping explains similarities in constellations across cultures} 
\shorttitle{Constellations across cultures}

\authorsnames[1,2,1,1]{Charles Kemp, Duane W.\ Hamacher, Daniel R.\ Little, Simon J.\ Cropper} 
\authorsaffiliations{{Melbourne School of Psychological Sciences, University of Melbourne},{School of Physics, University of Melbourne}}

\abstract{
Cultures around the world organise stars into constellations, or asterisms, and these groupings are often considered to be arbitrary and culture-specific. Yet there are striking similarities in asterisms across cultures, and groupings such as Orion, the Big Dipper, the Pleiades and the Southern Cross are widely recognized across many different cultures. Psychologists have informally suggested that these shared patterns are explained by Gestalt laws of grouping, but there have been no systematic attempts to catalog asterisms that recur across cultures or to explain the perceptual basis of these groupings. Here we compile data from 27 cultures around the world and show that a simple computational model of perceptual grouping accounts for many of the recurring cross-cultural asterisms. Our results suggest that basic perceptual principles account for more of the structure of asterisms across cultures than previously acknowledged and highlight ways in which specific cultures depart from this shared baseline.
}

\keywords{perceptual grouping, Gestalt principles, clustering, cultural astronomy \\
~\\
\noindent\textbf{Statement of Relevance:}
Throughout history, people from many cultures have organized the night sky into constellations and embedded these constellations in stories. Psychologists have informally suggested that constellations result from a process of perceptual grouping, and here we systematically explore the extent to which this idea accounts for constellations across cultures. Using data from 27 cultures, we establish which constellations appear frequently across cultures and find that the list of recurring constellations extends beyond familiar examples such as Orion and the Big Dipper.  We then present and evaluate a computational model that aims to capture how humans group stars into constellations. The model groups stars based on proximity and brightness, and these factors alone are enough to account for many of the constellations that recur across cultures. Although constellations are clearly shaped by culture-specific knowledge, our results reveal that basic perceptual factors account for a large set of similarities in constellations across cultures.}

\authornote{\addORCIDlink{Charles Kemp}{0000-0001-9683-8737}\\ 
\addORCIDlink{Duane W.\ Hamacher}{0000-0003-3072-8468}\\ 
\addORCIDlink{Daniel R.\ Little}{0000-0003-3607-5525}\\
\addORCIDlink{Simon J.\ Cropper}{0000-0002-3574-6414} 

Code and data are available at \protect\url{https://github.com/cskemp/constellations}

We acknowledge the Indigenous custodians of the traditional astronomical knowledge used in this paper, and thank Joshua Abbott, Celia Kemp, Bradley Schaefer and Yuting Zhang for comments on the manuscript. This work was supported in part by ARC FT190100200, ARC DE140101600, the McCoy Seed Fund, the Laby Foundation, the Pierce Bequest, and by a seed grant from the Royal Society of Victoria.  

Correspondence concerning this article should be addressed to Charles Kemp, 
Melbourne School of Psychological Sciences, University of Melbourne, Victoria 3010, Australia.  E-mail: \texttt{c.kemp@unimelb.edu.au}

}

\begin{document}
\maketitle

Anyone who has tried to learn the full set of 88 Western constellations will sympathize with \citet[p 156]{herschel1842}, who wrote that ``the constellations seem to have been almost purposely named and delineated to cause as much confusion and inconvenience as possible,'' and that ``innumerable snakes twine through long and contorted areas of the heavens, where no memory can follow them.'' Yet \citet[p 4]{herschel1841}  and others also point out that there are ``well-defined natural groups of conspicuous stars'' that have been picked out and named by multiple cultures around the world~\citep{krupp00,aveni08,kelley11}. For example, the Southern Cross is recognized as a cross by multiple cultures~\citep{urton05,roe05}, and is identified as a stingray by the Yolngu of northern Australia~\citep{mountford56}, an anchor by the Tainui of Aotearoa/New Zealand~\citep{best22}, and as a curassow bird by the Lokono of the Guianas~\citep{maganaj82}.

Asterisms (e.g.\ the Southern Cross) are sometimes distinguished from constellations (e.g.\ the region of the sky within which the Southern Cross lies), but in cross-cultural work these two terms are often used interchangeably. It is widely acknowledged that asterisms reflect both universal perceptual principles and culture-specific traditions.  For example, \citet[p 5]{urton81}  notes that  ``almost every culture seems to have recognized a few of the same celestial groupings (e.g., the tight cluster of the Pleiades, the V of the Hyades, the straight line of the belt of Orion), but the large constellation shapes of European astronomy and astrology simply are not universally recognized; the shapes were projected onto the stars because the shapes were important objects or characters in the Western religious, mythological, and calendrical tradition.'' 
Even groupings as apparently salient as the Southern Cross are not inevitable---some Australian cultures have many names for individual stars but tend not to ``connect the dots'' to form figured constellations~\citep{johnson14,maegraith32,cairnsh04}. 

Although cultural factors are undeniably important, we will argue that perceptual factors explain more of the inventory of asterisms across cultures than has previously been recognized. \citet[p 58]{krupp00}  suggests that a ``narrow company'' of asterisms is common across cultures and lists just four: Orion's Belt, the Pleiades, the Big Dipper, and the Southern Cross. Here we draw  on existing resources to compile a detailed catalog of asterisms across cultures, and find that the list of recurring asterisms goes deeper than the handful of examples typically given by Krupp and 
others~(Aveni,1980; Krupp, 2000a, 2000b)\nocite{aveni80, krupp00, krupp00_tales}. To demonstrate that these asterisms are mostly consistent with universal perceptual principles, we present a computational model of perceptual grouping and show that it accounts for many of the asterisms that recur across cultures. 

\begin{figure}[t]
\begin{center}
\includegraphics[width=\textwidth]{{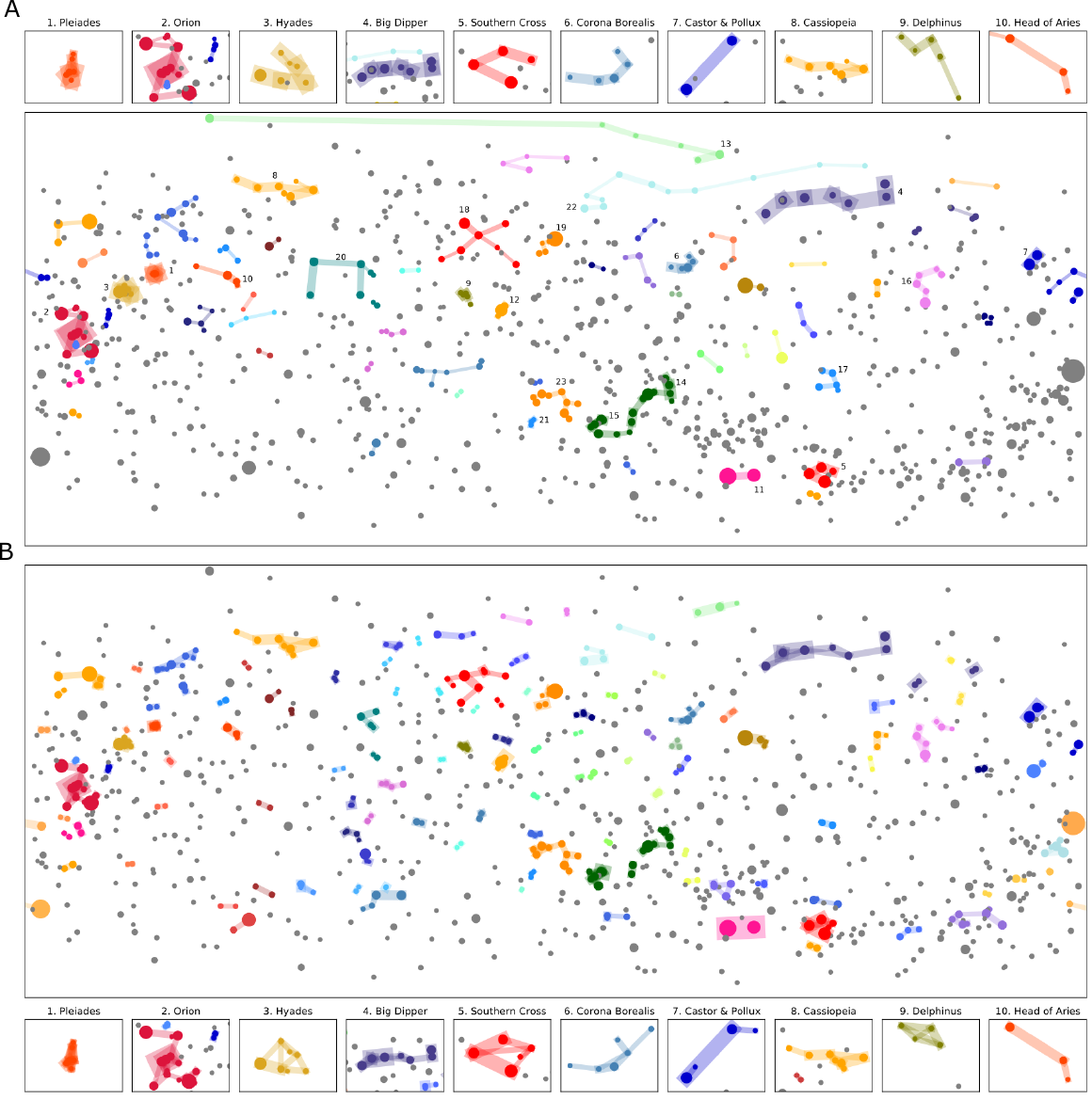}}
\caption{Common asterisms across cultures compared with model asterisms.  (\captionletter{A}) Consensus system created by overlaying minimum spanning trees for all asterisms in our data set of \systemcnt~cultures. Edge widths indicate the number of times an edge appears across the entire dataset, and edges that appear three or fewer times are not shown. Node sizes indicate apparent star magnitudes, and only stars with magnitudes brighter than 4.5 have been included. Insets show 10 of the most common asterisms across cultures, and numbers greater than 10 identify additional asterisms mentioned in the text or Table 1:  Southern Pointers (11), shaft of Aquila (12), Little Dipper (13), head of Scorpius (14), stinger of Scorpius (15), sickle in Leo (16), Corvus (17), Northern Cross (18), Lyra (19), Square of Pegasus (20), Corona Australis (21), head of Draco (22) and the teapot in Sagittarius (23). 
 (\captionletter{B}) Asterisms according to the Graph Clustering (GC) model with $n=320$. The model assigns a strength to each edge in a graph defined over the stars, and here the strongest 320 edges are shown. Edge widths are proportional to the strengths assigned by the model. 
}
\label{humanmodelsystem}
\end{center}
\end{figure}
\afterpage{\clearpage}

\section{A catalog of asterisms across cultures}

Our data set includes 22 systems drawn from the Stellarium software package~\citep{stellarium} and 5 from the ethnographic literature (all sources are listed in the captions of Figures S16-S42). To allow us to focus on the brighter stars, each system was pre-processed by removing stars fainter than 4.5 in magnitude then removing all asterisms that included no stars or just one star after filtering.  The data span six major regions (Asia, Australia, Europe, North America, Oceania, and South America), and include systems from both oral (e.g.\ Inuit) and literate cultures (e.g.\ Chinese). Stellarium currently includes a total of 42 systems, and we excluded 20 because they were closely related to a system already included or because their documentation was not sufficiently grounded in the scholarly literature. Most of our sources specify asterism figures in addition to the stars included in each asterism, but we chose not to use these figures because they can vary significantly within a culture and because they were not available for all cultures. Some of our analyses do not require asterism figures, and for those that do we used minimum spanning trees computed over the stars within each asterism.  

\begin{table*}[t]
\footnotesize
\begin{tabular}{rp{0.5in}p{0.5in}p{0.5in}p{3.0in}l}
 & \textbf{Human Score} & \textbf{Weighted Human Score} & \textbf{Model Score} & \textbf{Stars} & \textbf{Description} \\
1 & 0.66 & 0.73 & 1.0 & 25EtaTau, 17Tau, 19Tau, 20Tau, 23Tau, 27Tau & Pleiades \\
2 & 0.65 & 0.64 & 1.0 & 34DelOri, 46EpsOri, 50ZetOri &  Orion's Belt \\
3 & 0.61 & 0.52 & 1.0 & 87AlpTau, 54GamTau, 61Del1Tau, 74EpsTau, 78The2Tau & Hyades \\
4 & 0.58 & 0.47 & 0.88 & 50AlpUMa, 48BetUMa, 64GamUMa, 69DelUMa, 77EpsUMa, 79ZetUMa, 85EtaUMa &  Big Dipper \\
5 & 0.43 & 0.4 & 1.0 & Alp1Cru, BetCru, GamCru, DelCru & Southern Cross \\
6 & 0.38 & 0.27 & 0.83 & 5AlpCrB, 3BetCrB, 8GamCrB, 13EpsCrB, 4TheCrB &  Corona Borealis \\
7 & 0.35 & 0.28 & 1.0 & 66AlpGem, 78BetGem &  Castor and Pollux \\
8 & 0.35 & 0.34 & 0.45 & 58AlpOri, 24GamOri, 34DelOri, 46EpsOri, 50ZetOri &  \\
9 & 0.31 & 0.35 & 0.6 & 34DelOri, 46EpsOri, 50ZetOri, 44IotOri, 42Ori &  Orion's Belt and Sword \\
10 & 0.31 & 0.25 & 0.56 & 50AlpUMa, 48BetUMa, 64GamUMa, 69DelUMa, 77EpsUMa, 79ZetUMa, 85EtaUMa, 1OmiUMa, 29UpsUMa, 63ChiUMa, 23UMa &  \\
11 & 0.3 & 0.22 & 0.71 & 18AlpCas, 11BetCas, 27GamCas, 37DelCas, 45EpsCas & Cassiopeia \\
12 & 0.3 & 0.35 & 1.0 & 9AlpDel, 6BetDel, 12Gam2Del, 11DelDel & Delphinus \\
13 & 0.28 & 0.23 & 0.38 & 11AlpDra, 50AlpUMa, 48BetUMa, 64GamUMa, 69DelUMa, 77EpsUMa, 79ZetUMa, 85EtaUMa, 23UMa, 26UMa, 12Alp2CVn &  \\
14 & 0.27 & 0.25 & 0.75 & 46EpsOri, 50ZetOri, 48SigOri &  \\
15 & 0.24 & 0.21 & 1.0 & 13AlpAri, 6BetAri, 5Gam2Ari & Head of Aries \\
16 & 0.24 & 0.22 & 1.0 & 53AlpAql, 60BetAql, 50GamAql & Shaft of Aquila \\
17 & 0.24 & 0.34 & 1.0 & Alp1Cen, BetCen & Southern Pointers \\
18 & 0.23 & 0.33 & 1.0 & AlpCrA, BetCrA, GamCrA & Corona Australis \\
19 & 0.23 & 0.15 & 0.75 & 1AlpUMi, 7BetUMi, 13GamUMi, 23DelUMi, 22EpsUMi, 16ZetUMi &  Little Dipper \\
20 & 0.22 & 0.19 & 1.0 & 50AlpUMa, 48BetUMa &  \\
21 & 0.22 & 0.22 & 1.0 & 8Bet1Sco, 7DelSco, 6PiSco & Head of Scorpius \\
22 & 0.21 & 0.21 & 0.04 & 54AlpPeg, 53BetPeg &  \\
23 & 0.21 & 0.2 & 1.0 & 35LamSco, 34UpsSco & Stinger of Scorpius \\
24 & 0.21 & 0.21 & 1.0 & Iot1Sco, KapSco, 35LamSco, 34UpsSco &  \\
25 & 0.2 & 0.17 & 0.75 & 32AlpLeo, 41Gam1Leo, 17EpsLeo, 36ZetLeo, 30EtaLeo, 24MuLeo &  Sickle \\
26 & 0.2 & 0.15 & 0.83 & 1AlpCrv, 9BetCrv, 4GamCrv, 7DelCrv, 2EpsCrv &  Corvus\\
27 & 0.19 & 0.18 & 0.56 & 21AlpSco, 8Bet1Sco, 7DelSco, 6PiSco, 20SigSco &  \\
28 & 0.19 & 0.19 & 0.0 & 21AlpAnd, 88GamPeg &  \\
29 & 0.18 & 0.16 & 0.58 & 21AlpSco, 8Bet1Sco, 7DelSco, 26EpsSco, Zet2Sco, Mu1Sco, 6PiSco, 20SigSco, 23TauSco &  \\
30 & 0.18 & 0.13 & 0.33 & 50AlpCyg, 6Bet1Cyg, 37GamCyg, 18DelCyg, 53EpsCyg, 21EtaCyg & Northern Cross \\
31 & 0.17 & 0.16 & 0.56 & 26EpsSco, Zet2Sco, EtaSco, TheSco, Iot1Sco, KapSco, 35LamSco, Mu1Sco, 34UpsSco & Tail of Scorpius \\
32 & 0.17 & 0.17 & 1.0 & 21AlpSco, 20SigSco, 23TauSco &  \\
33 & 0.17 & 0.09 & 0.83 & 3AlpLyr, 10BetLyr, 14GamLyr, 12Del2Lyr, 6Zet1Lyr &  Lyra \\
34 & 0.16 & 0.13 & 0.01 & 21AlpAnd, 54AlpPeg, 53BetPeg, 88GamPeg & Square of Pegasus \\
35 & 0.16 & 0.08 & 0.56 & 6Alp2Cap, 9BetCap, 40GamCap, 49DelCap, 34ZetCap, 23TheCap, 32IotCap, 16PsiCap, 18OmeCap & Capricornus \\
36 & 0.16 & 0.12 & 0.35 & 21AlpSco, 8Bet1Sco, 7DelSco, 26EpsSco, EtaSco, TheSco, Iot1Sco, KapSco, 35LamSco, Mu1Sco, 6PiSco, 23TauSco &  Scorpius \\
\end{tabular}
\caption{Common asterisms across cultures. Human scores roughly indicate how often an asterism is found in our data set, and weighted human scores adjust for the unequal representation of geographic regions within the data set. The model scores roughly indicate how well these asterisms are 
captured by the Graph Clustering (GC) model (1.0 indicates a perfect match).}
\label{commonasterismtt}
\end{table*}
\afterpage{\clearpage}

Figure~\humanmodelsystem A shows a consensus system generated by overlaying minimum spanning trees for asterisms from all \systemcnt~cultures. The thick edges in the plot join stars that are grouped by many cultures. The most common asterisms include familiar groups such as Orion's belt, the Pleiades, the Hyades, the Big Dipper, the Southern Cross, and Cassiopeia. The plot also highlights asterisms such as Corona Borealis, Delphinus and the head of Aries that are discussed less often but nevertheless picked out by multiple cultures.  All of these asterisms and more are listed in Table~\commonasterismmt, which ranks 36 asterisms based on how frequently they recur across cultures (an extended version of the table appears as Table~\commonasterism). 

The ranking in Table~\commonasterismmt~is based on a quantitative approach that allows for partial matches between asterisms in different cultures. We first define the 
match between an asterism $a$ and a reference asterism $r$ as
\begin{align}
\text{match}(a,r) &= \max \left( \frac{| a \cap r| - |a \setminus r|}{ |r| } , 0 \right),
\label{match}
\end{align}
where $ | a \cap r| $ is the number of stars shared by $a$ and $r$, 
$|a \setminus r|$ is the number of stars in $a$ that are not shared by $r$, and 
$ |r| $ is the number of stars in $r$. The function attains its maximum value of 1 when $a$ and $r$ are identical.  There are two ways in which $a$ can differ from $r$: it can include extraneous stars, and it can fail to include some of the stars in $r$. The match function penalizes the first of these failings more heavily than the second. This property is especially useful when comparing an asterism against a reference that includes a relatively large number of stars. For example, the teapot asterism includes 8 of the brightest stars in Sagittarius, and the version of Sagittarius in our Western system includes 17 stars after thresholding at magnitude 4.5. Intuitively, the teapot matches Sagittarius fairly well, and the function in Equation \ref{match} assigns a match of 0.47 between the teapot ($a$) and Sagittarius ($r$). If we used an alternative match function where the numerator included penalties for both $|a \setminus r|$ and $|r \setminus a|$, then the match between the teapot and Sagittarius would be 0.

The match between asterism $a$ and an entire system of asterisms $S$ is defined as
\begin{align}
\text{match}(a,S) &= \max_{r \in S} \left( \text{match}(a, r) \right).
\label{matchsystem}
\end{align}
Equation~\ref{matchsystem} captures the idea that $a$ matches $S$ well if there is at least one asterism $r$ in $S$ such that the match between $a$ and $r$ is high.
Finally, the human scores in Table 1 are calculated using
\begin{align}
\text{human}(a, \mathcal{S}_\text{human}) &= \text{mean}_{S \in \mathcal{S}_\text{human}  }\left(\text{match}(a, S) \right),
\label{humanscoremean}
\end{align}
where $\mathcal{S}_\text{human}$ is the set of all \systemcnt~systems in our data set. We computed scores for all asterisms in the entire data set, but to avoid listing variants of the same basic asterism, an asterism $a$ is included in Table 1 only if $\text{match}(a,r) < 0.5$ for all asterisms $r$ previously listed in the table. To establish a somewhat arbitrary threshold, we will say that an asterism recurs across cultures if it achieves a human score of 0.2 or greater. 28\% of the 605 asterisms in our data set meet this criterion, and these recurring asterisms are the subset most likely to be explained by principles of perceptual grouping.


Our data set is tilted towards cultures from the Northern Hemisphere and cultures with historical relationships to the Western system, and both factors may have distorted the human scores in Table 1. To adjust for this imbalance we sorted the 27 cultures into six geographic regions that are listed in the caption of Figure 3. We then computed weighted human scores for each asterism by replacing the mean in Equation~\ref{humanscoremean} with a weighted mean that gives equal weight to each of the six geographic regions.  Weighted human scores are included in Table 1, and these scores suggest that southern asterisms including the Southern Pointers and Corona Australis deserve to be listed alongside the ten recurring asterisms singled out at the top of Figure~\humanmodelsystem A. Regardless of whether we consider human scores or weighted human scores, we find that convergences in asterisms across cultures go beyond the handful of prominent examples typically cited in the literature.



\section{A computational model of the grouping of stars into asterisms}

To explain shared patterns in the night sky, scholars from multiple disciplines have suggested that asterisms are shaped in part by universal perceptual principles, including the principle that bright objects are especially salient, and that nearby objects are especially likely to be grouped~\citep{metzger36, yantis92,hutchins08}. \citet[p 325]{yantis92}, for example, writes that ``certain stellar configurations are `seen' by virtually all cultures (e.g. the Big Dipper)'', and that `these constellations are universal in that they satisfy certain of the classic Gestalt laws of proximity, good continuation, similarity (in brightness) and Pragnanz.'' Claims that Gestalt principles account for star grouping across cultures are mostly anecdotal, but the principles themselves have been studied in detail by 
psychologists~(Wagemans et al., 2012a, 2012b; Elder, 2015)\nocite{wagemans12I,wagemans12II,elder15} and have inspired the development of formal models of perceptual grouping~\citep{comptonl93, kubovyhw98, dry09, vandenberg98, froyenfm15,imzh16}. We build on this tradition by using a computational model (the Graph Clustering model, or GC model for short) to explore the extent to which the factors of brightness and proximity account for asterisms across cultures. 

The GC model constructs a graph with the stars as nodes, assigns strengths to the edges based on proximity and brightness, and thresholds the graph so that only the $n$ strongest edges remain.  Figure~\humanmodelsystem B shows the model graph when the threshold $n$ is set to 320. The connected components of this thresholded graph are the asterisms formed by the model.  There is a strong resemblance between these model asterisms and the consensus system in Figure~\humanmodelsystem A. The model picks out groups that correspond closely to the ten frequently-occurring asterisms highlighted in the inset panels of Figure~\humanmodelsystem A. Beyond these ten asterisms the model also picks out the Southern Pointers, the teapot in Sagittarius, the head of Draco, the head and stinger of Scorpius,  Lyra, the sickle in Leo, the shaft of Aquila, and more.  Table~\modelasterism~lists all groups found by the model and indicates which of them are similar to human asterisms attested in Table~\commonasterism.

\begin{figure}[t]
\begin{center}
\includegraphics[width=5in]{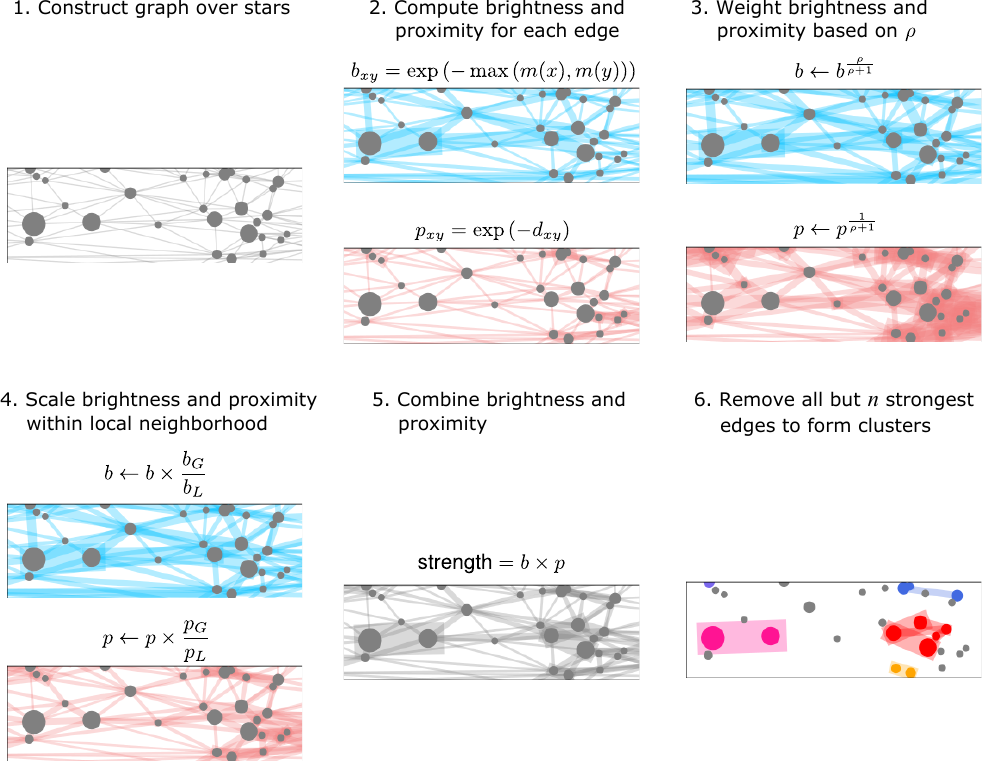}
\caption{Steps carried out by the graph clustering (GC) model. Each step is illustrated using a region of the sky that includes the Southern Cross and the Southern Pointers. $b_{xy}$ and $p_{xy}$ denote brightness weights (blue) and proximity weights (red) associated with the edge between $x$ and $y$. $m(x)$ and $m(y)$ are the apparent magnitudes of stars $x$ and $y$, and $d_{xy}$ is the angular separation between these stars. $b_G$ denotes the median brightness weight across the entire graph, $b_L$ denotes the median brightness weight within 60\protect\textdegree~of a given edge, and $p_G$ and $p_L$ are defined similarly.  In steps 2 through 6 edge widths are proportional to edge weights.}
\label{modeldemo}
\end{center}
\end{figure}

The steps carried out by the model are summarized by Figure \modeldemo. The first step is to construct a graph over stars. Existing graph-based clustering models typically operate over a graph corresponding to a minimal spanning tree~\citep{zahn71} or Delaunay Triangulation~\citep{ahuja82,vandenberg98}, and the GC model uses the union of three Delaunay triangulations defined over stars with apparent magnitudes brighter than 3.5, 4.0 and 4.5. Delaunay-like representations are hypothesized to play a role in early stages of human visual processing~\citep{dry09}, and combining Delaunay triangulations at multiple scales ensures that the resulting graph includes both edges between bright stars that are relatively distant and edges between fainter stars that are relatively close. The second step assigns a brightness and proximity to each edge. For an edge joining two stars, proximity is inversely related to the angular distance between the stars, and brightness is based on the apparent magnitude of the fainter of the two stars. The third step weights brightness and proximity based on a parameter $\rho$. For all analyses we set $\rho=3$, which means that brightness is weighted more heavily than proximity. The fourth step scales brightness and proximity so that the distribution of these variables within a local neighborhood of 
60\textdegree~is comparable with the distribution across the entire celestial sphere. Scaling in this way allows the impact of brightness and proximity to depend on the local context. For example, the Southern Cross lies in a region that contains many stars in close proximity, and we propose that stars need to be especially close to stand out in this context. Previous psychological models of perceptual grouping incorporate analogous local scaling steps~\citep{comptonl93,vandenberg98}, and the neighborhood size of 60\textdegree~was chosen to match the extent of mid-peripheral vision. The fifth step multiplies brightness and proximity to assign an overall strength to each edge, and the final step thresholds the graph so that only the strongest $n$ edges remain.  

We compared the GC model to several alternatives, including variants that lack one of its components, and variants that rely on either collinearity or good continuation in addition to brightness and proximity. We also evaluated a pair of models that rely on k-means clustering, and the CODE model of perceptual grouping~\citep{comptonl93}. The results reveal that the GC model performs better than all of these alternatives, and full details are provided in the supplemental material. Although adding good continuation to the model did not improve its performance, future work may be able to improve on our efforts in this direction. For example, the current model combines Corona Borealis with an extraneous star and does not connect the tail of Scorpius into a single arc, and finding the right way to incorporate good continuation may resolve both shortcomings.

Each human asterism can be assigned a score between 0 and 1 that measures how well it is captured by the GC model.  For this purpose we created a set $\mathcal{S}_{\text{GC}}$ that includes model systems for all values of the threshold parameter $n$ between 1 and 2000. 
The model score for each human asterism $a$ is defined as
\begin{align}
\text{modelscore}(a, \mathcal{S}_{\text{GC}}) &= \max_{S \in \mathcal{S}_{\text{GC}}} \left( \text{match}(a, S) \right),
\label{modelscore}
\end{align}
which is 1 if $a$ belongs to some system in $\mathcal{S}_{\text{GC}}$.  This approach makes it possible for nested asterisms (e.g.\ Orion's belt and Orion) to both be captured by the model, even though a single setting of $n$ could capture at most one of these asterisms. Model scores are included in Table~\commonasterismmt , and we will say that
an asterism is captured by the model if it achieves a model score of 0.2 or higher. By this criterion 98\% of the asterisms that recur across cultures and 80\% of the entire data set are captured by the model.

Distributions of model scores for each culture in our data set are plotted in Figure~\perculturescores . The model accounts for some cultures well --- for example, 13 of 20 Arabic asterisms, 
19 of 38 Marshall Islands asterisms and 55 of 161 Chinese asterisms are captured perfectly by the model.
The systems captured well by the model are drawn from a diverse set of geographical regions, suggesting that genealogical relationships between cultures are not enough to explain the recurring patterns captured by the model.  Yet there are also many asterisms that are not captured by the model, and the Chinese and Western systems in particular both include many asterisms with a model score of 0. Both systems partition virtually all of the visible sky into asterisms, and achieving this kind of comprehensive coverage may require introducing asterisms (including Herschel's ``innumerable snakes'') that do not correspond to natural perceptual units. 

\begin{figure}[t]
\begin{center}
\includegraphics[]{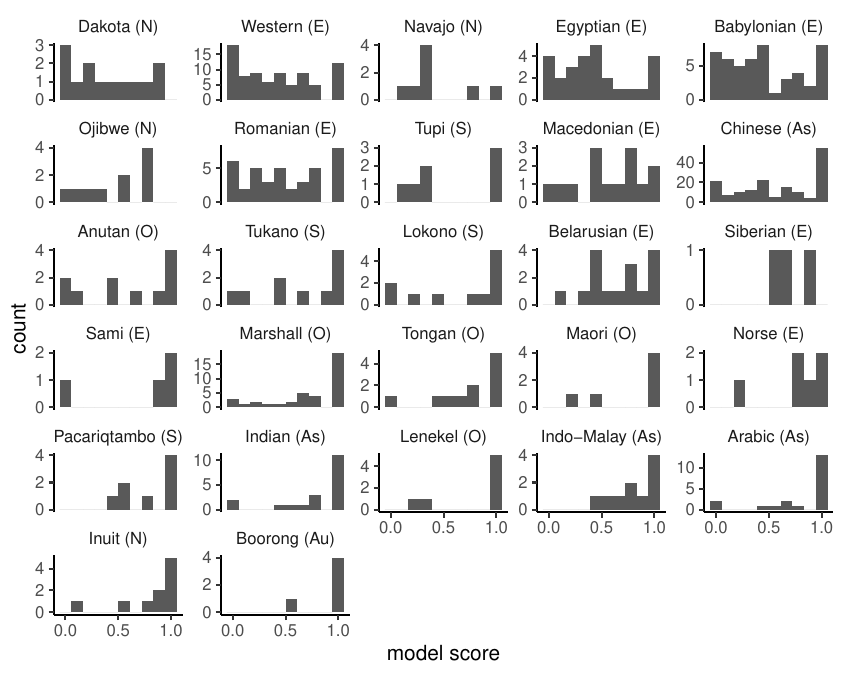}
\caption{Model results for individual cultures in our data set. Scores of 1 indicate asterisms that are perfectly captured by the GC model for some value of the threshold $n$, and each distribution includes scores for all asterisms that remain after filtering at a stellar magnitude of 4.5. The cultures are ordered based on the means of the distributions, and panel labels show the geographic regions used to compute the weighted scores in Table 1: AS (Asian), AU (Australian), E (European), N (North American), O (Oceanic), S (South American).}
\label{perculturescores}
\end{center}
\end{figure}

Although some attested asterisms missed by the GC model will probably resist explanation by any model of perceptual grouping, others can perhaps be captured by extensions of the model.  For example, the model tends not to group stars separated by a relatively large distance. As a result it misses the lower arm of the Northern Cross (Cygnus) and misses the Great Square of Pegasus entirely. These errors could perhaps be addressed by developing a multi-scale approach that forms groups at different levels of spatial resolution~\citep{froyenfm15,estradae06}. Another possible extension is to incorporate additional grouping cues such as symmetry and parallelism, which are known to influence human judgments~\citep{feldman07,machilsenpw09} and have been explored in previous computational work~\citep{jacobs03,stahlw08}.

In addition to scoring each system in our data relative to the GC model, we also examined how closely each system resembles other systems in our data set (see Figure \perculturescoresvsothers ). The system most different from all others is the Chinese system, which includes more than 300 asterisms, many of which are small and have no counterparts in records for other cultures.  In future work, the model may prove useful for evaluating hypotheses about historical relationships between systems from different cultures~\citep{gibbon64,berezkin05}. For example, the model could be used to ask whether Oceanic constellations are more similar to Eurasian constellations than would be expected based on perceptual grouping alone. 

\section{Discussion}

For around a century, constellations have been informally used by Gestalt psychologists and their successors to illustrate basic principles of perceptual grouping~\citep{kohler29,metzger36}.
To our knowledge, however, our work is the first to systematically explore the extent to which perceptual grouping can account for constellations across cultures. We began by asking which asterisms appear frequently across cultures, and our data suggest that lesser-known asterisms such as Delphinus and the head of Aries should be included alongside more familiar asterisms such as the Pleaides and the Big Dipper. We then presented a computational analysis which suggests that perceptual grouping based on brightness and proximity is enough to account for many of the asterisms that recur across cultures. Previous discussions of convergence in asterisms across cultures typically focus on a handful of familiar examples including the Pleiades and the Big Dipper, but our work suggests that similarities in asterisms across cultures go deeper than previously recognized. 

Our computational model aimed to explain how the night sky is clustered into groups of stars but did not address how the stars in each group are organized into figures. Our approach therefore complements the approach of \citet{dry09}, who focused on the organization of star groups into figures but did not explain how these star groups might initially have been picked out of the full night sky. Both of these approaches use a Delaunay triangulation to capture proximity relationships between stars, and future work may be able to combine them into a single computational model that both picks out groups of stars and organizes them into figures. 

We focused throughout on similarities in star groups across cultures, but there are also striking similarities in the names and stories associated with these groups~\citep{gibbon64,baity73,culver08}. For example, in Greek traditions Orion is known as a hunter pursuing the seven sisters of the Pleiades, and versions of the same narrative are shared by multiple Aboriginal cultures of Australia~\citep{johnson11,leamanh19}. Our work suggests that perceptual grouping helps to explain which patterns of stars are singled out for attention but understanding the meanings invested in these asterisms requires a deeper knowledge of history, cognition and culture.

\section{Author Contributions}

C.K.\ and D.W.H.\ compiled the data, and C.K.\ implemented the models and analyses and wrote the paper.  All authors discussed the models and analyses and commented on the manuscript.

\bibliographystyle{apalike}
\bibliography{stars}

\end{document}


\baselineskip24pt

\maketitle 

\setcounter{table}{0}
\renewcommand{\thetable}{S\arabic{table}}

\renewcommand{\thefigure}{S\arabic{figure}}

\singlespacing
\tableofcontents
\doublespacing

\section{Cross-cultural data}

Appendix~\ref{systemplots} shows asterisms for all \systemcnt~cultures in our data set. The majority of the systems were drawn from the Stellarium software package, and we compiled the remainder using sources given in the figure captions in Appendix~\ref{systemplots}.  Stellarium includes multiple systems for some cultures: for example, there are early and later versions of the Babylonian sky culture, and three versions of the Chinese sky culture. In cases like these we removed all but a single representative of each culture. We also removed a number of additional Stellarium systems for reasons documented in Table~\ref{stellariumexclusions}.  Our final data set includes 22 of the 42 Stellarium systems available as of May 25, 2020.

\begin{table*}[]
\begin{center}
\begin{tabular}[t]{lp{4.8in}}
 \textbf{Sky culture} & \textbf{Reason for exclusion} \\
Almagest & Lists 48 constellations of the Greeks, which are the source of the Western system \\
Arabic & Based on the 48 constellations of the Greeks, which are the source of the Western system \\
Armintxe & Identifications not sufficiently grounded in the published literature\\
Aztec & Identifications not sufficiently grounded in the published literature\\
Boorong & Already included in the data set\\
Chinese contemporary & Only one Chinese system is included\\
Chinese medieval & Only one Chinese system is included \\
Hawaiian starlines &  Identifications not sufficiently grounded in the published literature \\
Indian&  Already included in the data set\\
Japanese moon stations &  Identifications not sufficiently grounded in the published literature \\
Kamilaroi & Includes names of single stars only\\
Korean &  Closely related to the Chinese system\\
Maya & Identifications not sufficiently grounded in the published literature \\
Mongolian &  Identifications not sufficiently grounded in the published literature \\
Northern Andes &  Identifications not sufficiently grounded in the published literature \\
Sardinian &  Identifications not sufficiently grounded in the published literature \\
Seleucid &  Only one Babylonian system is included\\
Western (Sky \& Telescope) &  Only one Western system is included \\
Western (Hlad) &  Only one Western system is included \\
Western (Rey) &  Only one Western system is included \\
\end{tabular}
\caption{Stellarium systems excluded from our analysis.}
\label{stellariumexclusions}
\end{center}
\end{table*}

\afterpage{\clearpage}

\begin{figure}
\begin{center}
\includegraphics[]{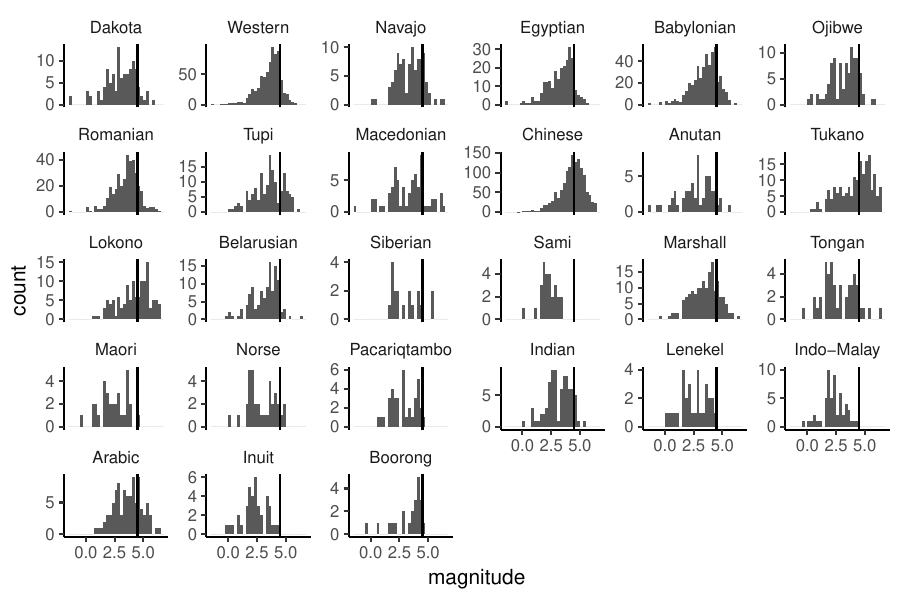}
\caption{ Distributions of star magnitudes for all systems in our data set. The vertical line in each plot shows the threshold value of 4.5, and the order of the systems matches Figure \perculturescores .}
\label{filtering}
\end{center}
\end{figure}

Stellar magnitude is conventionally measured using a scale on which fainter stars have higher magnitudes. Before carrying out our analyses we pre-processed each system by removing stars fainter than 4.5 in magnitude then removing all asterisms that included no stars or just one star after filtering.  For example, the constellation Mensa is removed from the Western system because the brightest star in this constellation has a magnitude of 5.08. Figure~\ref{filtering} shows the distribution of magnitudes for each system in our data set. Filtering at 4.5 removes around 25\% of the stars across the entire set of systems, but the proportion of faint stars varies across systems. Nearly 50\% of the stars in the Tukano system have magnitudes greater than 4.5, but for around half of the systems, 10\% or fewer of the stars have magnitudes greater than 4.5. Many of the cultures in our data have systems of asterisms that have not been documented in full, and the ethnoastronomical accounts that do exist naturally tend to focus on brighter stars. The distributions in Figure~\ref{filtering} therefore may not reflect the full set of asterisms that would be identified by expert astronomers from the cultures in question.

\begin{figure}
\begin{center}
\includegraphics[]{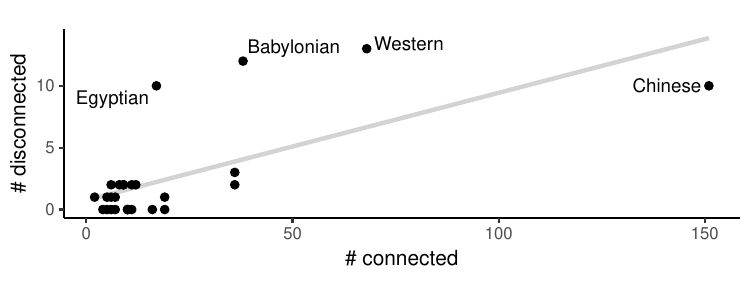}
\caption{  Counts of asterisms that are connected and disconnected with respect to the GC model graph. Each point corresponds to a system in our data, and systems with more than 3 disconnected asterisms have been labelled. }
\label{connectedness}
\end{center}
\end{figure}

The plots in Figures~\ref{anutan} through \ref{western} show all asterisms remaining after the initial filtering step. In each plot, the constellation figures are minimum spanning trees computed over the model graph using angular distance as the edge weight. In some cases an asterism does not correspond to a connected subset of the model graph, and in these cases minimal spanning forests are shown instead. For example, the Dakota system in Figure~\ref{dakota} includes a large asterism called ``Ki Inyanka Ocanku'' (The Race Track) that groups stars from Gemini, Canis Minor, Canis Major, Orion, Taurus and Auriga into a large circle. The scale of this asterism is larger than the scale of the model graph, and as a result Figure~\ref{dakota} shows the asterism as a collection of 5 disconnected components.  Figure~\ref{connectedness} shows the number of disconnected asterisms for each culture in our data. Around 90\% of asterisms in the filtered data are connected with respect to the model graph, and the Egyptian, Babylonian and Western systems stand out as having relatively high proportions of disconnected asterisms.

\section{Stellar data}
\label{stellardata}

We used stellar data from version 5.0 of the Yale Bright Star catalog~\cite{hoffleitj91},
which includes information about magnitude and position (right ascension and declination) for 9110 stars. Star positions in our data use J2000 coordinates, and are therefore correct for Jan 1, 2000. Star positions change over time due to precession, nutation, and proper motion. Precession and nutation do not affect our analyses because they do not affect the relative positions of stars with respect to each other.  Proper motion does affect the shapes of asterisms over long periods of time --- for example, Hamacher~\cite{hamacher12} describes how the shape of the Southern Cross has changed over the past 10,000 years. J2000 coordinates are suitable for our purposes because the systems analyzed in this paper are based on records from the last few thousand years, and because the stars with the greatest proper motion are at or below the threshold of visibility. 

We filtered the data to retain only stars brighter than 6.5 in magnitude, which roughly corresponds to the faintest magnitude still visible to the naked eye. Some stars (e.g.\ double stars) are very close to each other, and if two stars had positions that matched up to 5 decimal places we replaced them with a single star with magnitude equal to the combined magnitude of the pair. These initial pre-processing steps yielded a set of 8258 stars. For all analyses we filtered the set further and considered only the 918 stars brighter than 4.5 in magnitude. 

\section{Convergence in asterisms across cultures}

The most common asterisms across our data set are listed in Table~\ref{commonasterism} in Appendix~\ref{commonasterismsec}, which extends Table 1 by including additional asterisms. 



\section{The Graph Clustering (GC) Model}
\label{model}

\begin{figure}
\begin{center}
\includegraphics[width=\textwidth]{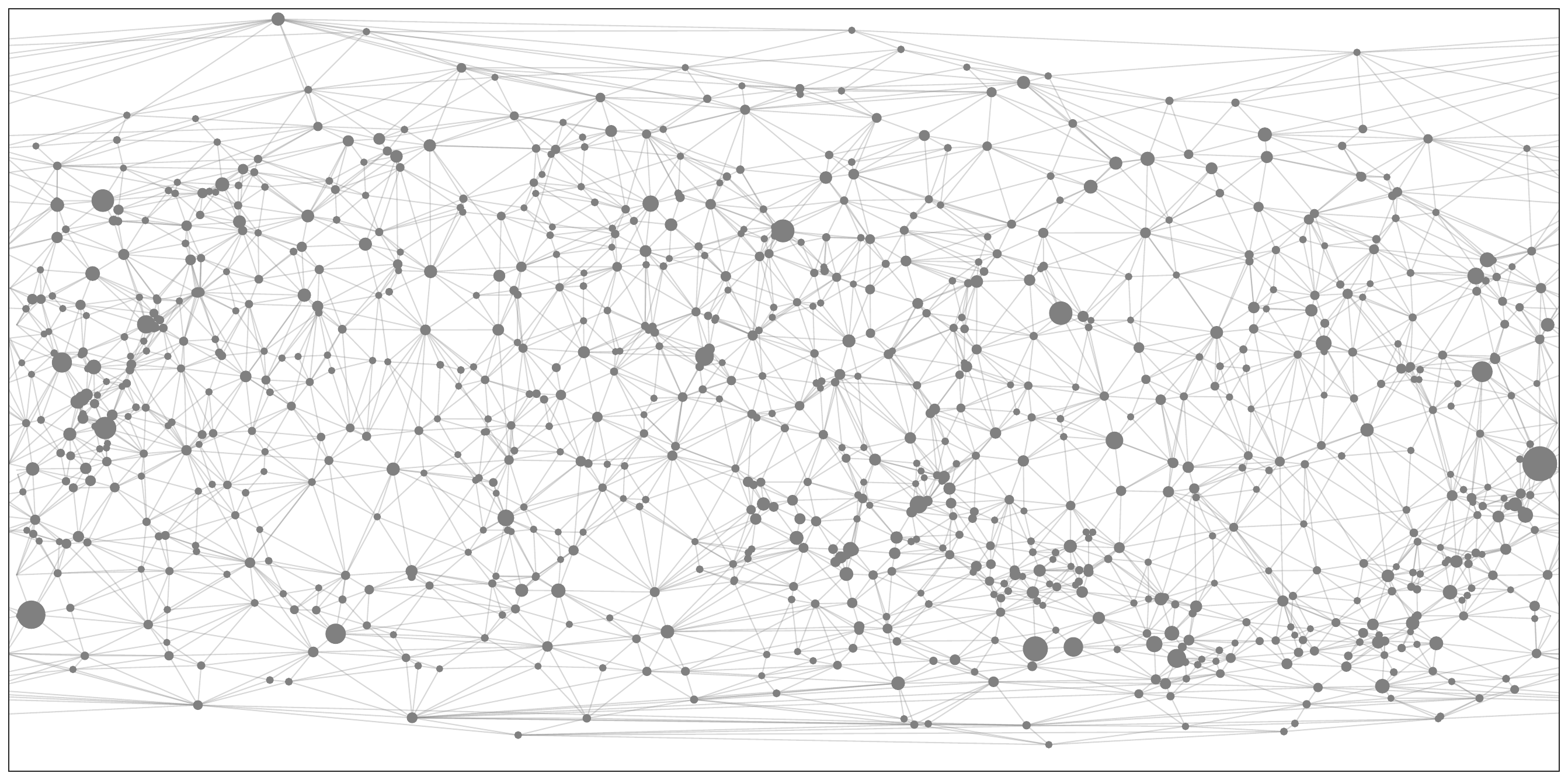}
\caption{Graph over stars used by the GC model.}
\label{modelgraph}
\end{center}
\end{figure}

\afterpage{\clearpage}

The Graph Clustering (GC) model begins by building a graph over the 918 stars that remained after pre-processing. We construct three Delaunay triangulations over stars with magnitudes less than 3.5, 4.0 and 4.5 respectively, and the final graph $G$ (shown in Figure~\ref{modelgraph}) is the union of all three.

An edge in $G$ that joins stars $x$ and $y$ is labelled with two attributes: $m_{xy}$, the apparent magnitude of the fainter of the two stars, and $d_{xy}$, the angular distance between the stars.\footnote{ $m_{xy}$ could be defined as the average magnitude of the two stars, but this approach favours stars that are connected to a bright star in the neighborhood graph, which seems undesirable. If proximity is effectively removed from the model by setting all proximities to 1, then the GC model reduces to a model that uses brightness alone. Our definition of $m_{xy}$ yields a reduced model that is equivalent to specifying a brightness threshold, then retaining only edges in the neighborhood graph connecting stars that are both brighter than the threshold. Defining $m_{xy}$ as an average yields a reduced model that seems less sensible.}
The two attributes are on different scales:  $m$ lies between -1.46 and 4.5, and $d$ lies between 0.1 and 41.6 degrees.
In both cases higher values are ``worse:'' distant stars are relatively unlikely to be grouped, and faint stars are relatively unlikely to be included in groupings. We convert each magnitude $m$ to a brightness $b$, and each distance $d$ to a proximity $p$:
\begin{align}
\begin{split}
b_{xy} &= \exp{(- m_{xy})} \\
p_{xy} &= \exp{ (-d_{xy})} \\
\end{split}
\label{transformation}
\end{align}
The negative exponential transformation means that higher values are now ``better.''\footnote{Brightness could be defined as flux (i.e.\ $b_{xy} = 10^\frac{-m_{xy}}{2.5}$) but the natural exponential formulation is simpler and equally good for our purposes. Changing from base $e$ to base 10 leaves the model unchanged if the parameter $\rho$ is adjusted accordingly.} 
We then weight brightness $b_{xy}$ and proximity $p_{xy}$ based on a parameter $\rho$:
\begin{align}
\begin{split}
b &\leftarrow b ^{\frac{\rho}{\rho+1}} \\
p &\leftarrow p ^{\frac{1}{\rho+1}} \\
\end{split}
\label{rho}
\end{align}
where we have dropped the subscripts of both $b_{xy}$ and $p_{xy}$.  When $\rho = 1$ proximity and brightness are weighted equally, and when $\rho > 1$ brightness is weighted more than proximity. When $\rho=0$ brightness is effectively discarded, and when $\rho = \infty$ proximity is effectively discarded.\footnote{Elder and colleagues have developed models of contour grouping that avoid free parameters like $\rho$ by leveraging natural scene statistics~\cite{elderkj03,estradae06}, and we evaluated a closely related model. This model relies on likelihood ratios that capture the probability that two stars belong to the same constellation as opposed to remaining ungrouped, and the key components of these likelihood ratios can be estimated from statistics computed over our data set such as the probability that two stars separated by a given distance (e.g.\ 5 degrees of visual angle) belong to the same asterism. The resulting model is appealing because it has no free parameters, but we found that it performed substantially worse than the GC model, perhaps because our data set is too small to allow reliable estimates of the relevant statistics. }

The next step is to scale the brightness and proximity values within a local neighborhood of 60\textdegree. For each edge $(s_1, s_2)$ joining stars $s_1$ and $s_2$, the local neighborhood $L$ is the subgraph of the full model graph $G$ that includes all stars that lie within 60\textdegree~of either  $s_1$ or $s_2$. The $p$ value of the edge $(s_1, s_2)$ is then scaled by the factor
\begin{equation}
\frac{p_G}{p_L} = \frac{\underset{e_g\in G}{\mathrm{median}} \{e_g(p)\}}
{\underset{e_l \in L}{\mathrm{median}} \{e_l(p)\}}
\label{scaling}
\end{equation}
where $e_g$ is an edge in the full model graph $G$, $e_l$ is an edge that lies within the local neighborhood $L$, and $e_i(p)$ is the $p$ value of edge $e_i$. Scaling $p$ in this way means that the distribution of $p$ values within any local neighborhood becomes comparable to the distribution over the entire graph. For example, consider a neighborhood that includes many close stars. Before scaling, most edges in the neighborhood will have high values of $p$. After scaling, only pairs of stars that are especially close relative to the neighborhood will have high values of $p$.  
The same approach in Equation~\ref{scaling} is used to scale the brightness values $b$. When scaling both attributes a neighborhood size of 60\textdegree~was chosen so that the neighborhood corresponds roughly to the extent of mid-peripheral vision. 

After scaling, the proximity and brightness values for each edge are combined multiplicatively to produce a single strength $s = bp$ for each edge. We then threshold the graph by removing all but the top $n$ edges in the graph, and the clusters returned by the model correspond to connected components of the thresholded graph. 

To assess the contribution made by different components of the GC model we will compare the  model to three variants.  First is a model that omits the local scaling step. This GC (no scaling) model can also be viewed as a variant in which neighborhood $L$ in Equation~\ref{scaling} expands to encompass the entire graph $G$.  The second and third variants set $\rho = 0$ and $\rho=\infty$ respectively, and we refer to them as the GC (no brightness) and GC (no proximity) models. These labels indicate whether or not brightness and proximity contribute to the final edge strengths $s$, but in both cases brightness and proximity are still used when constructing the original graph $G$: each Delaunay triangulation uses proximity, and combining the three triangulations means that only stars brighter than 4.5 in magnitude are included. 

\subsection{Fitting $\rho$}

The GC model has two parameters: $\rho$, which determines the relative contributions of proximity and brightness, and $n$, which determines the number of edges in the thresholded graph. We fit $\rho$ based on the idea that stars connected by strong edges in the model graph should be frequently grouped across cultures. The first step is to assemble a set of \emph{human edges} by computing minimum spanning trees (MSTs) for each asterism in our data set. All MSTs were computed over the model graph $G$ using raw angular distance as the edge weight. The human edges included all edges in these MSTs, and the \emph{human strength} of each edge was defined as the number of times it appeared in the set. For example, the edge joining the Southern Pointers appears in 9 of the MSTs and therefore has a human strength of 9. 

We then assembled a set of \emph{model edges} that included all of the strongest edges according to the model. The model edges include all edges in the MST of $G$,  where the MST is computed using model strengths $s = bp$ rather than angular distance. Parameter $\rho$ can then be set to the value that maximizes the correlation between the model edge strengths and the human edge strengths. The best values of $\rho$ for the GC and GC (no scaling) models were 3.5 and 3.2, which yield correlations of 0.72 and 0.66 respectively. Both models achieve almost identical correlations for $\rho=3$, and for simplicity we set $\rho=3$ for all subsequent analyses. Because $\rho > 1$, this setting means that the edge strengths in the model are influenced more by brightness than by proximity.

\begin{figure}
\begin{center}
\includegraphics[width=5in]{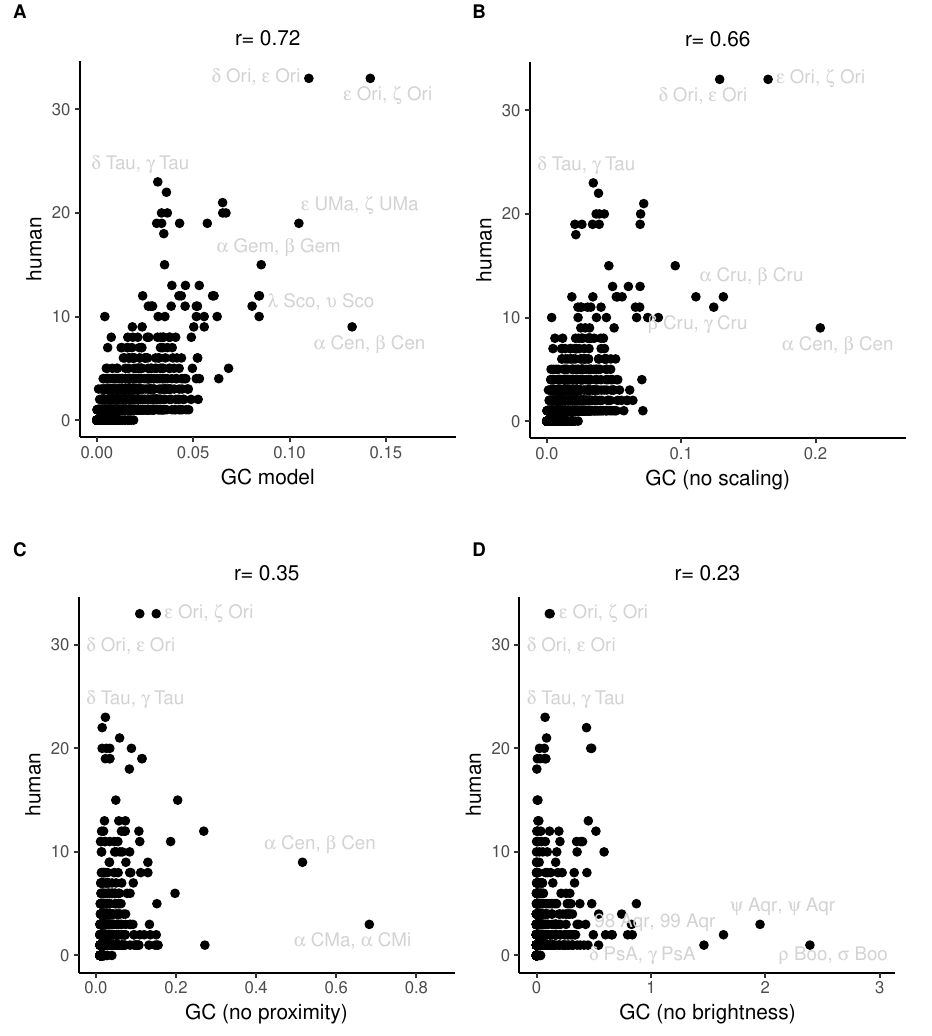}
\caption{ Strengths of star pairs according to the cross-cultural data and four models: (\captionletter{A}) the GC model, (\captionletter{B}) the GC model without local scaling, (\captionletter{C}) the GC model with edge strengths based on brightness only, and (\captionletter{D}) the GC model with edge strengths based on proximity only. In each panel each point corresponds to a pair of stars joined by an edge in the model graph, and selected edges are labelled in gray. The y-axis of each panel shows counts across the entire data set, and the x-axis shows strengths according to the model.}
\label{humanmodelscatterrest}
\end{center}
\end{figure}

Figure~\ref{humanmodelscatterrest}a compares human strengths with strengths according to the GC model.  The two edges with greatest human strengths join the three stars in Orion's belt ($\delta$, $\epsilon$ and $\zeta$ Ori), and these edges have human strengths of 33 because some of the \systemcnt~systems in our data include Orion's belt in more than one asterism. The same two edges are the strongest and third-strongest edges according to the model, and the second strongest model edge joins the Southern Pointers ($\alpha$ and $\beta$ Cen). This second edge appears as an outlier in Figure~\ref{humanmodelscatterrest}a, and one possible reason is that these stars lie relatively far south and our data set is tilted towards cultures from the Northern Hemisphere.

Corresponding plots for the three model variants are shown in Figure~\ref{humanmodelscatterrest}.  All three perform worse than the full GC model, suggesting that local scaling helps to account for human groupings and confirming that both brightness and proximity are important.

\subsection{Collinearity and Good Continuation}

The GC model variants evaluated in Figure~\ref{humanmodelscatterrest} are all created by subtracting elements from the model, but elements can also be added to the model. A natural question is whether the model can be improved by incorporating additional Gestalt cues such as collinearity, good continuation, symmetry, and parallelism~\cite{feldman07, machilsenpw09,elder15}. Cues like these have been informally proposed to influence star grouping across cultures. For example, Macpherson~\cite{macpherson81} suggests that the asterisms formed by the Boorong of south east Australia reflect a sensitivity to triads of stars that are close to collinear. 

Here we consider models that incorporate collinearity and good continuation, and leave other Gestalt cues for future work. We focus on collinearity and good continuation because our graph-based approach allows simple formulations of both principles. The specific formulations described here closely follow the prior work of van den Berg~\cite{vandenberg98}, who used a graph-based approach to model people's judgments about the organization of random dot patterns. 

\begin{figure}
\begin{center}
\includegraphics{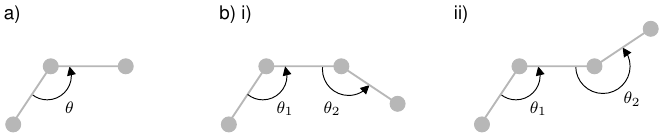}
\caption{a) The collinearity score for a triad is high if  the angle between the three stars ($\theta$)  is close to 180\textdegree. b) The good continuation score for a tetrad is high if the angles $\theta_1$ and $\theta_2$ are very similar. Good continuation is higher if an agent walking from one end of the tetrad to another makes two turns in the same direction (i) than if the agent performs a zig zag (ii).} 
\label{goodcfig}
\end{center}
\end{figure}

Collinearity is defined with respect to triads of stars (Figure~\ref{goodcfig}a), and we consider all triads that can be formed over the neighborhood graph. The collinearity score $c$ for a triad is high if the angle between the three stars is close to 180\textdegree: 
\begin{equation}
c_{hij} = \exp{(\lvert \pi - \angle{hij} \rvert)},  
\label{collinearity}
\end{equation}
where $\angle{hij}$ is the angle in radians between the three stars. The collinearity score for each triad is then combined with a GC score $s_{hij}$ defined as 
\begin{equation}
s_{hij} = \min{(s_{hi}, s_{ij})},
\label{gctriadscore}
\end{equation}
where $s{ij}$ is the strength of the edge between $i$ and $j$ according to the GC model.\footnote{Taking a minimum rather than an average in Equation~\ref{gctriadscore} is supported by the same general argument previously used to explain why $m_{xy}$ is defined using a minimum.}  Before combining the scores they are weighted based on a parameter $\tau$, and the final combined score for each triad is $ s^{\frac{\tau}{\tau + 1}} c^{\frac{1}{\tau + 1}}$, where $s$ and $c$ are the scores in Equations~\ref{collinearity} and \ref{gctriadscore}. The output of the model is generated by ranking all triads using their combined scores, then returning those that lie above some threshold. 

A model that incorporates good continuation can be defined analogously over tetrads. The good continuation score $g$ for a tetrad is high if the difference between the two angles in the tetrad ($\theta_1$ and $\theta_2$ in Figure~\ref{goodcfig}b) is close to zero: 
\begin{equation}
g_{hijk} = \exp{(\lvert \angle{hij} - \angle{ijk} \rvert)},  
\label{goodc}
\end{equation}
where $\angle{ijk}$ is the angle in radians between stars $i$, $j$ and $k$. The good continuation score for each tetrad is then combined with a GC score $s_{hijk}$ defined as 
\begin{equation}
s_{hijk} = \min{(s_{hi}, s_{ij}, s_{jk})},
\label{gctetradscore}
\end{equation}
where $s{ij}$ is an edge strength according to the GC model. The good continuation and GC scores for each tetrad are combined according to a weighting parameter $\tau$, and the output of the model is generated by ranking all tetrads based on their combined scores and returning those that lie above some threshold.

For both models, we fit the weighting parameter $\tau$ using the same approach used to fit the weighting parameter $\rho$ of the GC model. This procedure involves comparing human edge strengths with model edge strengths, and identifying the parameter that maximizes the correlation between the two. For the collinearity model, the best fitting parameter was $\tau = \infty$, which means that collinearity scores are effectively discarded and the resulting model is identical to the GC model.\footnote{We considered $\tau \in \{0,1,2,3,4,5,7,9, \ldots, 27,30,35,40,80,\infty \}$.  } A similar result was obtained for the good continuation model. These findings suggest that incorporating collinearity and good continuation does not improve the GC model's ability to account for our data.

A useful perspective on these negative results is given by the work of van den Berg (1998)\nocite{vandenberg98}, who evaluated models similar to ours using data from a task in which participants organized displays of random dots into groups. van den Berg concluded that good continuation had a relatively small influence on responses to his task, and found no evidence that people's judgments were influenced by collinearity. These findings suggest that the effects of collinearity and good continuation on star grouping may be relatively subtle.

Despite our results, we continue to believe that good continuation (although perhaps not collinearity) has influenced the formation of asterisms across cultures. In particular, it seems likely that good continuation helps to explain why asterisms such as Corona Australis and Corona Borealis are picked out by multiple cultures. Future work can consider at least two ways to further explore the role of good continuation. First, Equation~\ref{goodc} is a reasonable first attempt, but alternative formulations may better capture the notion of good continuation. For example, an alternative approach might incorporate both angles and distances to capture the idea that good continuation is strongest when stars are relatively evenly spaced along a smooth contour.  Second, our analysis focused on relatively bright stars, but it seems possible that good continuation becomes especially important when considering stars that would otherwise be too faint to attract attention. Future work can aim to consider the entire set of visible stars, which might provide stronger statistical evidence for the role of good continuation.

\section{GC model results}
\label{modeleval}

Because we found no evidence that collinearity and good continuation improved the GC model, our subsequent evaluations will focus on the version of the model that includes brightness and proximity alone.  We began by testing that the idea that the most common asterisms in Table~\ref{commonasterism} should be relatively well captured by the model. To avoid having to choose a single value of the threshold parameter $n$, we created a set $\mathcal{S}_{\text{GC}}$ that includes model systems for all values of $n$ between 1 and 2000. 
Movie S1 includes a frame for each system and shows how model asterisms emerge as $n$ is increased.  We then computed model scores for each human asterism $a$ using Equation \modelscore~in the main text.  The model scores in Table~\ref{commonasterism} indicate that most of the common asterisms are captured fairly well by the model. The most notable exception is the Great Square of Pegasus.  Model scores for each culture in our data set are plotted in Figure \perculturescores.  Although Table~\ref{commonasterism} suggests that common asterisms are often captured by the model, Figure \perculturescores~shows that there are many less common asterisms that the model does not explain.

The model evaluations thus far do not depend on a specific setting of the threshold parameter $n$. The model system in Figure \humanmodel B, however, is based on setting $n=320$, and Table~\ref{modelasterism} in Appendix~\ref{secmodelasterism} lists all 124 asterisms in this system.  The column labeled \textbf{Score} indicates how well these asterisms match our data set, and this column is defined using
\begin{align}
\text{human score}(a, \mathcal{S}_\text{human}) &= \max_{S \in \mathcal{S}_{\text{human}}  } \left( \text{match}(a, S) \right),
\label{humanscoremax}
\end{align}
where $\mathcal{S}_{\text{human}}$ is the set of all \systemcnt~systems in our data set.  27 of the model asterisms are identical to asterisms from one or more cultures, and  105 of the model asterisms have scores of 0.2 or greater, indicating that they correspond at least partially to asterisms from at least one culture. Note that the scoring function is relatively strict, especially for larger asterisms with many variants that can be created by including or excluding fainter stars. For example, 
Figure \humanmodel B suggests that the model captures Orion relatively well, but the model version of Orion achieves a score of only 0.36. 

\section{Model comparisons}

Our model belongs to a family of graph-based clustering algorithms that rely on a graph defined over the items to be clustered~\cite{zahn71,ahuja82,vandenberg98}. The main alternative in the literature on perceptual grouping is the CODE model~\cite{vanoeffelenv82,comptonl93,comptonl99}
which uses a continuous spatial representation of the items to be clustered. A third possible approach is k-means clustering, which has been previously applied to the problem of grouping stars into asterisms~\cite{schreiner,xucz14}. We compared all three approaches using a set of three different scoring functions. Consistent with our previous analyses, the input to each model includes all stars brighter than 4.5 in magnitude.

\subsection{Scoring functions}

Suppose that $H$ (for \emph{human}) is a set of human clusters and $M$ is a set of model clusters. A good set $M$ should have high \emph{precision}: each cluster in $M$ should be similar to a cluster in $H$. A good set $M$ should also have high \emph{recall}: for each cluster $h$ in $H$ there should be some cluster in $M$ that is similar to $h$. We formalize precision and recall as follows:
\begin{align}
\text{precision}(M, H) = \frac{1}{|M|} \sum_{m \in M} \text{match}(m, H) 
\label{precision}
\end{align}
\begin{align}
\text{recall}(M, H) = \frac{1}{|H|} \sum_{h \in H} \text{match}(h, M) 
\label{recall}
\end{align}
where the $\text{match}(\cdot, \cdot)$ function is defined in Equation~\systemmatch~of the main text.

Precision and recall are typically combined using an $F$ measure:
\begin{align}
F_\beta = (1+\beta^2) \frac{\text{precision} \cdot \text{recall}}{ (\beta^2 \cdot \text{precision}) + \text{recall}  }
\label{Fmeasure}
\end{align}
The standard $F$ measure sets $\beta=1$, but we also consider an $F_{10}$ measure that sets $\beta=10$ and weights recall more heavily than precision. The $F_{10}$ measure captures the idea that a model system $M$ that includes just one or two clusters (i.e.\ recall is low) should not score highly regardless of how well the model clusters match attested clusters.

Our measures of precision and recall and their combination using a $F_\beta$ score are directly inspired by the literature on information retrieval. 
If $\text{match}(a, S)$ returned 1 if $a$ belonged to $S$ and 0 otherwise, then our formulations of precision and recall in Equations~\ref{precision} and \ref{recall} would be equivalent to the standard definitions. Our $\text{match}(\cdot, \cdot)$ function, however, is graded, which means that our formulations of precision and recall are extensions of the standard definitions.

Our third scoring function is the adjusted Rand index, which is a standard measure of the similarity between two partitions. Many of the cluster systems that we consider pick out a relatively small number of clusters against a background of unclustered stars. In order to apply the adjusted Rand index we assign all unclustered stars to an ``everything else'' category. 

Of the three scoring functions, the $F_{10}$ measure deserves the most attention. The standard $F$ measure (i.e. $F_1$) has the shortcoming of assigning high scores to model solutions with a very small number of clusters. The adjusted Rand index is undesirable because of the need to include an ``everything else'' category. We report results for both measures because they are standard in the literature, but will focus primarily on the $F_{10}$ measure.

In the information retrieval literature, instead of relying on a single $F$ measure models are sometimes compared by constructing an entire precision-recall curve for each model and scoring each model based on the area under its curve. Comparing models in this way is natural because each curve has the same support --- for each model, recall increases from zero (when the retrieval threshold is set sufficiently high) to one (when the retrieval threshold is set sufficiently low). Our situation is different: none of the models under consideration can produce a recall of 1, and in general the precision-recall curves for different models will have different supports, and may not even be functions (it is possible for two model solutions to have the same recall but different precisions). It is not clear how these curves should be compared, and we therefore focus on a single F measure (the $F_{10}$ measure) instead.

Each of the measures so far scores a model system $M$ relative to a single human system $H$. We evaluate a model solution relative to the full set $\mathcal{H}$ of systems for \systemcnt~cultures by computing the average score across this set.

\subsection{GC model}

\begin{figure}
\begin{center}
\includegraphics[width=\textwidth]{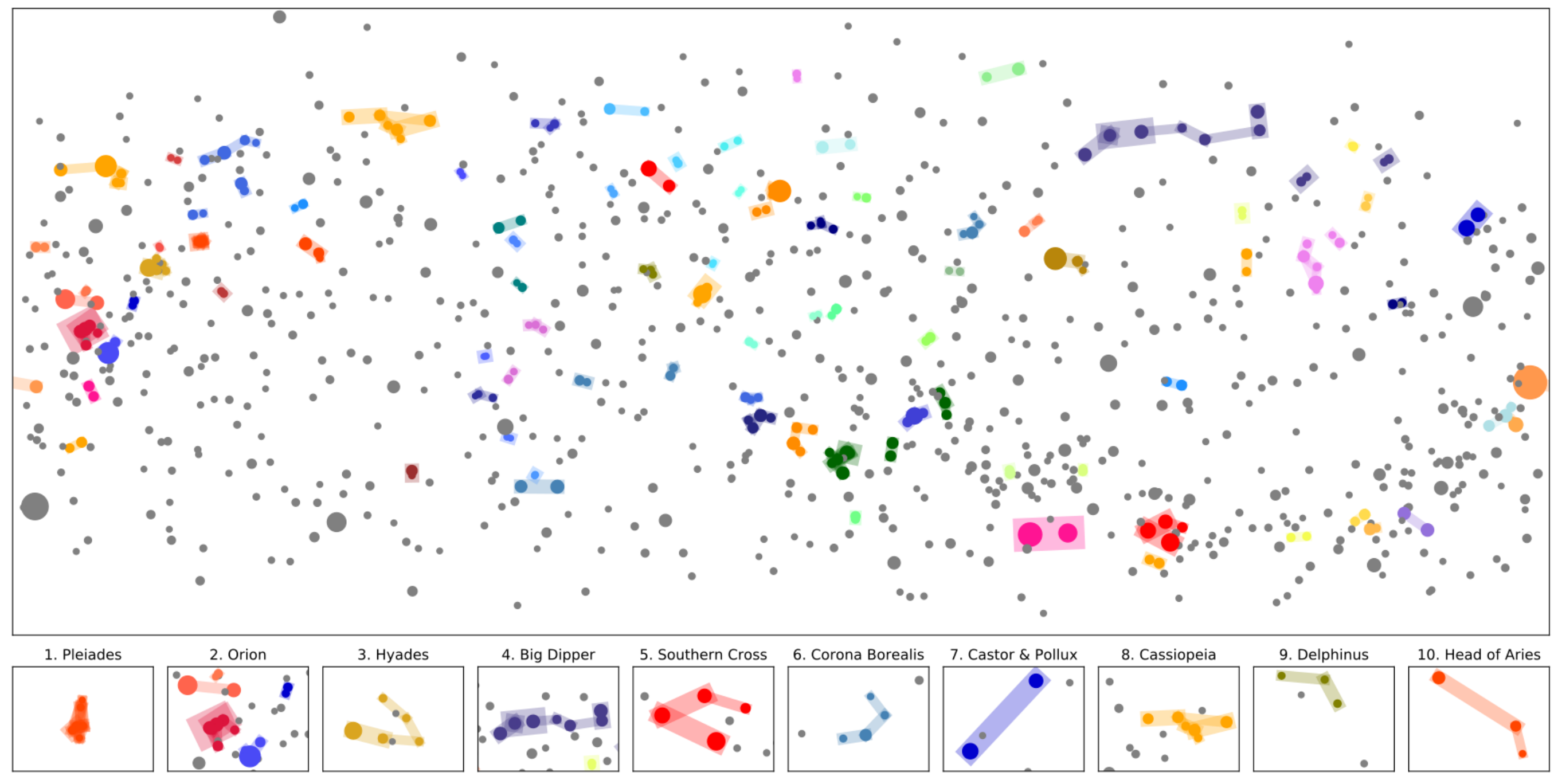}
\caption{Asterisms returned by the GC model ($n=165$).}
\label{model_n}
\end{center}
\end{figure}

\begin{figure}
\begin{center}
\includegraphics[width=\textwidth]{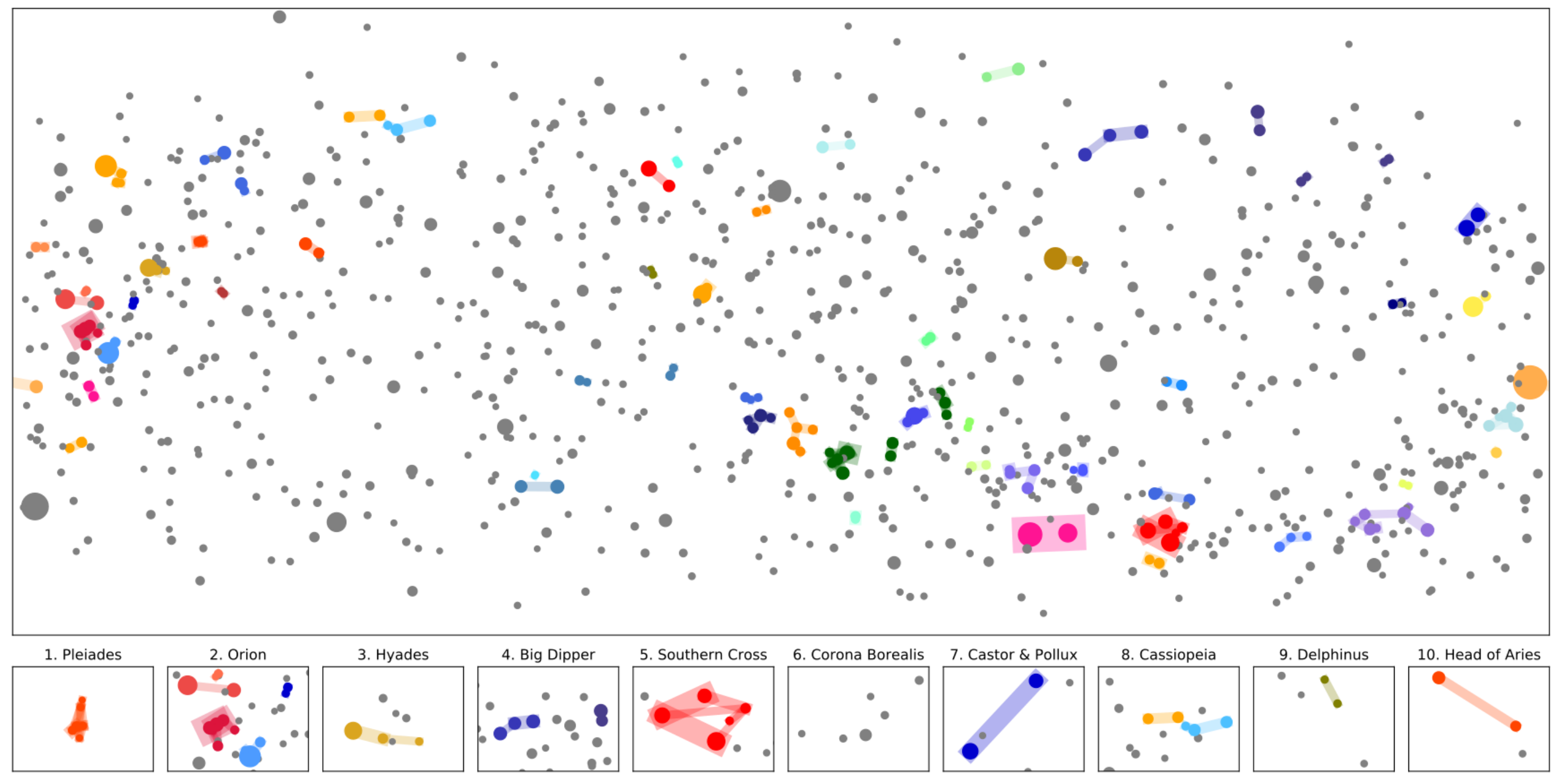}
\caption{Asterisms returned by the GC model (no scaling, $n=122$).}
\label{model_n_noscaling}
\end{center}
\end{figure}

\begin{figure}
\begin{center}
\includegraphics[width=\textwidth]{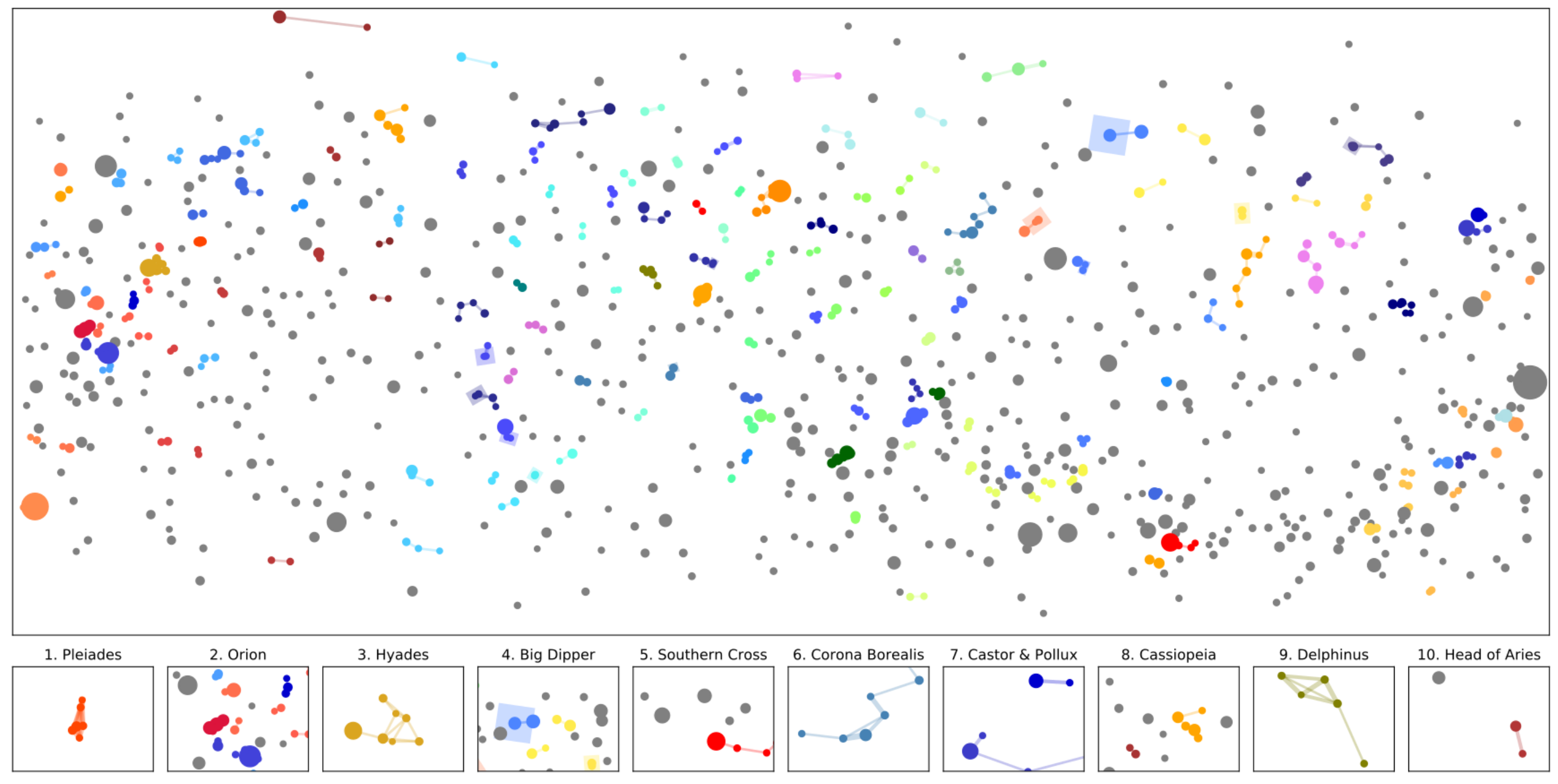}
\caption{Asterisms returned by the GC model (no brightness, $n=321$).}
\label{model_n_nobright}
\end{center}
\end{figure}

\begin{figure}
\begin{center}
\includegraphics[width=\textwidth]{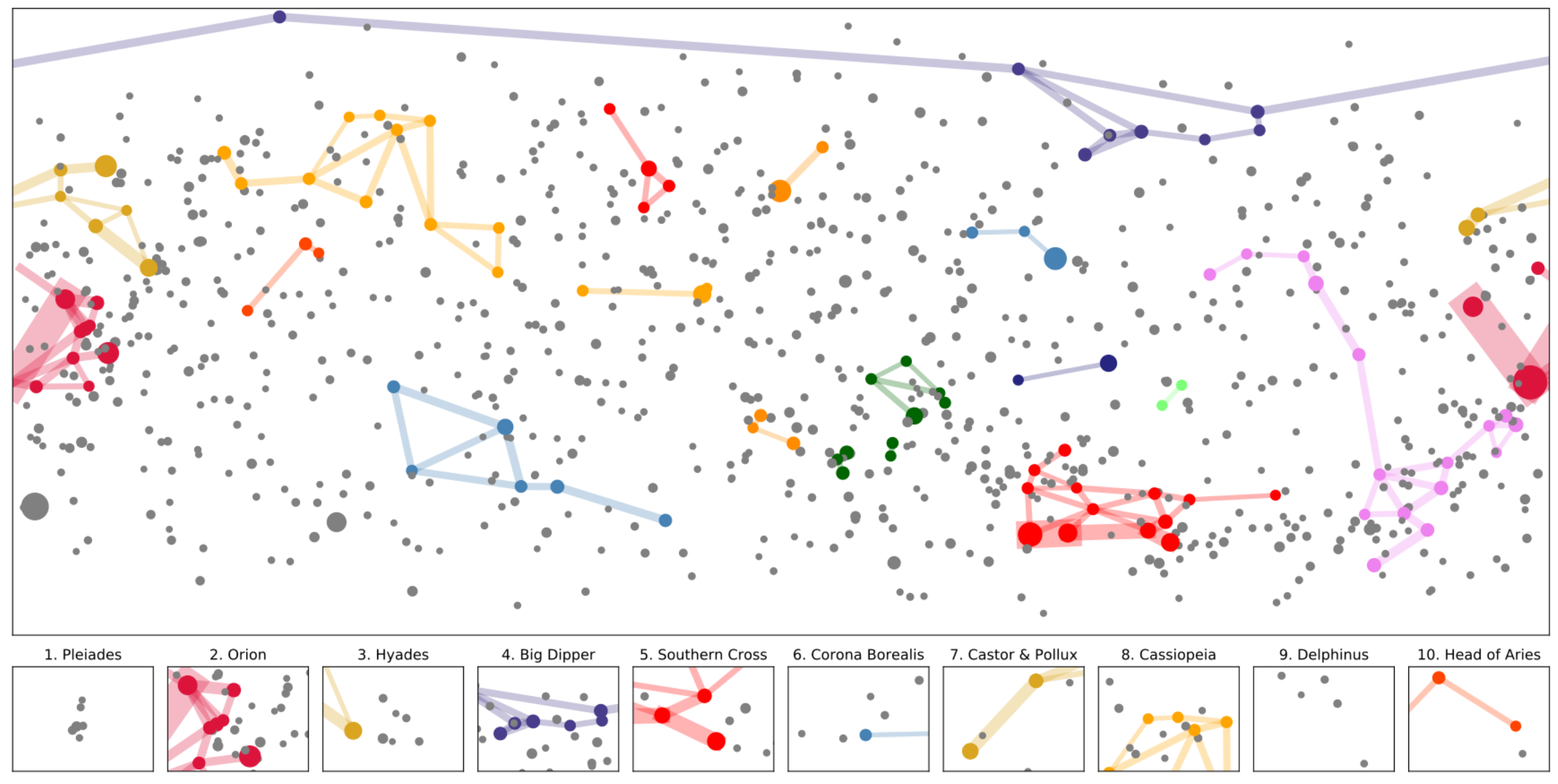}
\caption{Asterisms returned by the GC model (no proximity, $n=123$).}
\label{model_n_nodistance}
\end{center}
\end{figure}

The model includes two parameters: $\rho$, which controls the relative weights of brightness and proximity, and the threshold parameter $n$.  Previously $\rho$ was set to 3 based on the correlation analysis summarized by Figure~\ref{humanmodelscatterrest}a, and we retain that value here. The threshold is set to the value ($n=165$) that maximizes model performance according to the $F_{10}$ measure. In addition to the GC model we consider the three variants of the model previously evaluated in Figure~\ref{humanmodelscatterrest}, and the $n$ parameter is optimized separately for each one using the $F_{10}$ measure. Asterisms returned by all four models are shown in Figures~\ref{model_n} through \ref{model_n_nodistance}.

\subsection{CODE model}

\begin{figure}
\begin{center}
\includegraphics[width=\textwidth]{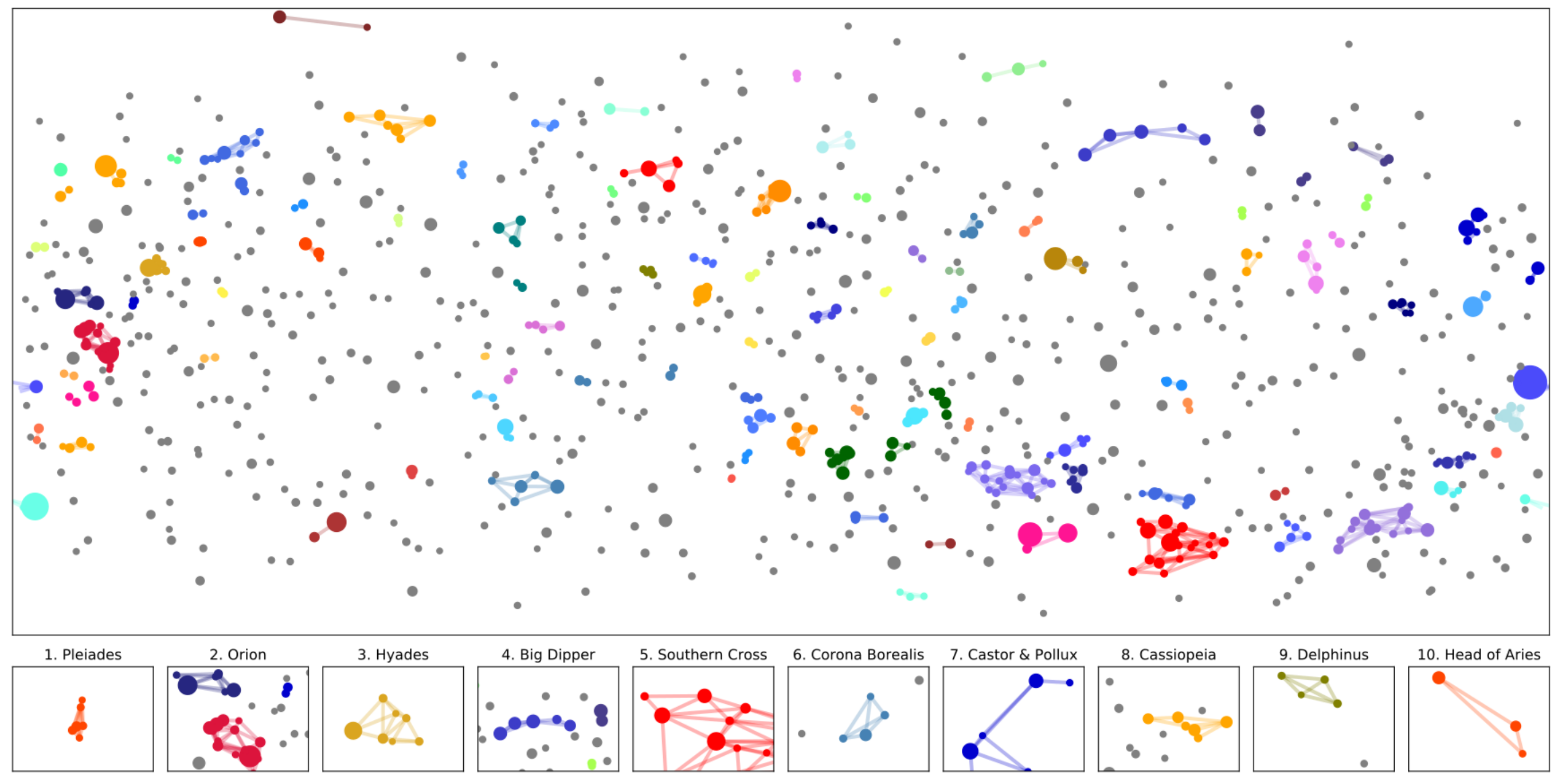}
\caption{Asterisms returned by the CODE model with rescaled Gaussian kernels,  local distances and the sum combination function ($t=0.86$, $\beta=0.86$, $h=0.5$ ).}
\label{model_code}
\end{center}
\end{figure}

The CODE model can be implemented by dropping a kernel function (e.g.\ a Gaussian) on each item, combining all of these kernel functions to form an ``activation surface,'' then cutting the activation surface at some threshold $t$ to produce clusters. Previous applications of the CODE model consider the problem of clustering a field of perceptually identical items, but for us the items are stars with different magnitudes. To capture the idea that brighter stars are more likely to be included in asterisms, we adapted the CODE model to allow taller kernel functions for brighter stars.  If two stars are extremely close to each other, the sum of the kernels on the two should be identical to a single kernel for a star with apparent magnitude equivalent to the two stars combined. To satisfy this condition we set kernel heights based on the flux (i.e.\ apparent brightness) of a star.  For a star with apparent magnitude $m$, the flux $F$ of the star in the visual band is 
\begin{equation}
F = F_0 \times 10^\frac{-m}{2.5}
\end{equation}
where $F_0$ is a normalizing constant. For our purposes we can drop the constant because scaling all kernels by a constant is equivalent to adjusting the threshold used by the CODE model.  We also introduce an additional parameter $\beta$ so that the height of the kernel on a star of magnitude $m$ is 
\begin{equation}
\left( 10^\frac{-m}{2.5} \right)^\beta
\label{kernelheight}
\end{equation}
When $\beta=1$ kernel height is proportional to flux, and when $\beta=0$ all stars have the same kernel height regardless of flux. 

Our formulation of the CODE model has two additional numeric parameters: the threshold $t$, and the nearest-neighbour coefficient $h$. The standard deviation of the kernel for star $i$ is
\begin{equation}
\sigma_i = hd_i 
\label{kernelsd}
\end{equation}
where $d_i$ is the distance between the star and its nearest neighbour. In addition to these numeric parameters Compton and Logan (1993) \nocite{comptonl93} consider several qualitative parameters of the model:
\begin{itemize}
\item \emph{Gaussian} vs \emph{Laplacian}: kernel functions may be Gaussian or Laplacian
\item \emph{sum} vs \emph{max}: kernels may be combined using a sum or a max function
\item \emph{local} vs \emph{global}: if \emph{global}, all distances $d_i$ in Equation~\ref{kernelsd} are replaced by the global mean of $d_i$
\item \emph{standard} vs \emph{rescaled}: if \emph{rescaled}, all kernel functions are rescaled to have the same height 
\end{itemize}
In our implementation the rescaling step specified by the fourth factor is carried out before adjusting the kernels for brightness as specified by Equation~\ref{kernelheight}. As a result, the kernels are equal in height at the end of the process only when rescaling is applied and $\beta=0$. 

We evaluated all 16 combinations of the four factors, and optimized the three numeric parameters ($\beta$, $t$ and $h$) separately for each combination using the $F_{10}$ measure. The optimization began with a grid search then used Powell's conjugate direction method initialized using the best values found in the grid search.  The best performing version used rescaled \emph{Gaussian} kernels, \emph{local} distances and the \emph{sum} combination function, and the best parameters for this model were $t=0.86$, $\beta=0.86$, and $h=0.5$.  The clusters returned by this model are shown in Figure~\ref{model_code}.

\subsection{k-means clustering}

\begin{figure}
\begin{center}
\includegraphics[width=\textwidth]{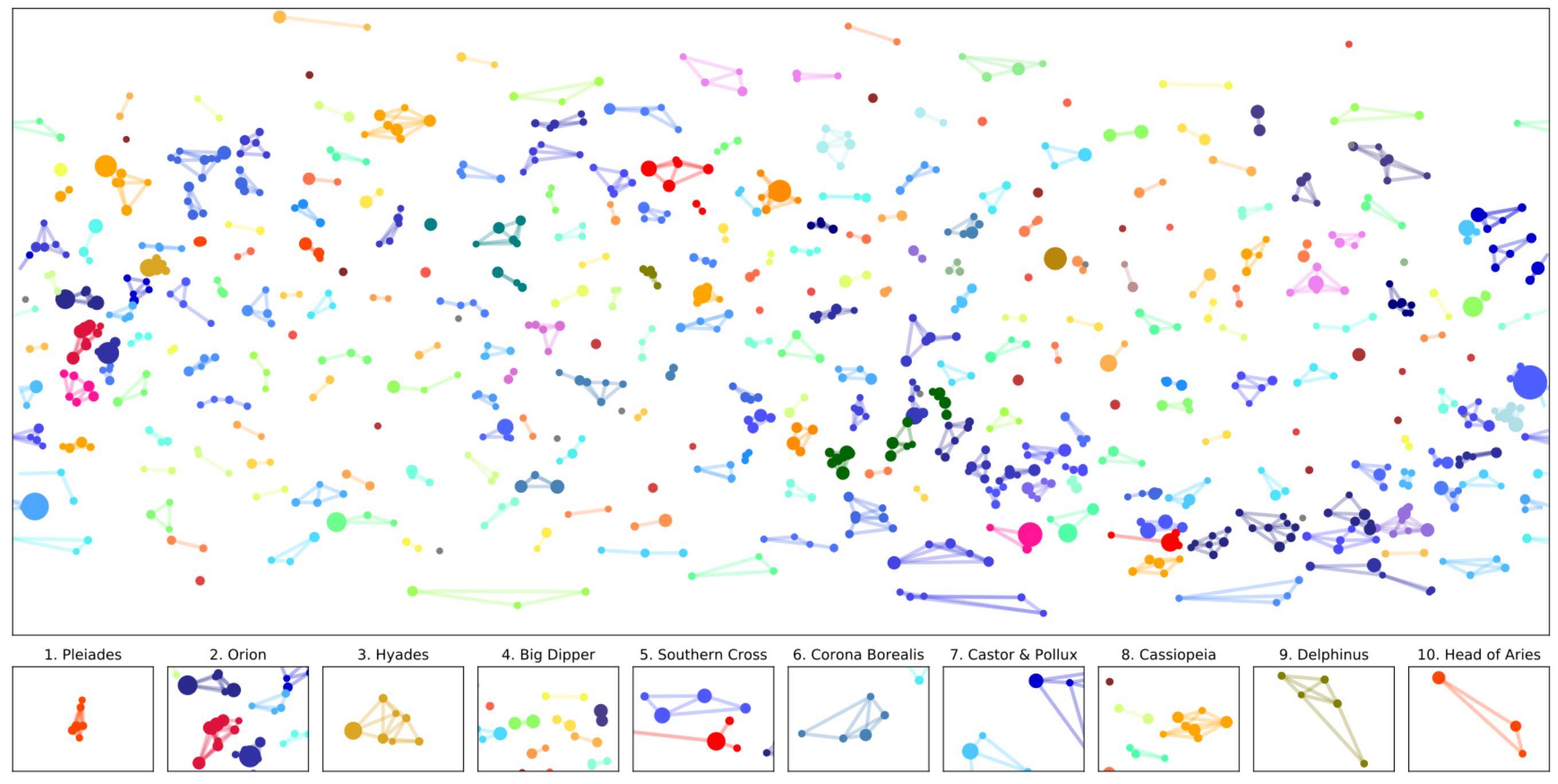}
\caption{Asterisms returned by k-means clustering ($k=277$).}
\label{model_knn}
\end{center}
\end{figure}

\begin{figure}
\begin{center}
\includegraphics[width=\textwidth]{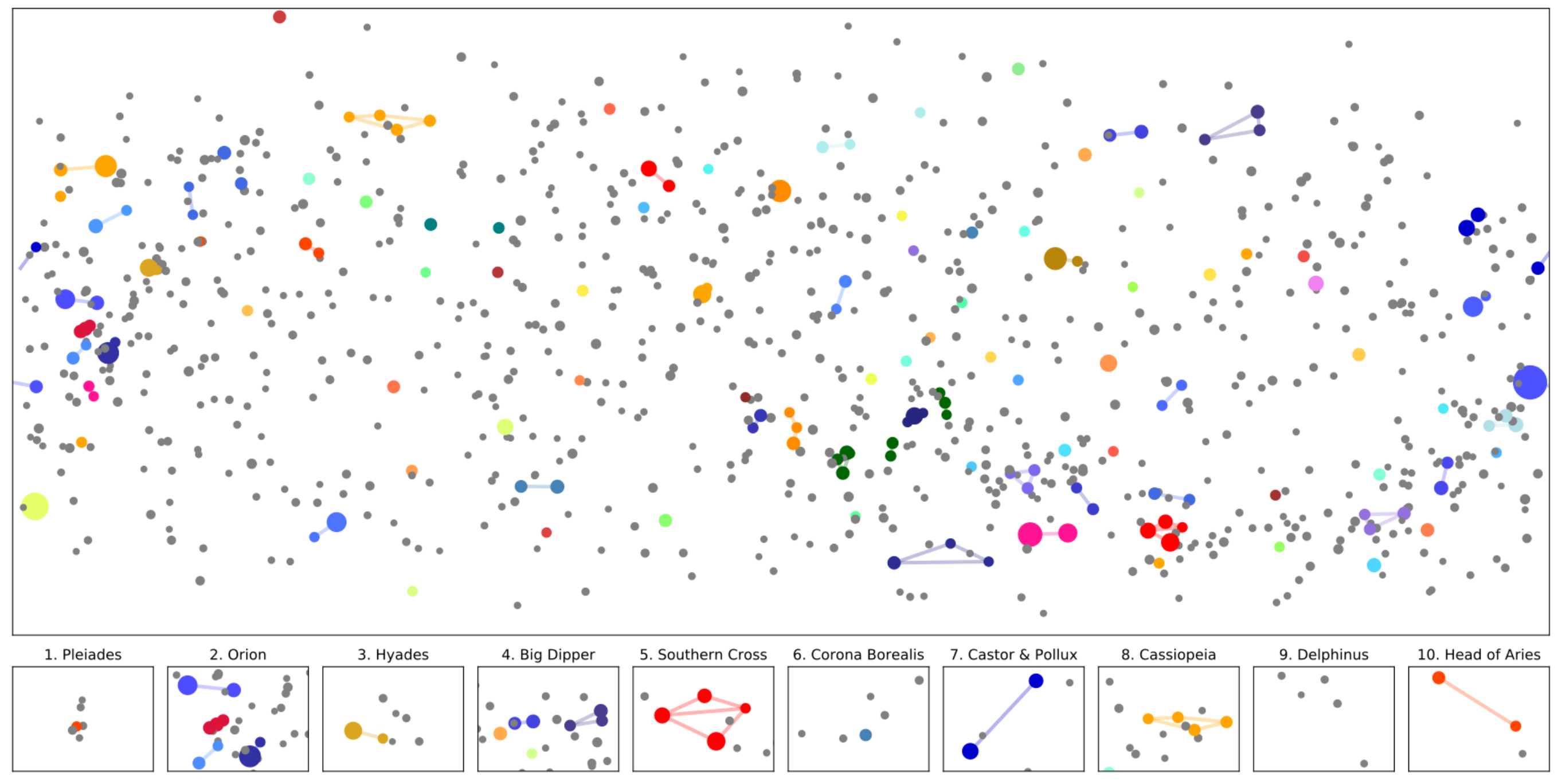}
\caption{Asterisms returned by k-means clustering ($k=100$) with a magnitude threshold of 2.9.}
\label{model_knn_thresh}
\end{center}
\end{figure}

k-means clustering begins by randomly choosing a set of $k$ cluster centers. The algorithm then repeatedly assigns items to the nearest cluster and recomputes the cluster centers based on these assignments until convergence. We used the spherecluster package in Python to implement k-means clustering with distances computed over the celestial sphere. 

We ran k-means clustering for all k between 1 and 300. For comparison, the largest system of asterisms in our data (Chinese) includes 318 asterisms, and 161 remain when we threshold the system at a stellar magnitude of 4.5. For each value of k we ran the algorithm using 100 different initial cluster assignments chosen using the package default (the k-means++ algorithm). Random initialization means that there is some noise in the results, but model performance according to the $F_{10}$ measure tended to increase monotonically with $k$. The best-scoring system in our run, however, $k=277$ and is shown in Figure~\ref{model_knn}. 

As Figure~\ref{model_knn} shows, k-means tends to partition the stars into clusters that are roughly equal in size. In contrast, the GC and CODE models both pick out a relatively small number of clusters against a background of ``unclustered'' stars. To parallel this behavior we consider a variant of k-means with a magnitude threshold $m$. This k-means threshold model runs regular k-means on all stars brighter than $m$, and all remaining stars are treated as unclustered. 
 Based on the $F_{10}$ measure the thresholded model achieves best performance at $m = 2.9$ and $k=100$, and a model result for these parameter values is shown in Figure~\ref{model_knn_thresh}. 

Other than the magnitude threshold, the k-means threshold model does not take brightness into account, and the same applies to the basic k-means model. The distance measure used by k-means could potentially be adjusted to take brightness into account, but our focus here is on k-means clustering as it is typically applied.   

\subsection{Additional baselines}

Two additional baselines were included in the model comparison, both of which rely on a magnitude parameter $m$. The ``one cluster'' model assigns all stars brighter than $m$ to a single cluster, and the ``singleton model'' assigns each of these stars to its own cluster. When $m=4.5$ the singleton model is the limit of k-means when $k$ approaches the total number of stars. For each baseline and each scoring metric we identified the best-performing value of $m$ using an exhaustive search over $m \in \{3, 3.1, \ldots, 4.5\}$.

\subsection{Model scores}

\begin{figure}
\begin{center}
\includegraphics[]{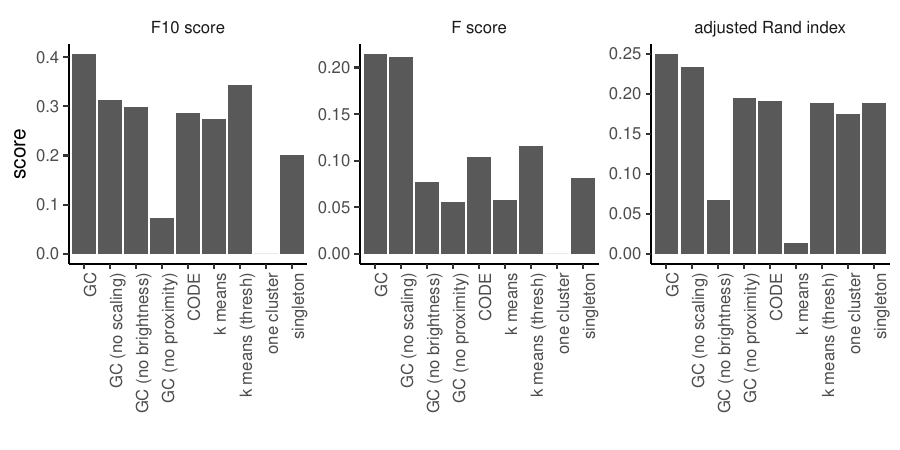}
\caption{Scores of nine models according to three measures: the $F_{10}$ measure, the $F$ measure, and the adjusted Rand index.}
\label{modelscores}
\end{center}
\end{figure}

Scores for all models according to the three scoring measures are shown in Figure~\ref{modelscores}. The GC model performs best regardless of which scoring measure is used, but we focus here on results for the $F_{10}$ measure. The second best model is k-means with a magnitude threshold. Figure~\ref{model_knn_thresh} shows that this model picks out asterisms including the Southern Cross and Orion's belt but the best magnitude threshold for the model ($m=2.9$) means that it misses the Big Dipper, which includes a star of magnitude 3.3.  Figure~\ref{model_knn} shows that k-means without the magnitude threshold produces a large number of compact groupings that cover the sky in a way that is qualitatively unlike any of the systems in our data. 

The CODE model performs worse than the GC model, and the asterisms in Figure~\ref{model_code} reveal at least two qualitative limitations of the model. First, the model misses asterisms (e.g.\ the Big Dipper) that include stars separated by relatively large distances. Second, in relatively dense regions (e.g.\ the area of the Milky Way surrounding the Southern Cross) the model tends to form groups containing relatively large numbers of fainter stars. If the CODE activation surface lies above the threshold in a given region, then all stars in the region are included, regardless of how faint they are. In contrast, human asterisms sometimes pick out a handful of bright stars without including fainter stars that lie nearby. For example, Betelgeuse and Bellatrix (the shoulders of Orion) are often grouped in our data without including a fainter star (32 Orionis) that lies between them.

\begin{figure}
\begin{center}
\includegraphics[]{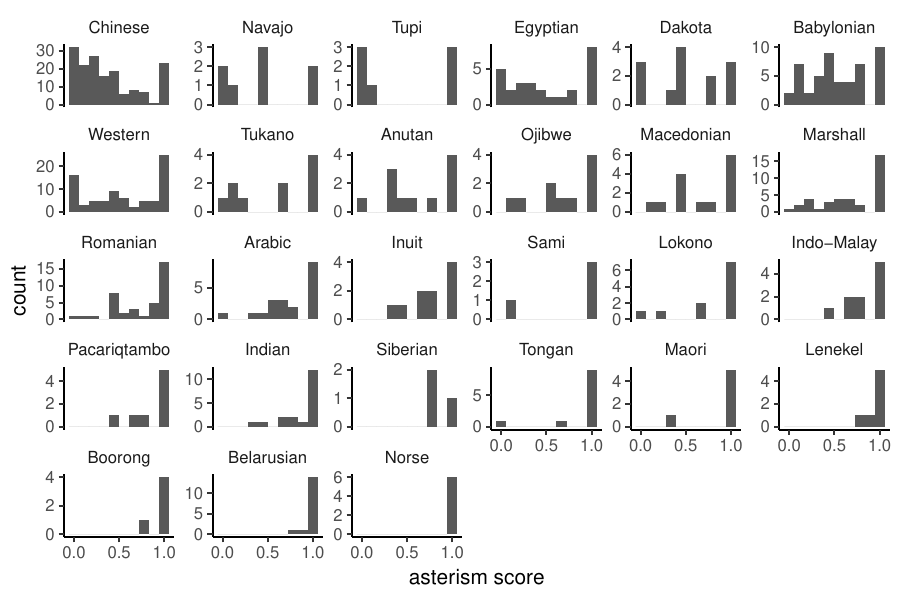}
\caption{Distributions of ``other culture'' scores for the asterisms in each culture. Scores of 1 indicate asterisms that are identical to asterisms in some other culture. The cultures are ordered based on the means of the distributions.}
\label{perculturescoresvsothers}
\end{center}
\end{figure}

\begin{figure}
\begin{center}
\includegraphics[width=3in]{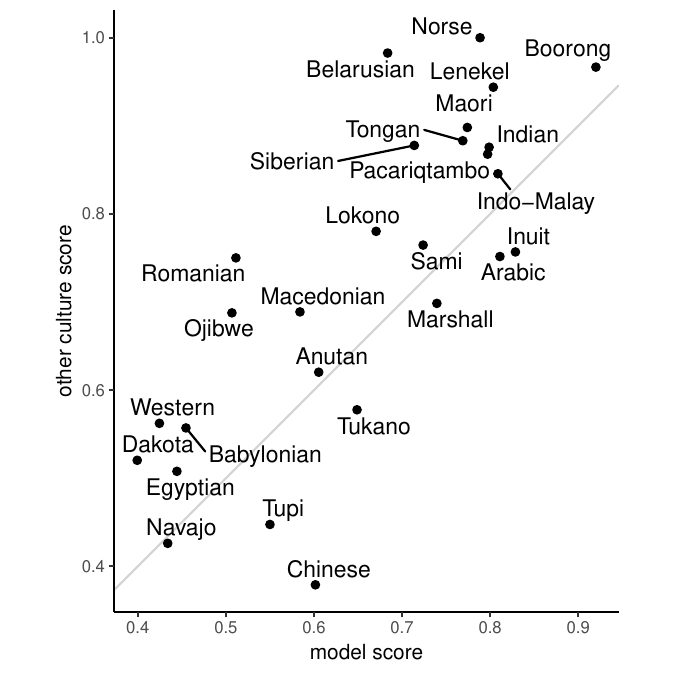}
\caption{``Other culture'' scores compared to model scores. Systems above the line match other cultures better than they match asterisms formed by the model.}
\label{perculturehumanvsmodel}
\end{center}
\end{figure}

\section{Comparisons across cultures}

In addition to comparing each system to the asterisms formed by the GC model (Figure \perculturescores ), we compared each system to other systems in the data set. For each system $S$ let 
$\mathcal{S}_{-S}$ be the set that includes all systems except for $S$. For each asterism $a$ in $S$ we used Equation \modelscore~to compute the extent to which $a$ resembled an asterism in some system belonging to $\mathcal{S}_{-S}$. An ``other culture''  score of zero indicates that asterism $a$ is dissimilar from all asterisms in all other cultures, and a score of 1 indicates that $a$ is identical to an asterism from at least one other culture.  Distributions of scores for each culture are shown in Figure~\ref{perculturescoresvsothers}. As mentioned in the main text, the system that differs most from all others is the Chinese system, but this result should be interpreted in light of genealogical relationships between cultures. There are strong historical relationships between some systems in our data set---for example, Western constellations are based in part on Babylonian tradition. One reason why the Chinese system stands out as different from the others is that the genealogical relationships between Chinese culture and most other cultures in our data are rather distant. 

Figure~\ref{perculturehumanvsmodel} explores whether systems that tend to resemble systems from other cultures also tend to match the GC model. The x and y coordinates of each point in the figure correspond to means of distributions plotted in Figures \perculturescores~and \ref{perculturescoresvsothers}.  The results are again influenced by genealogical relationships between cultures. In particular, the Western tradition is overrepresented in our data set, meaning that systems from this tradition have higher ``other culture'' scores than would otherwise be expected. A related bias arises because characterizations of other systems are often influenced by the Western system. For example, the Belarusian system achieves a very high ``other culture'' score partly because some of the asterisms in this system are assumed to be identical to Western constellations such as Draco and Gemini~\cite{avilin09}. 

Despite these limitations, Figure~\ref{perculturehumanvsmodel} suggests that model scores and ``other culture'' scores are highly correlated, which is expected given that the asterisms identified by the model tend to be shared across cultures. Most of the points fall above the line, indicating that systems tend to match systems from other cultures better than they match the model. This result is undoubtedly influenced by genealogical relationships between cultures, but may also indicate that there is scope to improve the model to better capture common patterns that recur across cultures.

\appendix 
\section{Asterism systems for 27 cultures}
\label{systemplots}

\begin{figure}[h!]
\begin{center}
\includegraphics[width=\textwidth]{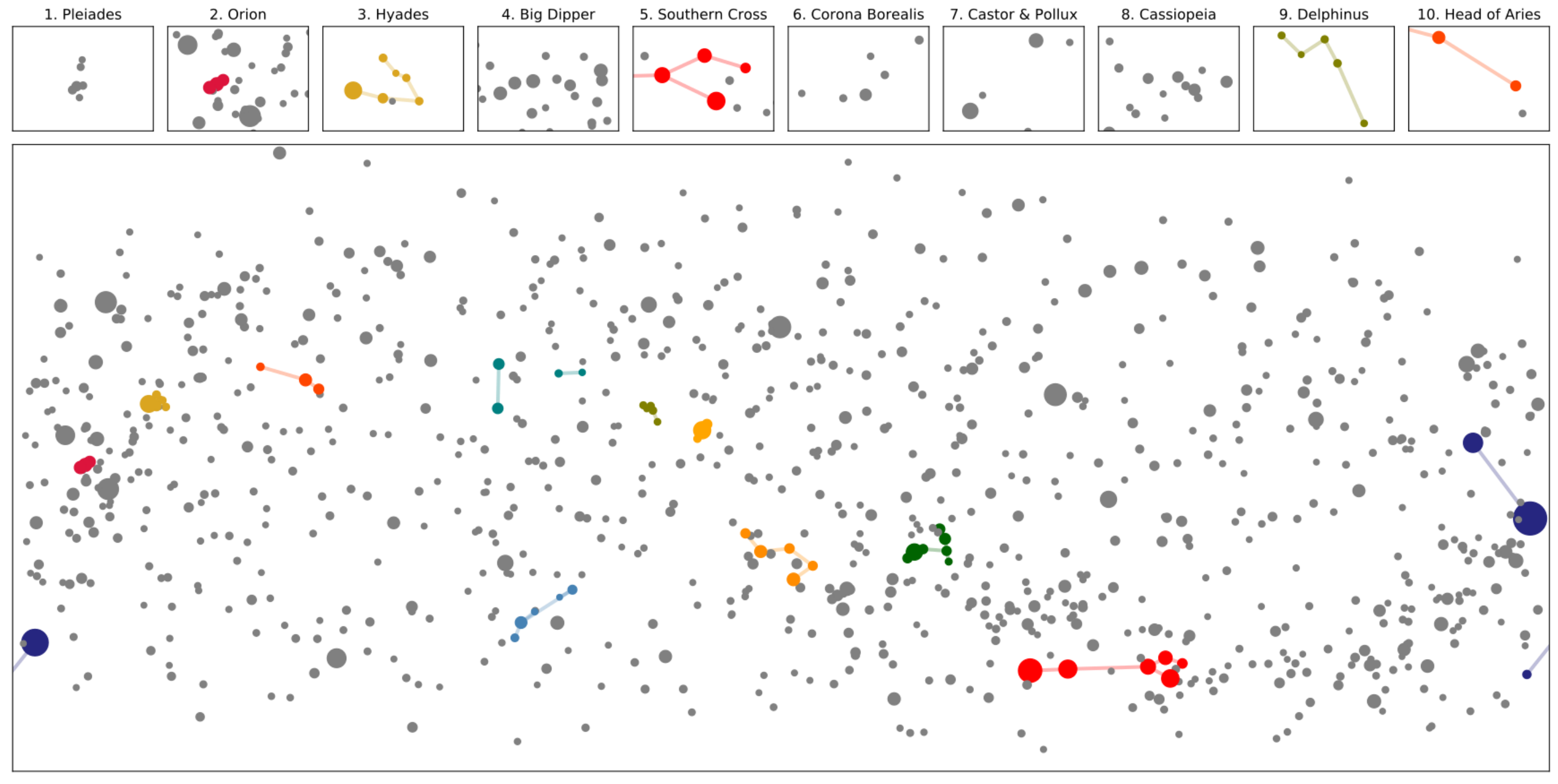}
\caption{Anutan (Stellarium)}
\label{anutan}
\end{center}
\end{figure}

\begin{figure}[h!]
\begin{center}
\includegraphics[width=\textwidth]{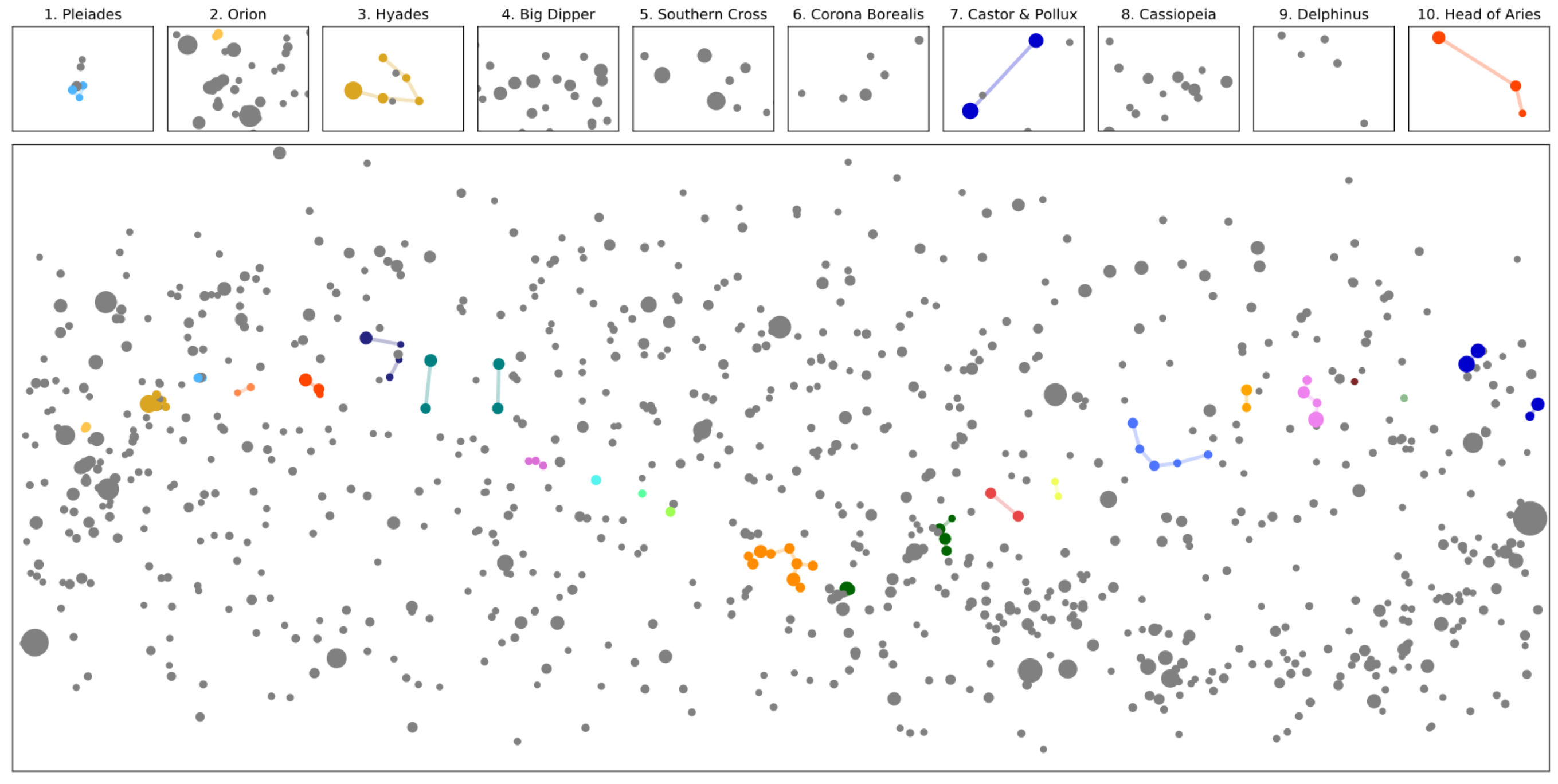}
\caption{Arabic moon stations (Stellarium)}
\label{arabic}
\end{center}
\end{figure}

\begin{figure}[h!]
\begin{center}
\includegraphics[width=\textwidth]{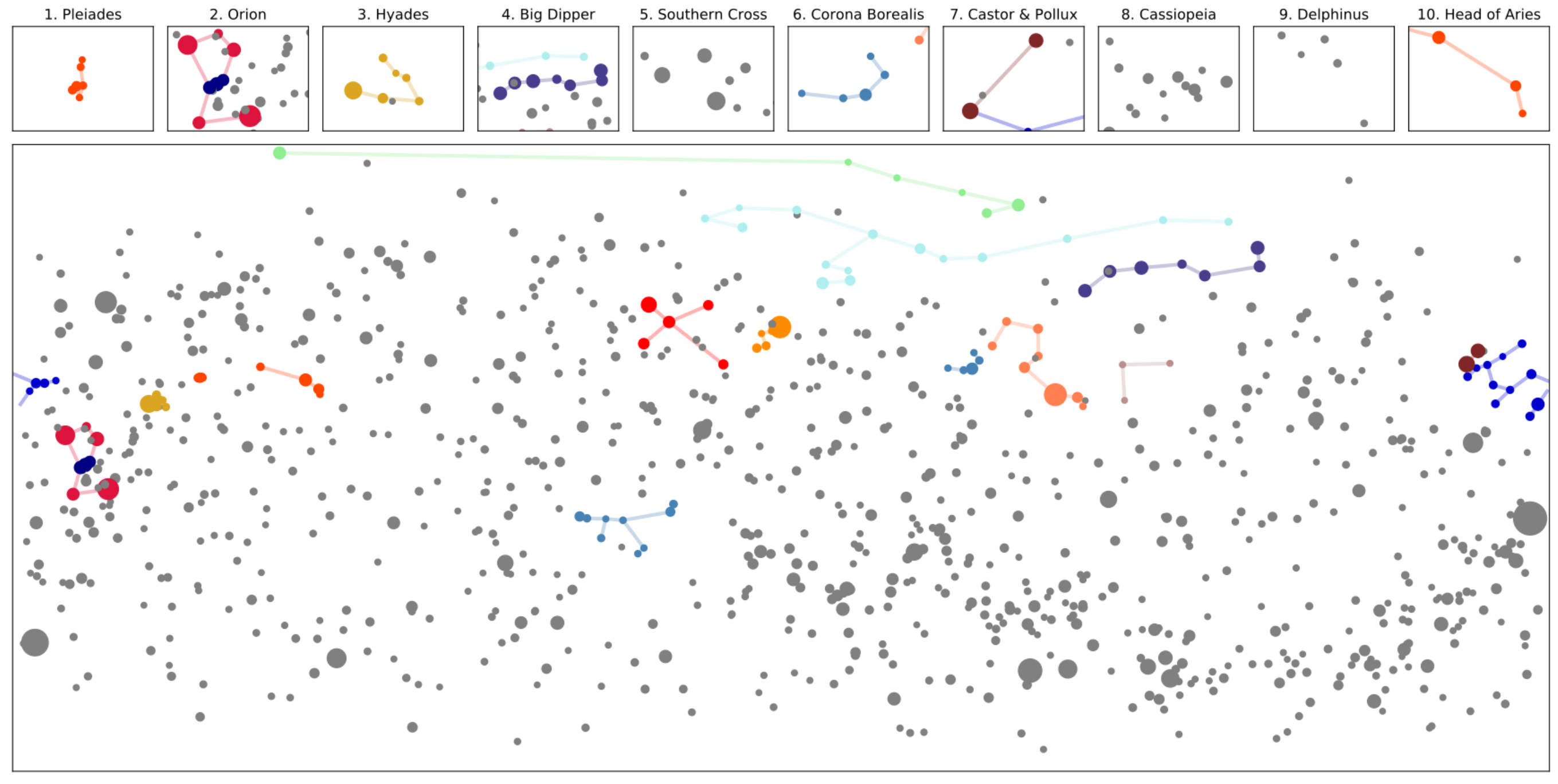}
\label{belarusian}
\caption{Belarusian (Stellarium)}
\end{center}
\end{figure}

\begin{figure}[h!]
\begin{center}
\includegraphics[width=\textwidth]{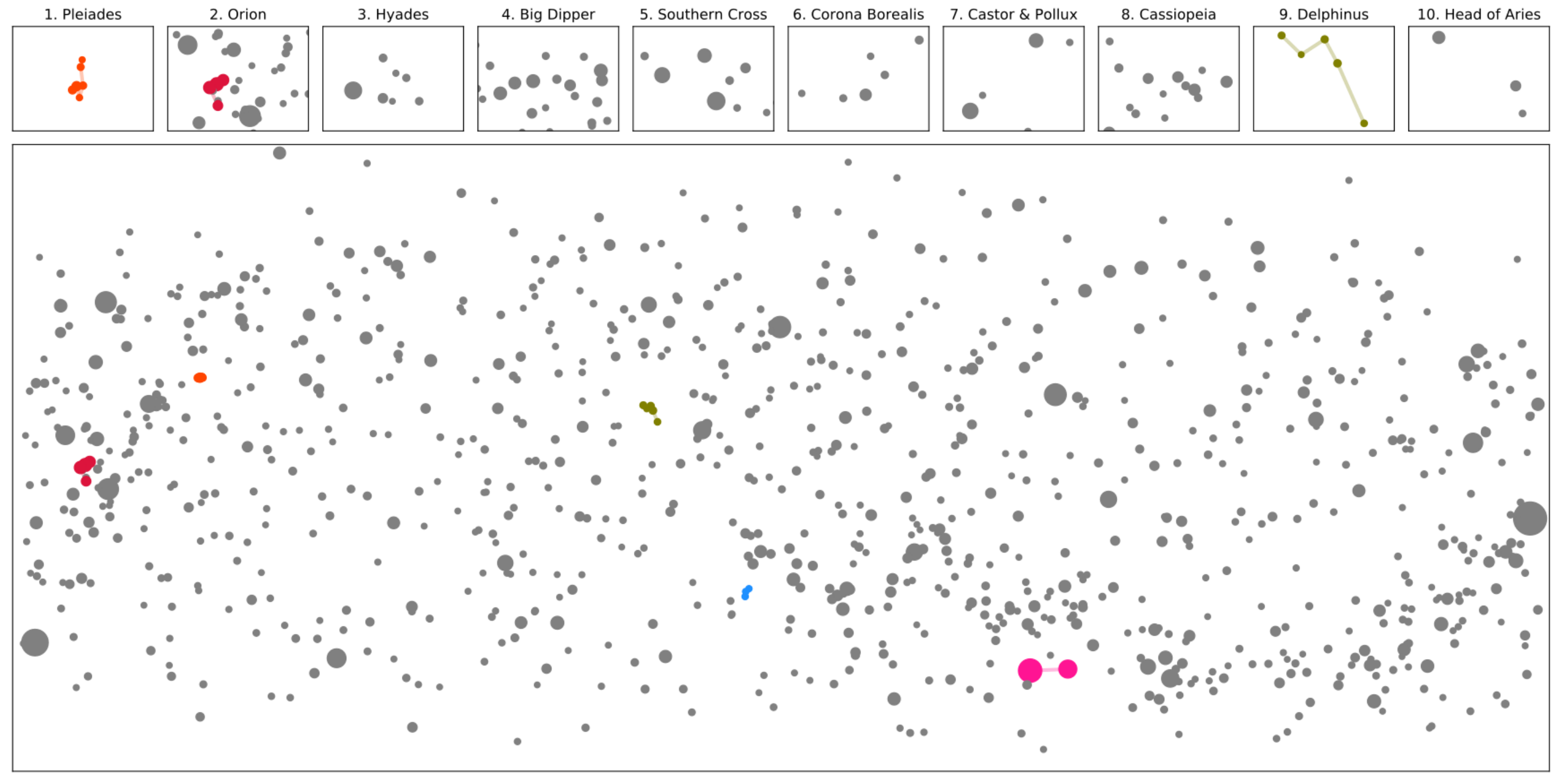}
\label{boorong}
\caption{Boorong~\cite{stanbridge1857,hamacher12}.}
\end{center}
\end{figure}

\begin{figure}[h!]
\begin{center}
\includegraphics[width=\textwidth]{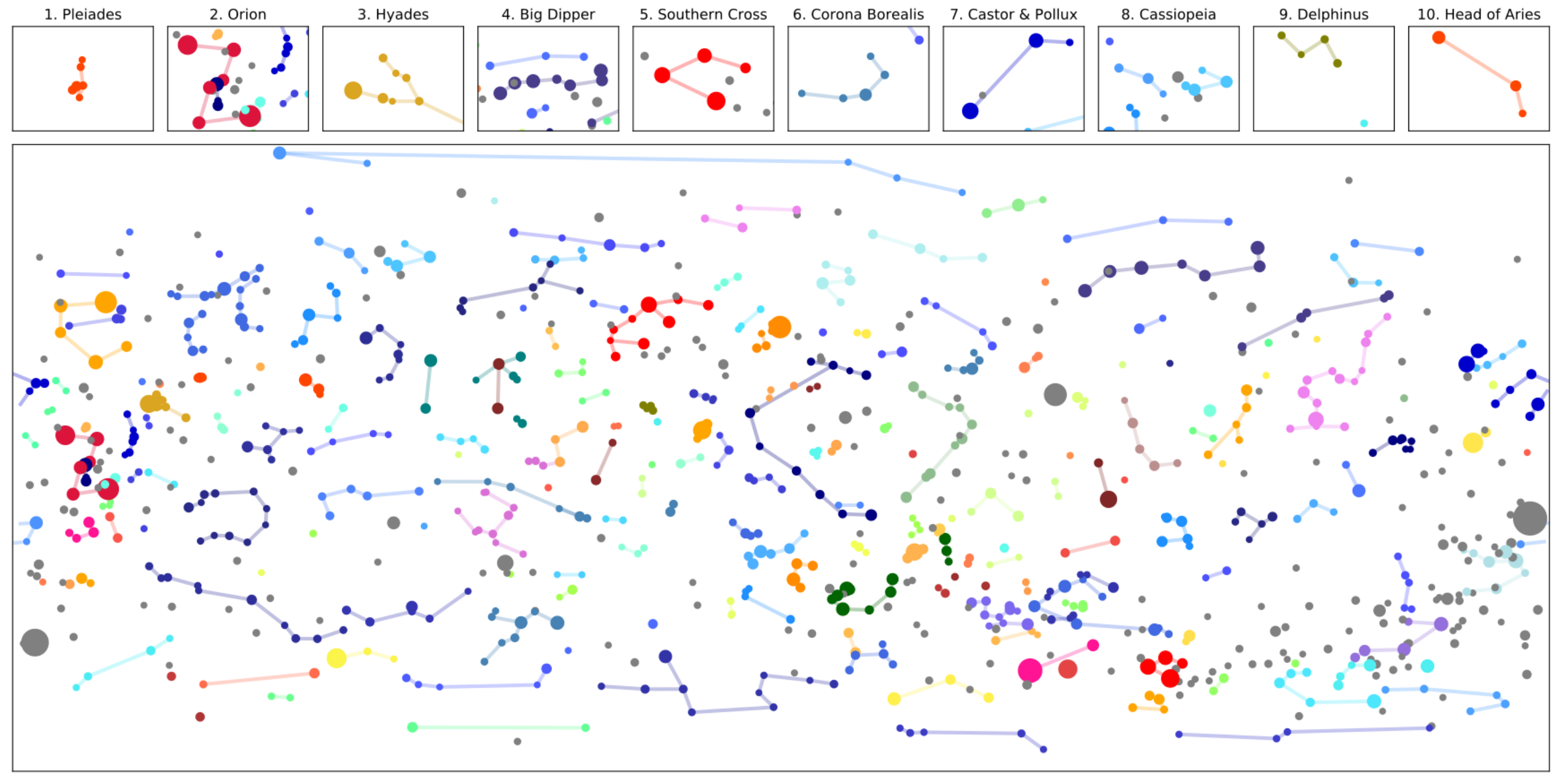}
\label{chinese}
\caption{Chinese (Stellarium).}
\end{center}
\end{figure}

\begin{figure}[h!]
\begin{center}
\includegraphics[width=\textwidth]{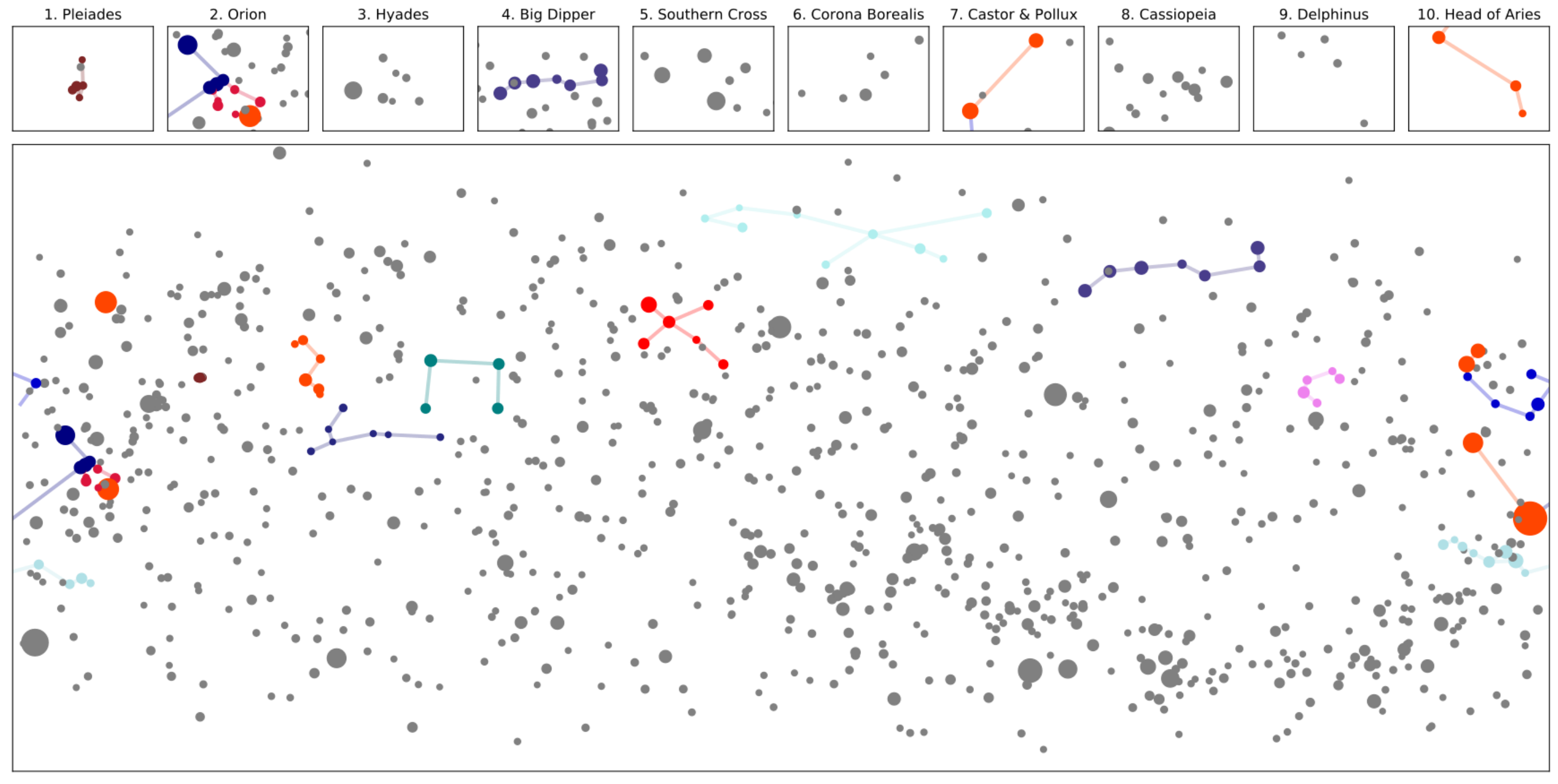}
\caption{Dakota (Stellarium).}
\label{dakota}
\end{center}
\end{figure}

\begin{figure}[h!]
\begin{center}
\includegraphics[width=\textwidth]{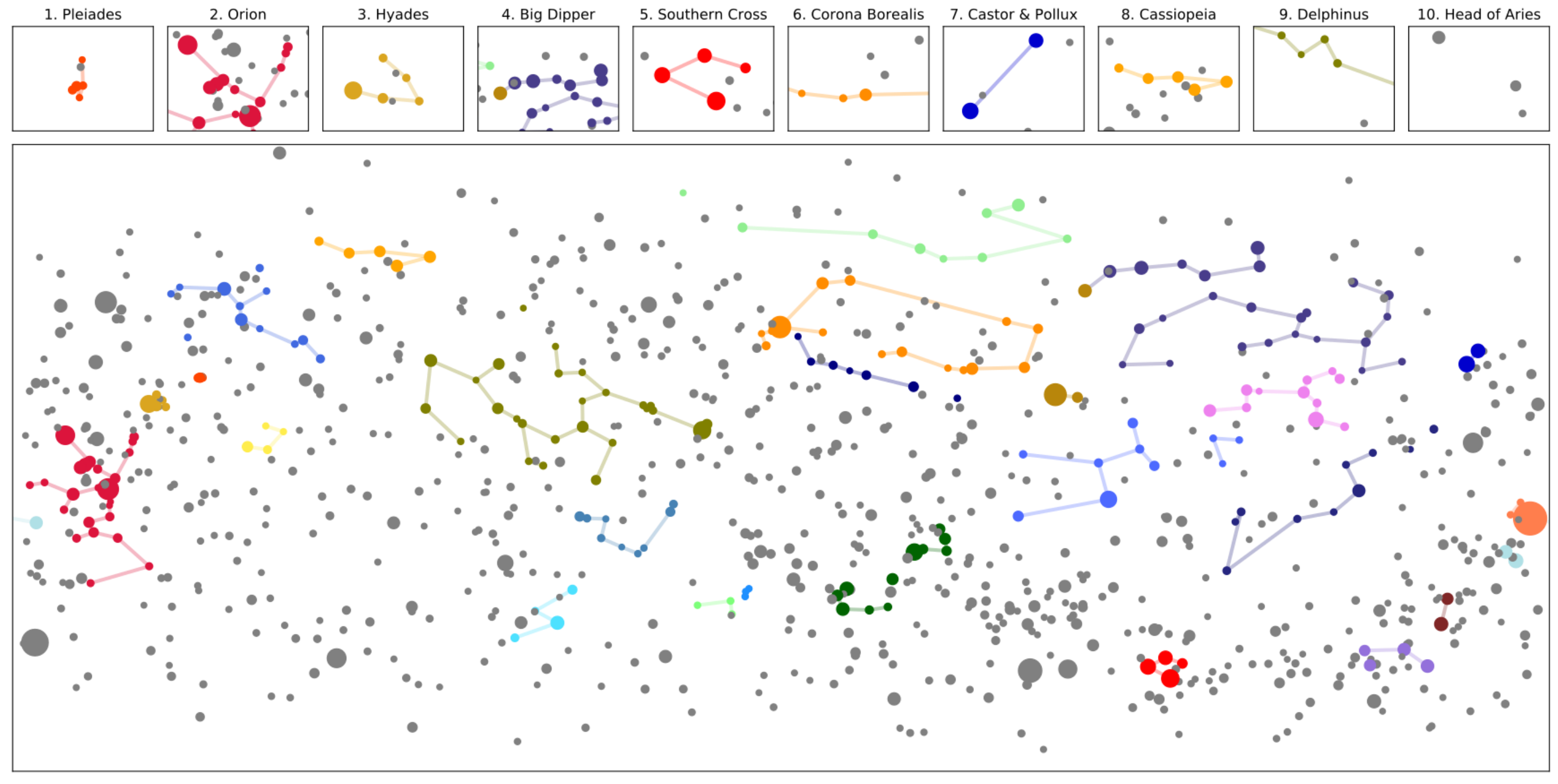}
\label{egyptian}
\caption{Egyptian (Stellarium).}
\end{center}
\end{figure}

\begin{figure}[h!]
\begin{center}
\includegraphics[width=\textwidth]{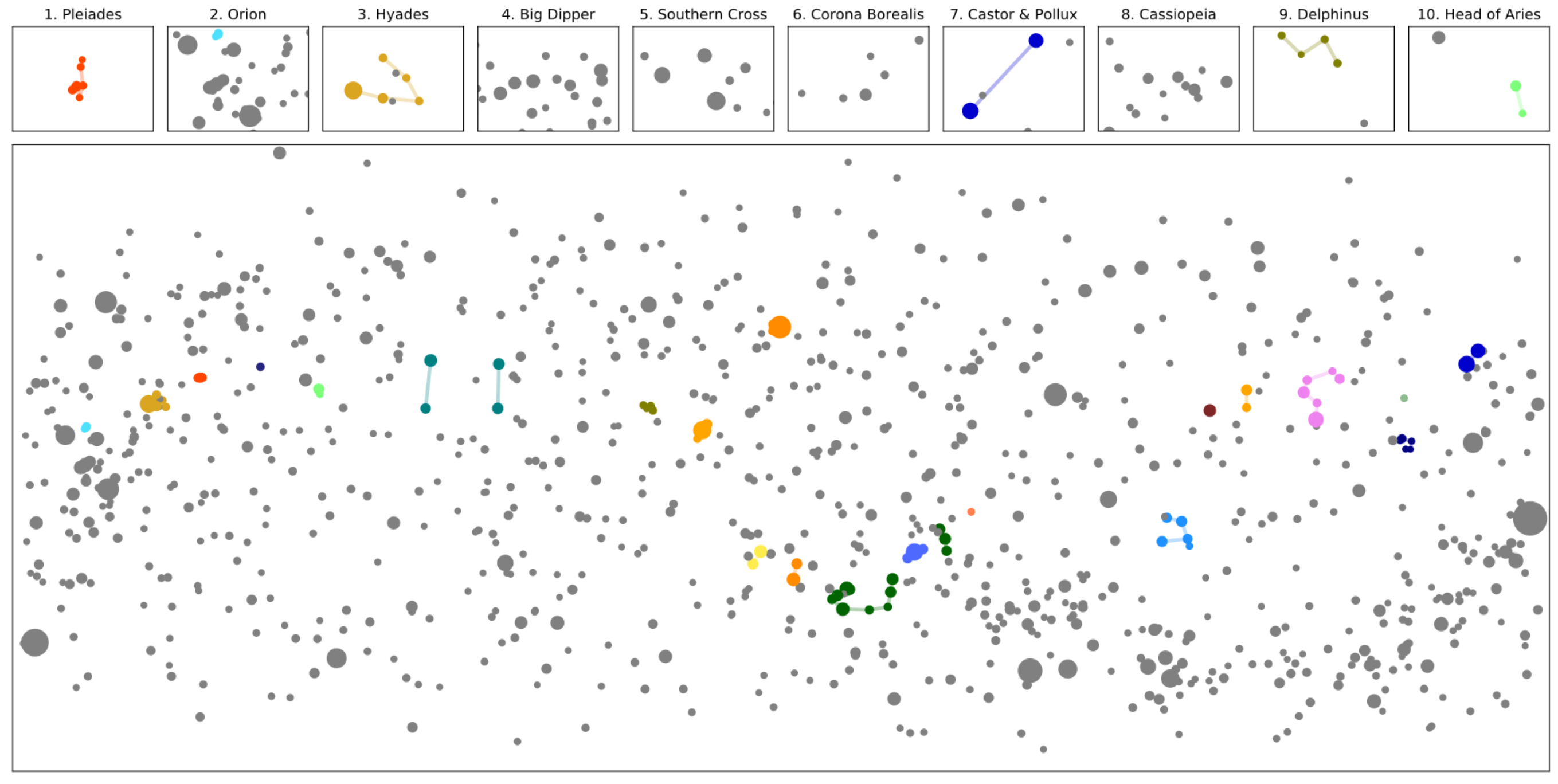}
\label{indian}
\caption{Indian~\cite{kaye81}.}
\end{center}
\end{figure}

\begin{figure}[h!]
\begin{center}
\includegraphics[width=\textwidth]{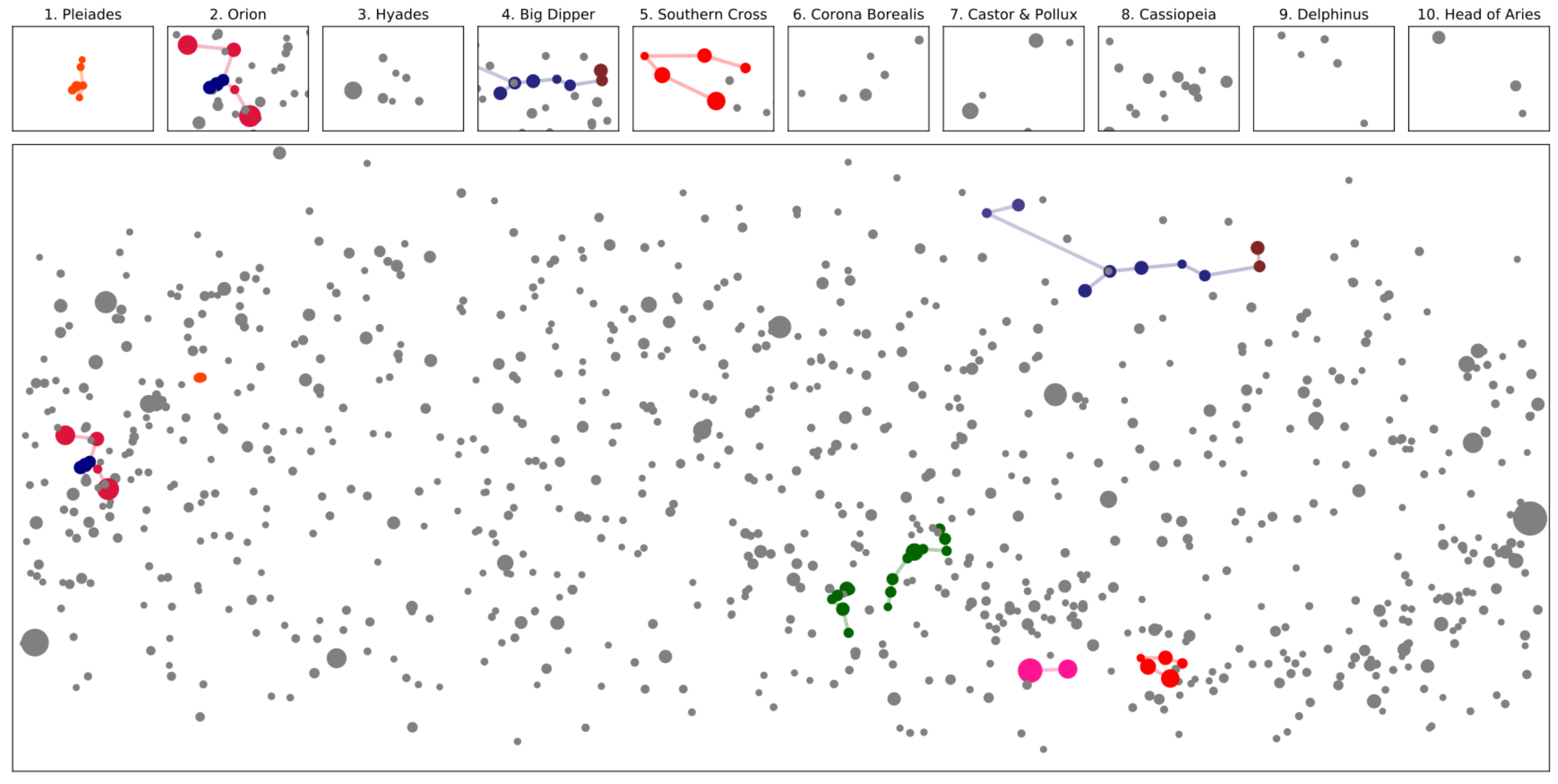}
\label{indomalay}
\caption{Indo-Malay~\cite{ammarell08}.}
\end{center}
\end{figure}

\begin{figure}[h!]
\begin{center}
\includegraphics[width=\textwidth]{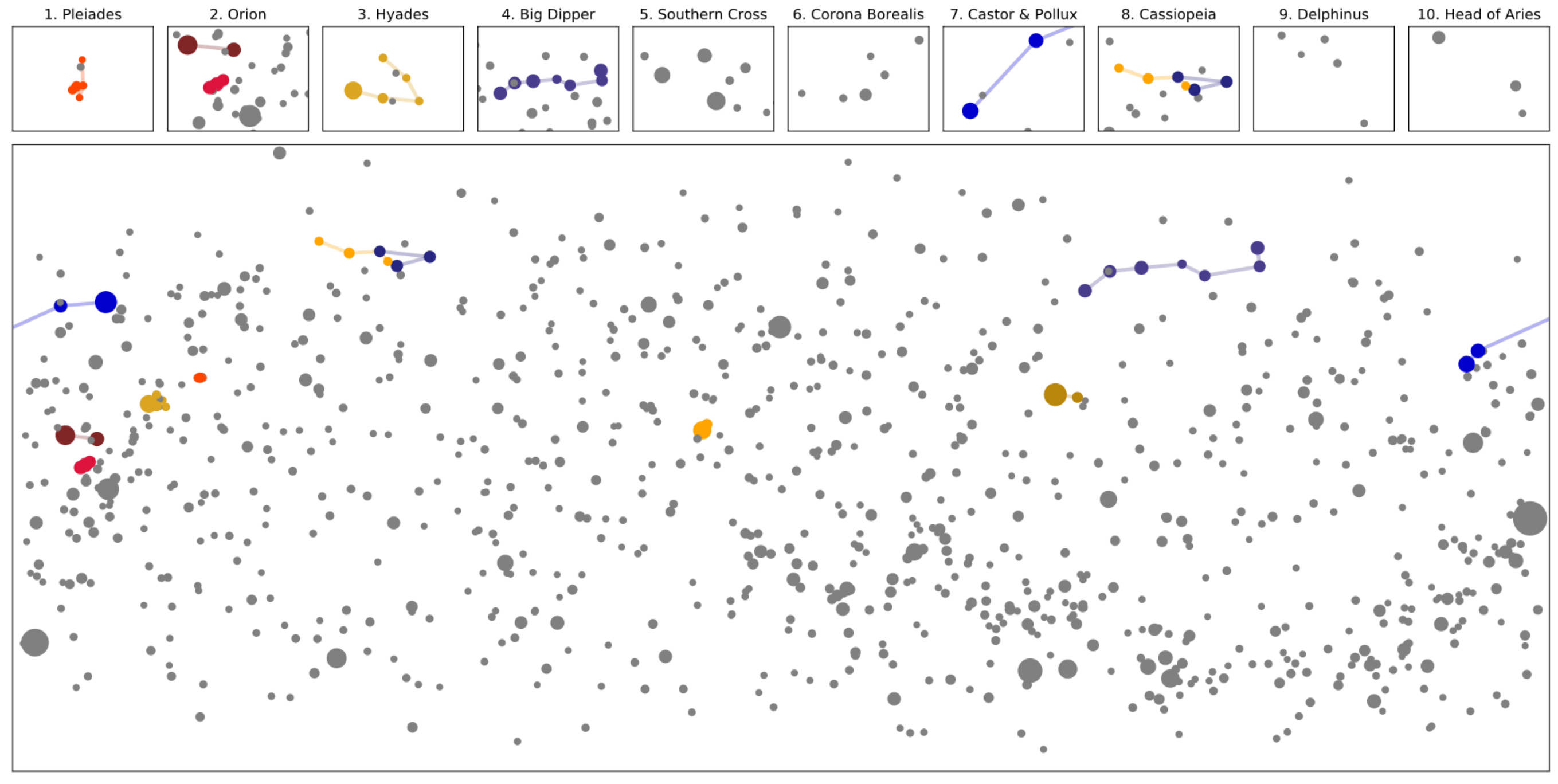}
\label{inuit}
\caption{Inuit (Stellarium).}
\end{center}
\end{figure}

\begin{figure}[h!]
\begin{center}
\includegraphics[width=\textwidth]{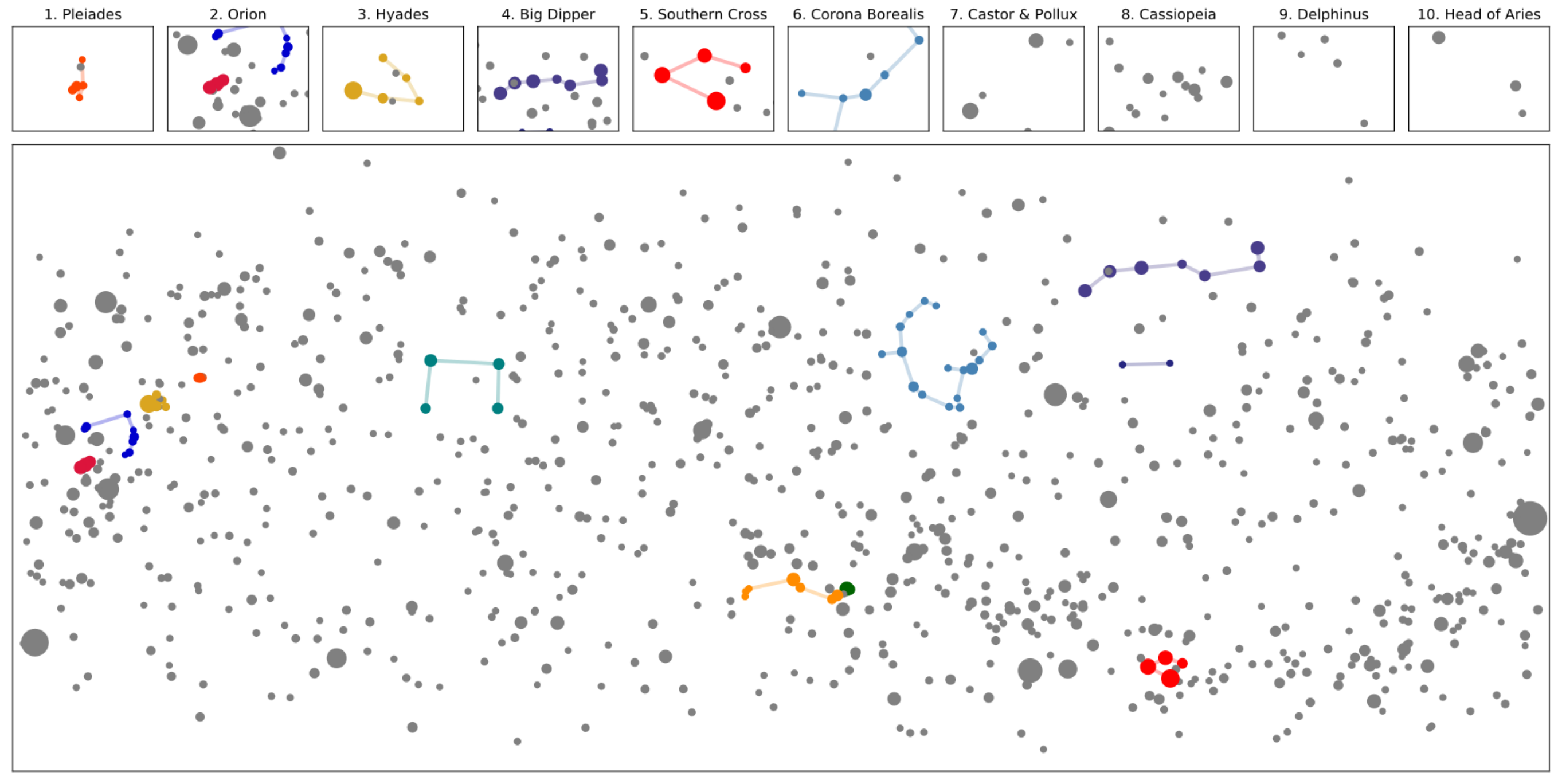}
\label{lokono}
\caption{Lokono (Stellarium).}
\end{center}
\end{figure}

\begin{figure}[h!]
\begin{center}
\includegraphics[width=\textwidth]{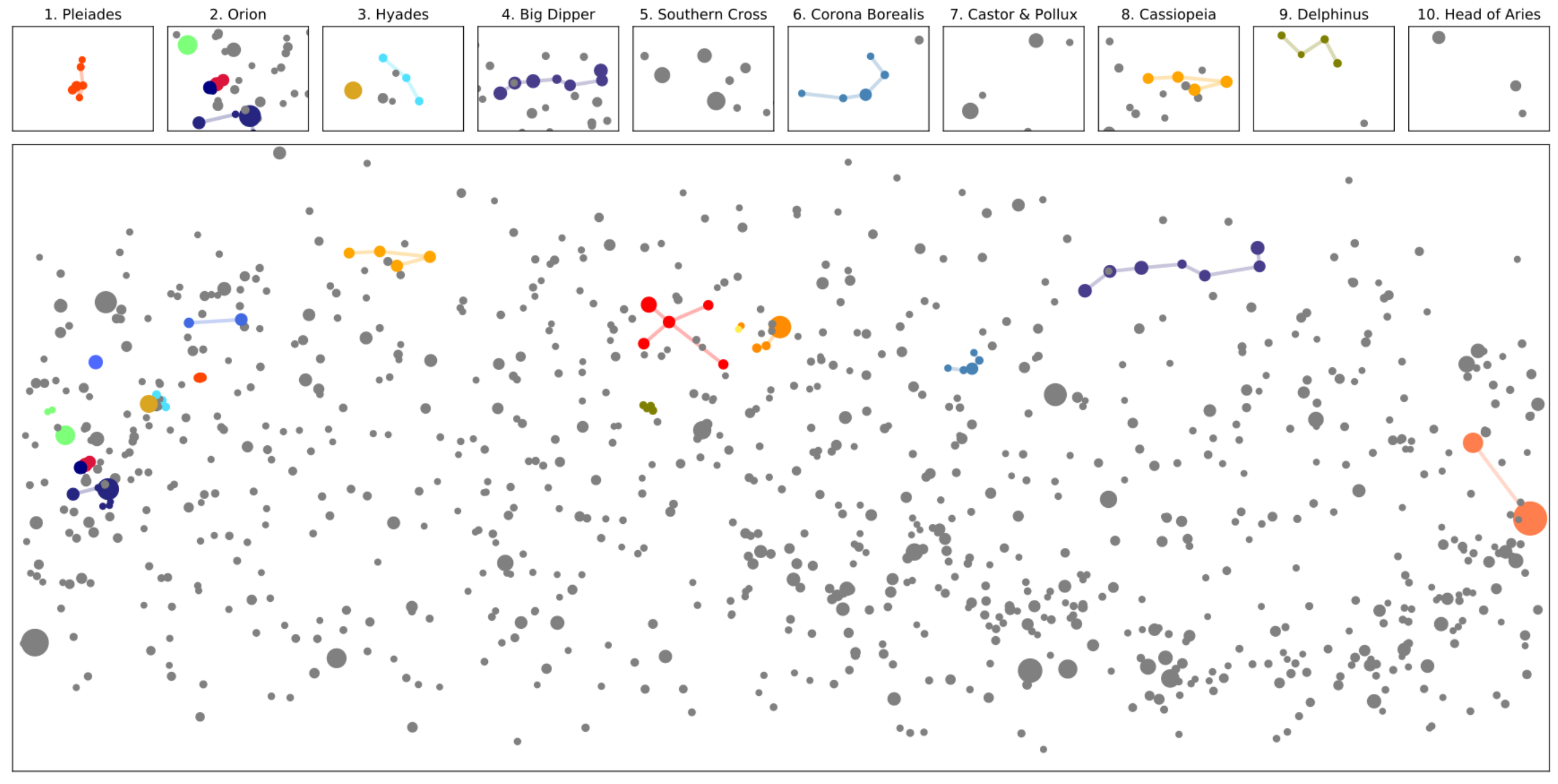}
\label{macedonian}
\caption{Macedonian (Stellarium).}
\end{center}
\end{figure}

\begin{figure}[h!]
\begin{center}
\includegraphics[width=\textwidth]{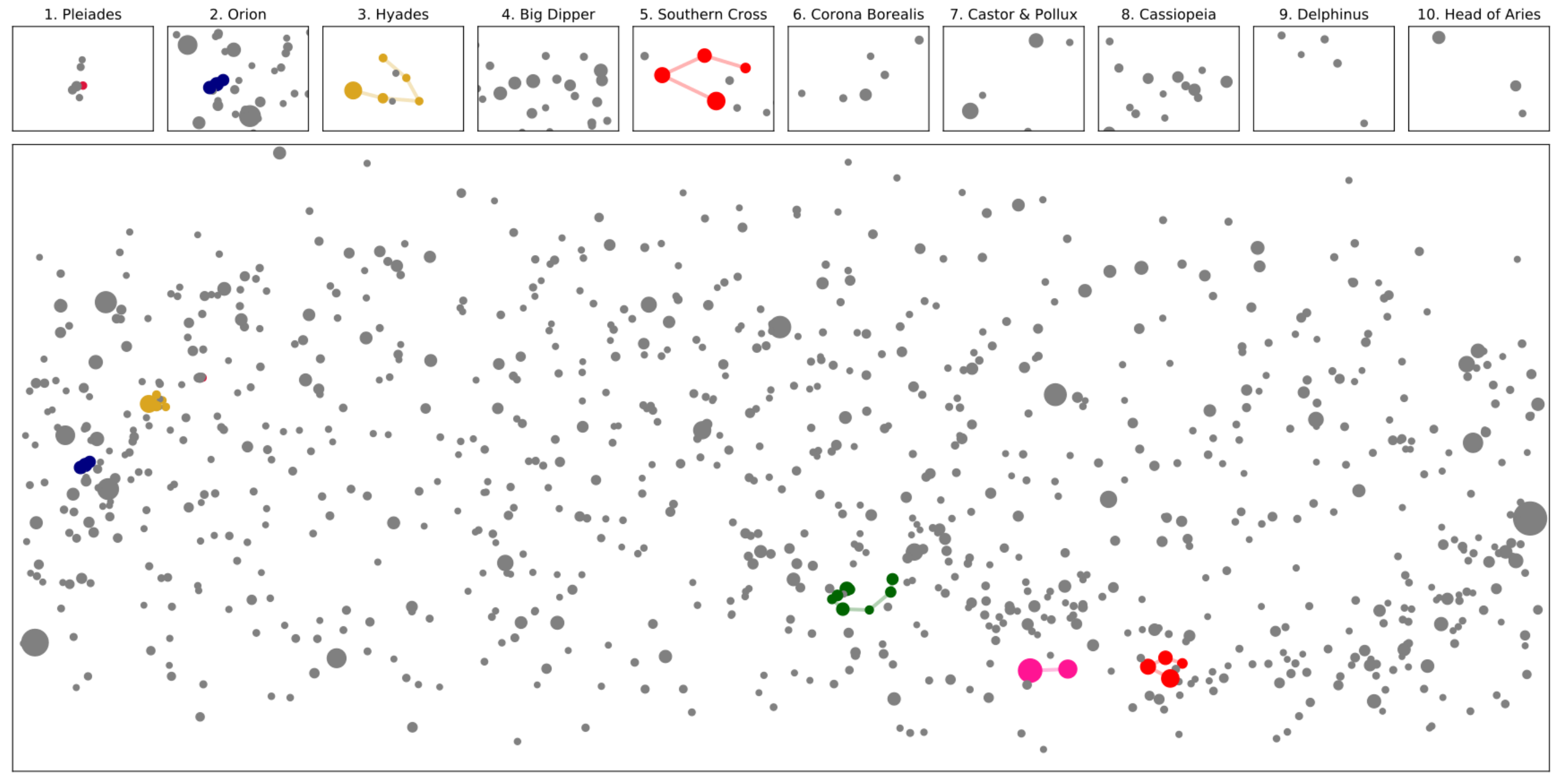}
\label{maori}
\caption{Maori (Stellarium).}
\end{center}
\end{figure}

\begin{figure}[h!]
\begin{center}
\includegraphics[width=\textwidth]{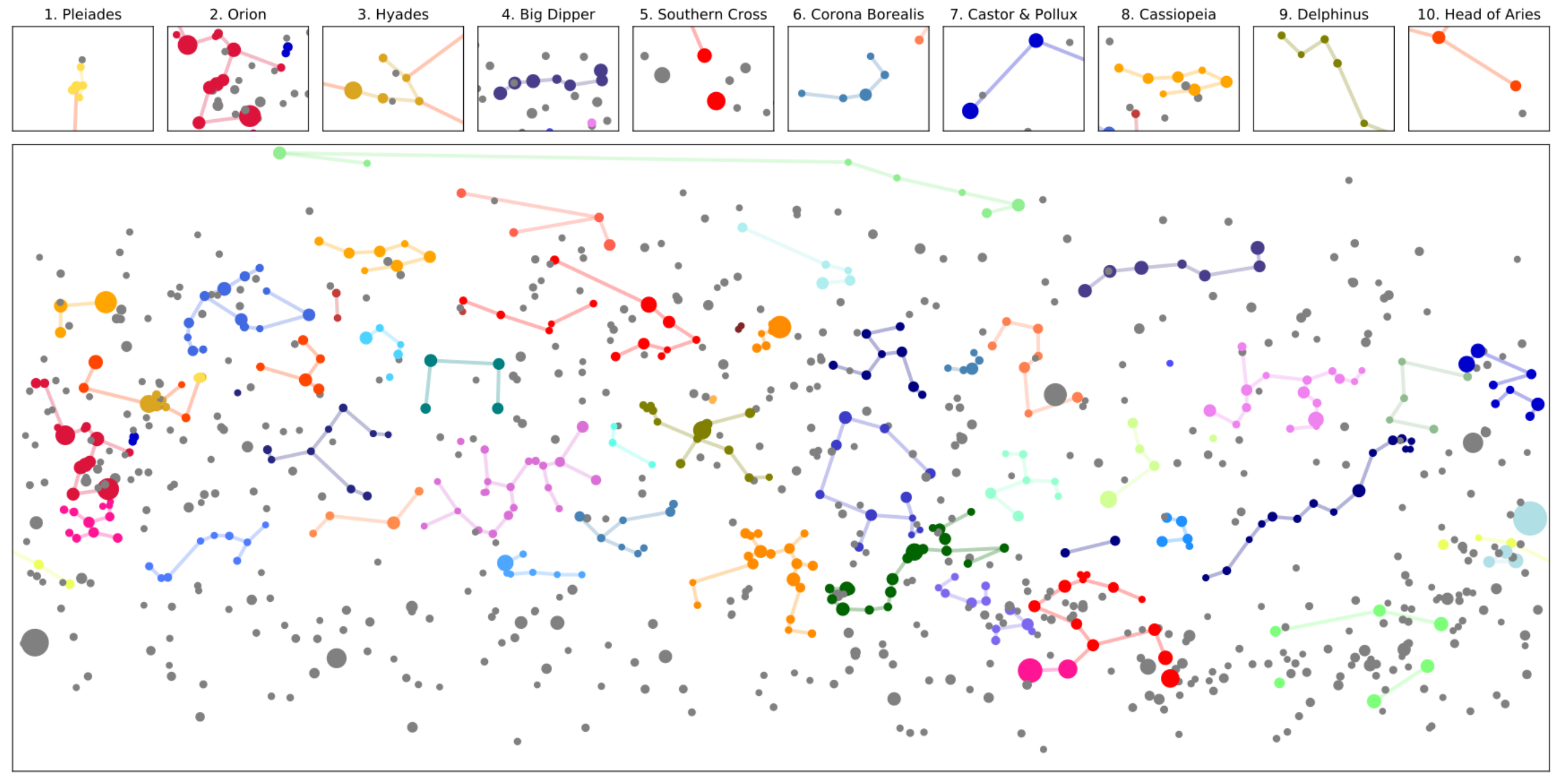}
\label{mulapin}
\caption{Babylonian (MUL.APIN sky culture in Stellarium).}
\end{center}
\end{figure}

\begin{figure}[h!]
\begin{center}
\includegraphics[width=\textwidth]{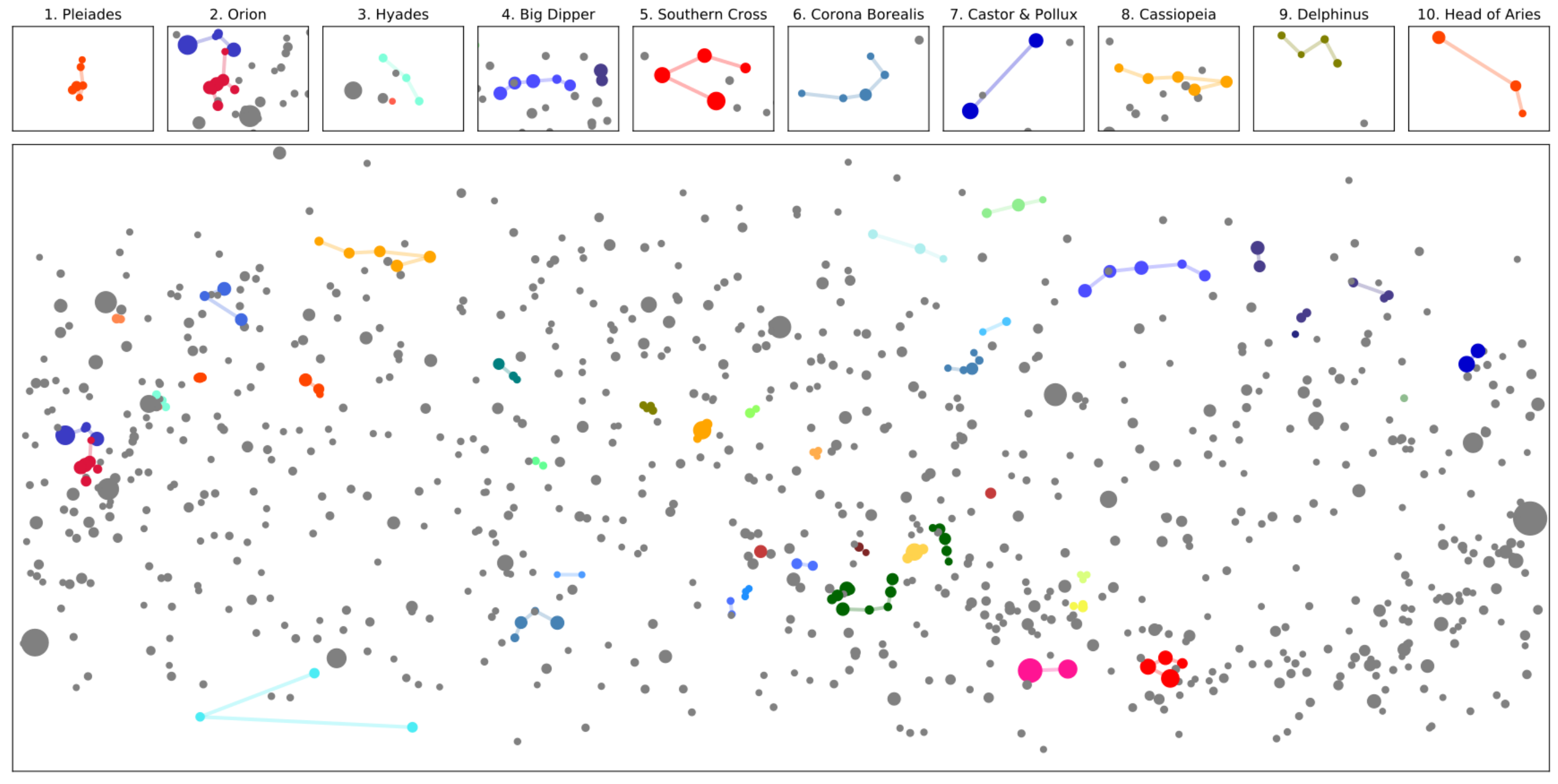}
\label{marshall}
\caption{Marshall Islands~\cite{erdland14}.}
\end{center}
\end{figure}

\begin{figure}[h!]
\begin{center}
\includegraphics[width=\textwidth]{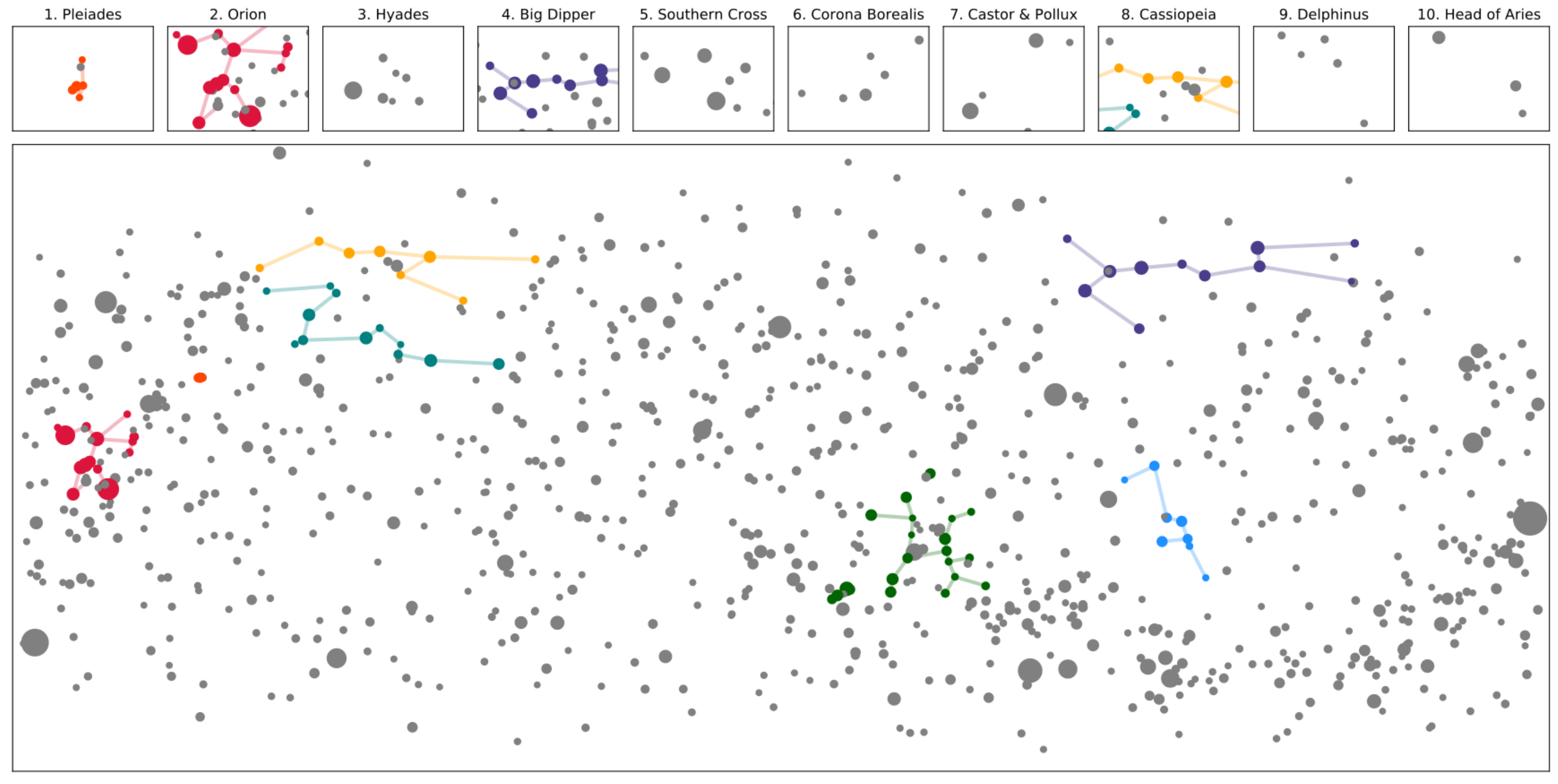}
\label{navajo}
\caption{Navajo (Stellarium).}
\end{center}
\end{figure}

\begin{figure}[h!]
\begin{center}
\includegraphics[width=\textwidth]{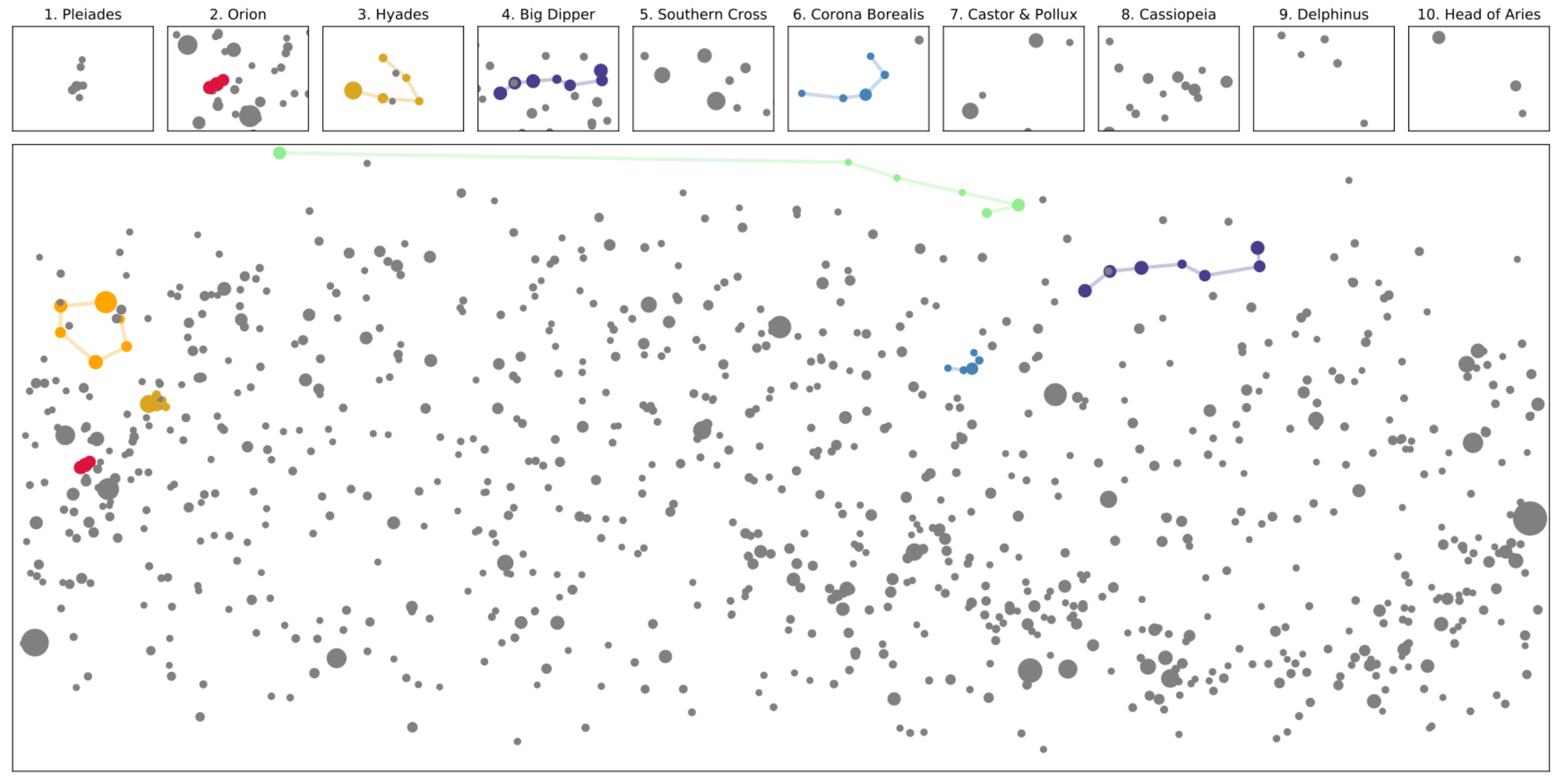}
\label{norse}
\caption{Norse (Stellarium).}
\end{center}
\end{figure}

\begin{figure}[h!]
\begin{center}
\includegraphics[width=\textwidth]{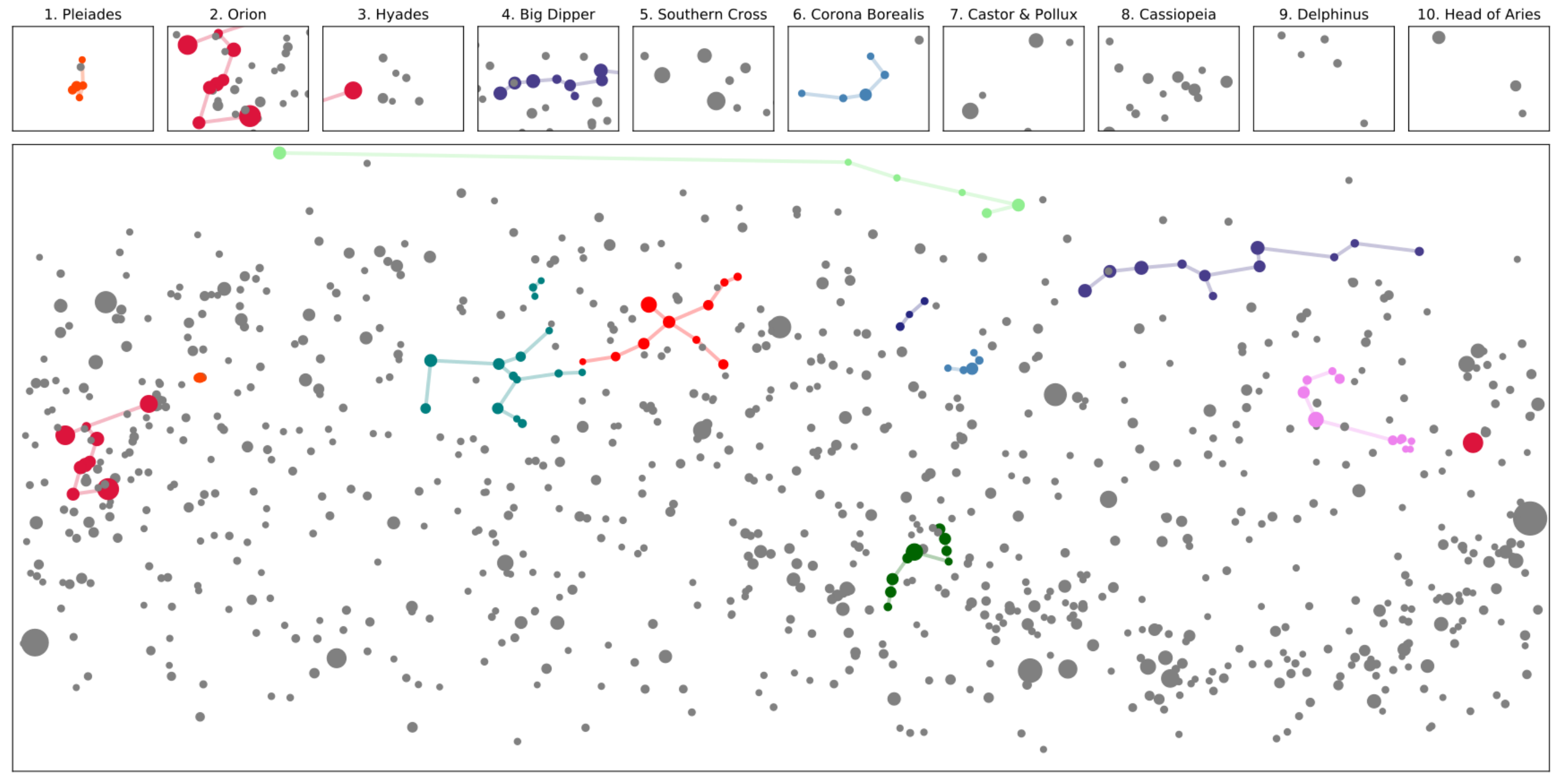}
\label{ojibwe}
\caption{Ojibwe (Stellarium).}
\end{center}
\end{figure}

\begin{figure}[h!]
\begin{center}
\includegraphics[width=\textwidth]{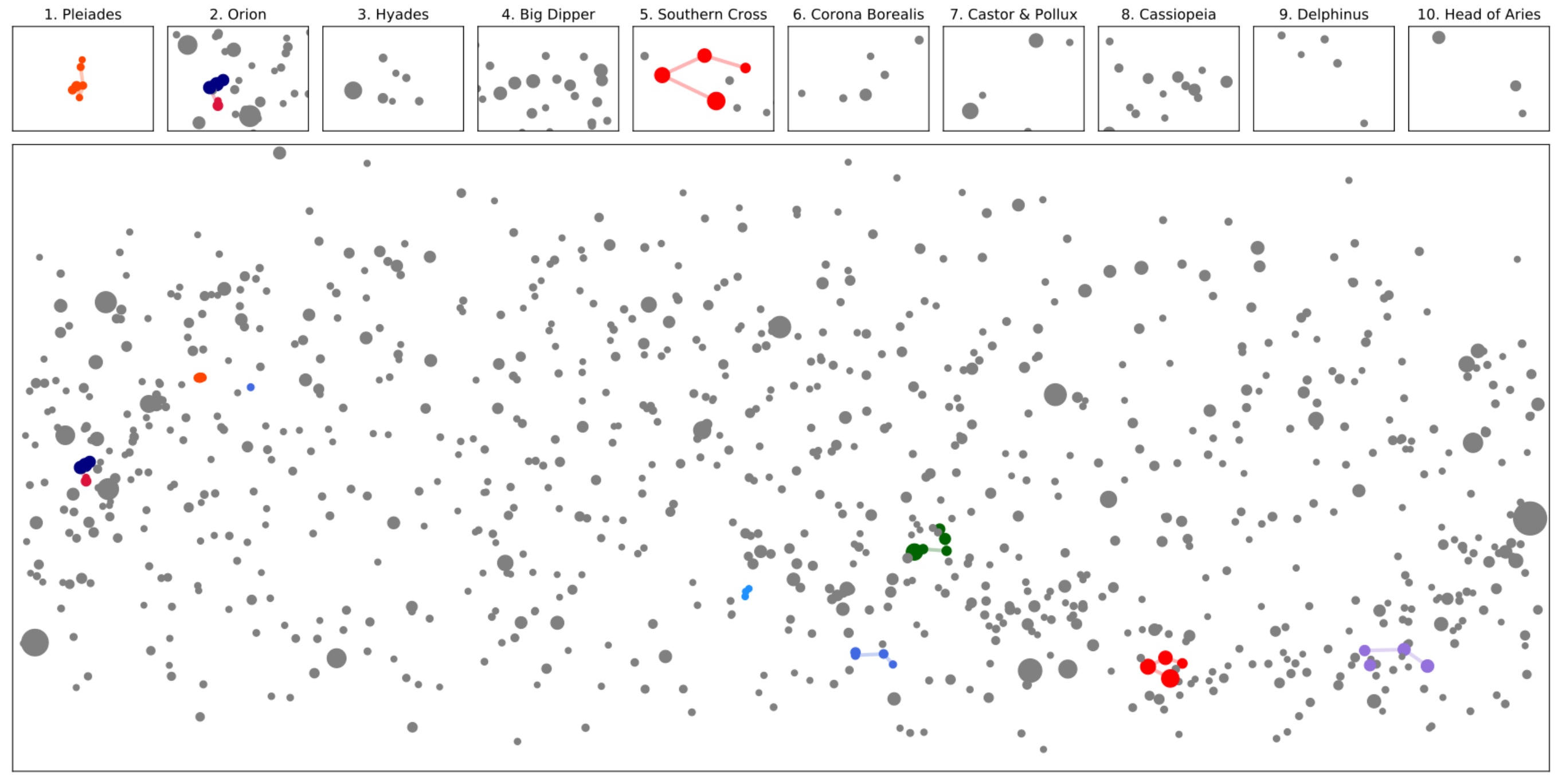}
\label{pacariqtambo}
\caption{Pacariqtambo~\cite{urton05}.}
\end{center}
\end{figure}

\begin{figure}[h!]
\begin{center}
\includegraphics[width=\textwidth]{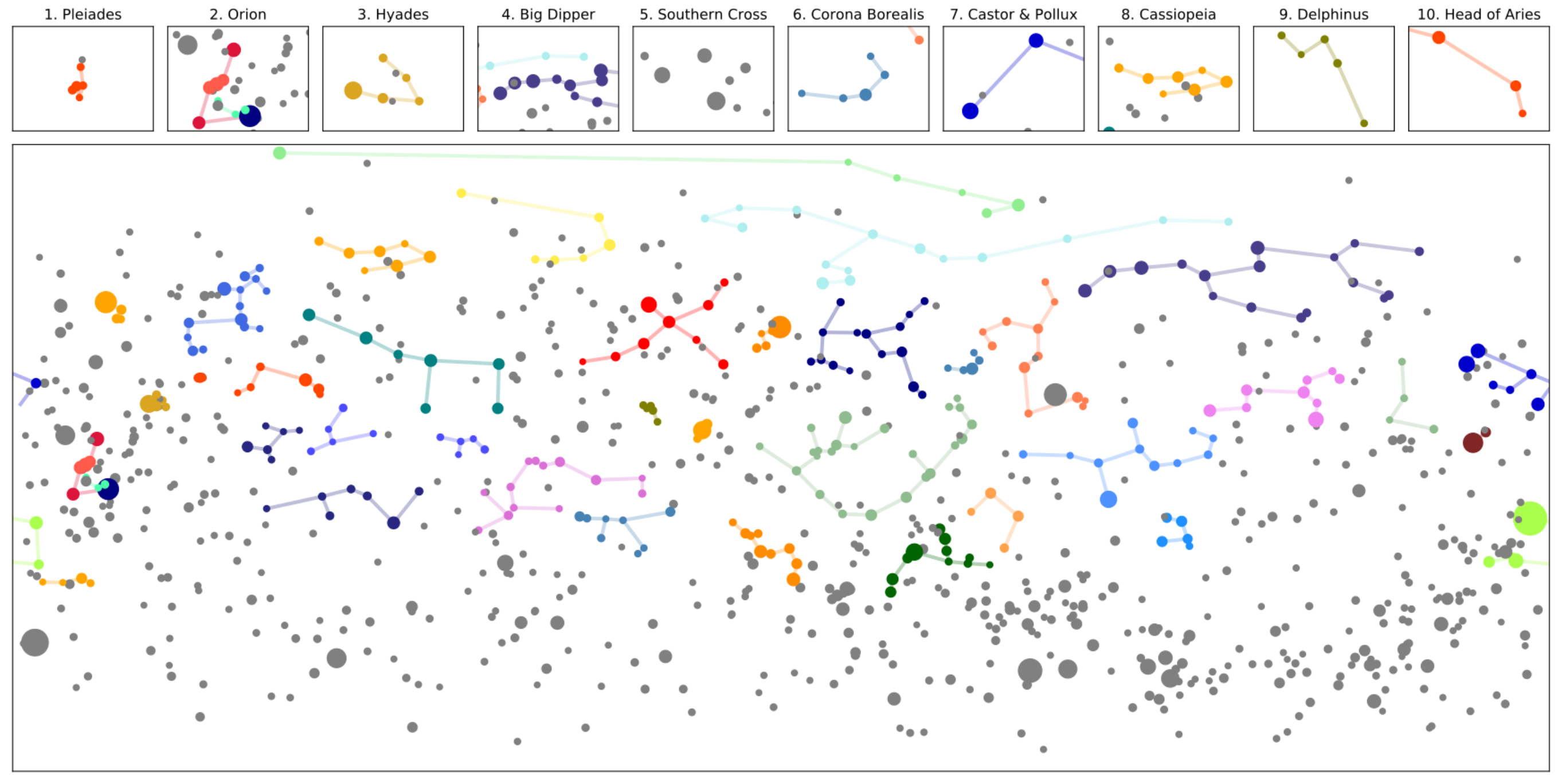}
\label{romanian}
\caption{Romanian (Stellarium).}
\end{center}
\end{figure}

\begin{figure}[h!]
\begin{center}
\includegraphics[width=\textwidth]{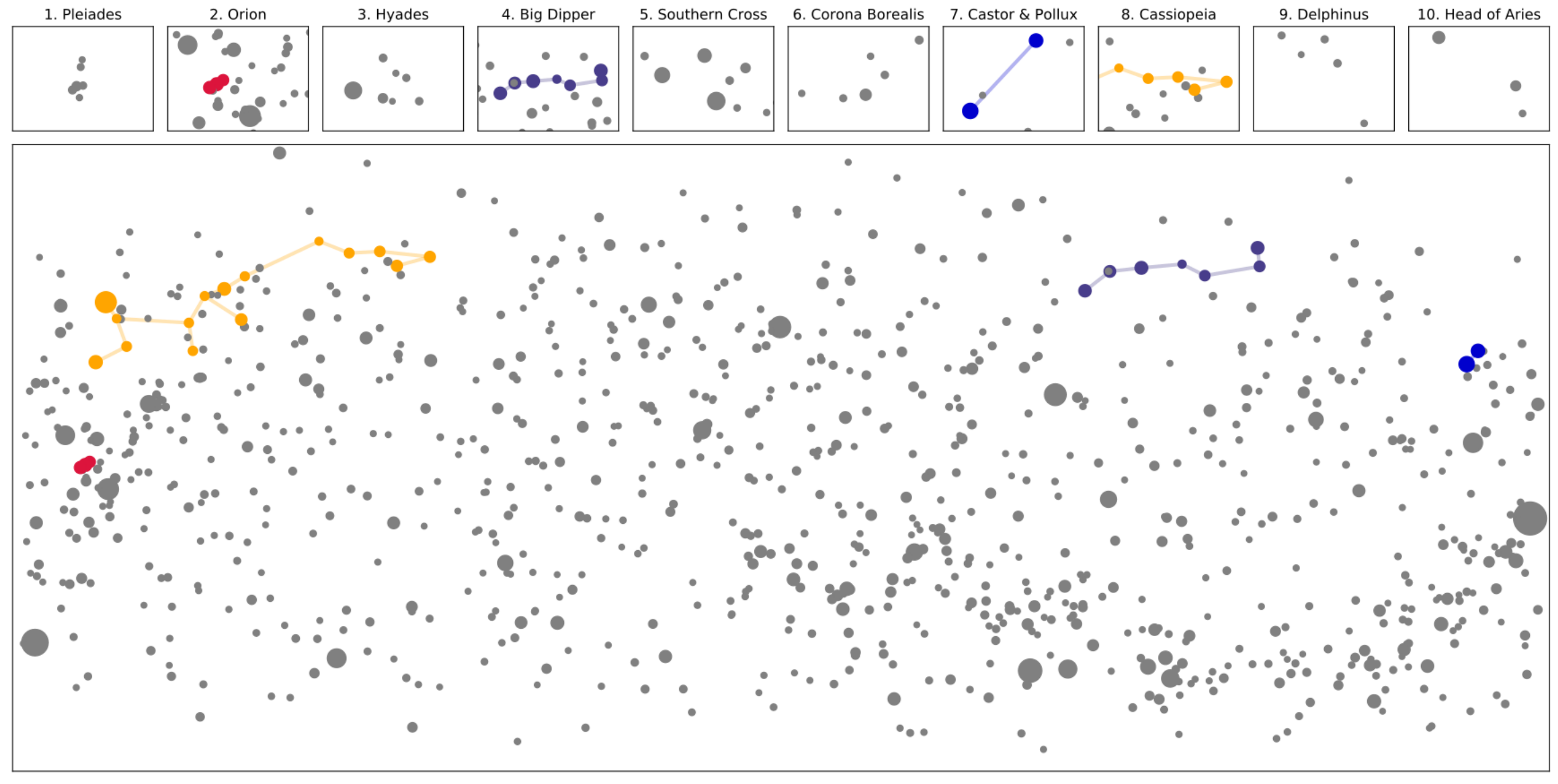}
\label{sami}
\caption{Sami (Stellarium).}
\end{center}
\end{figure}

\begin{figure}[h!]
\begin{center}
\includegraphics[width=\textwidth]{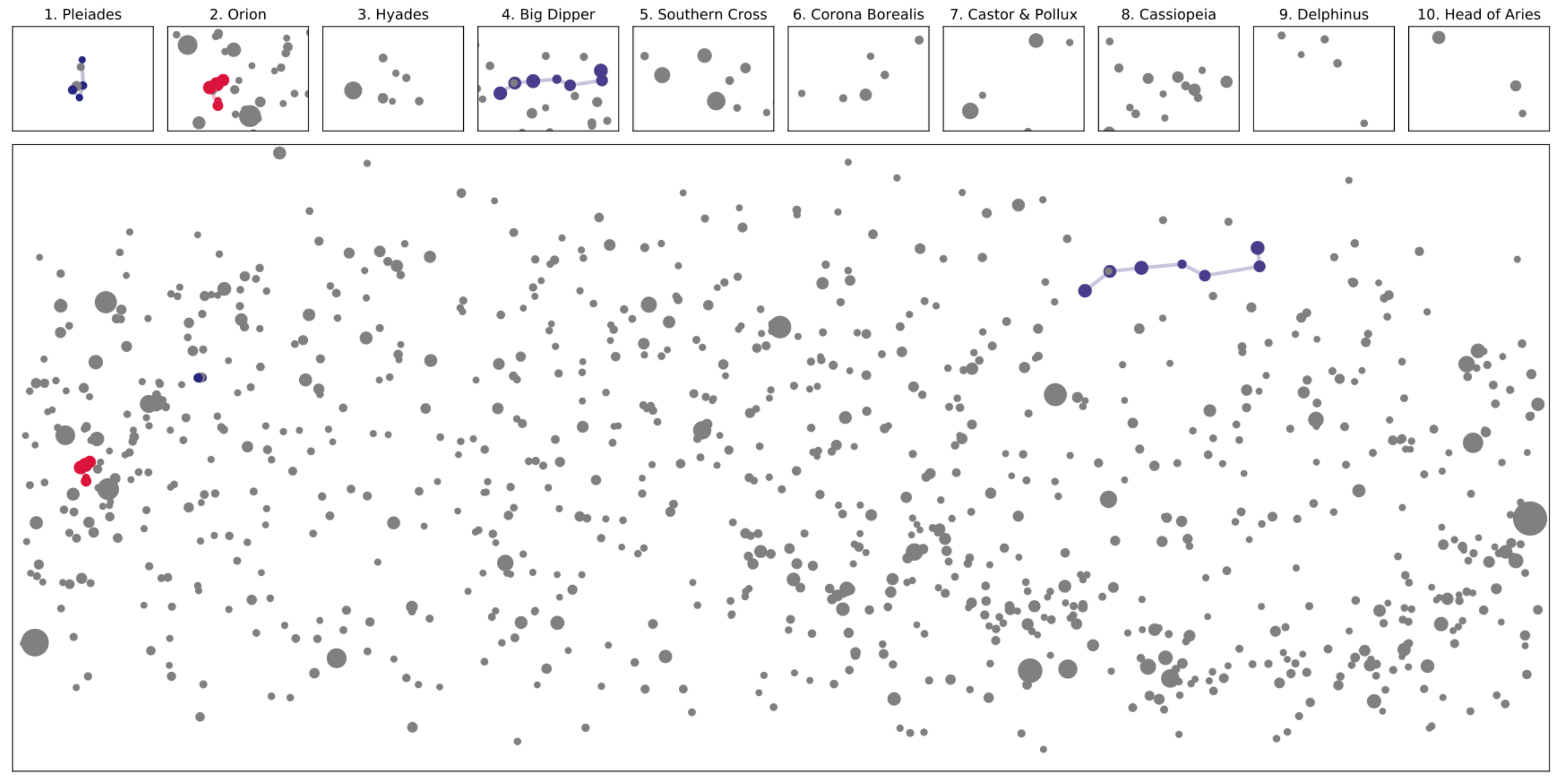}
\label{siberian}
\caption{Siberian (Stellarium).}
\end{center}
\end{figure}

\begin{figure}[h!]
\begin{center}
\includegraphics[width=\textwidth]{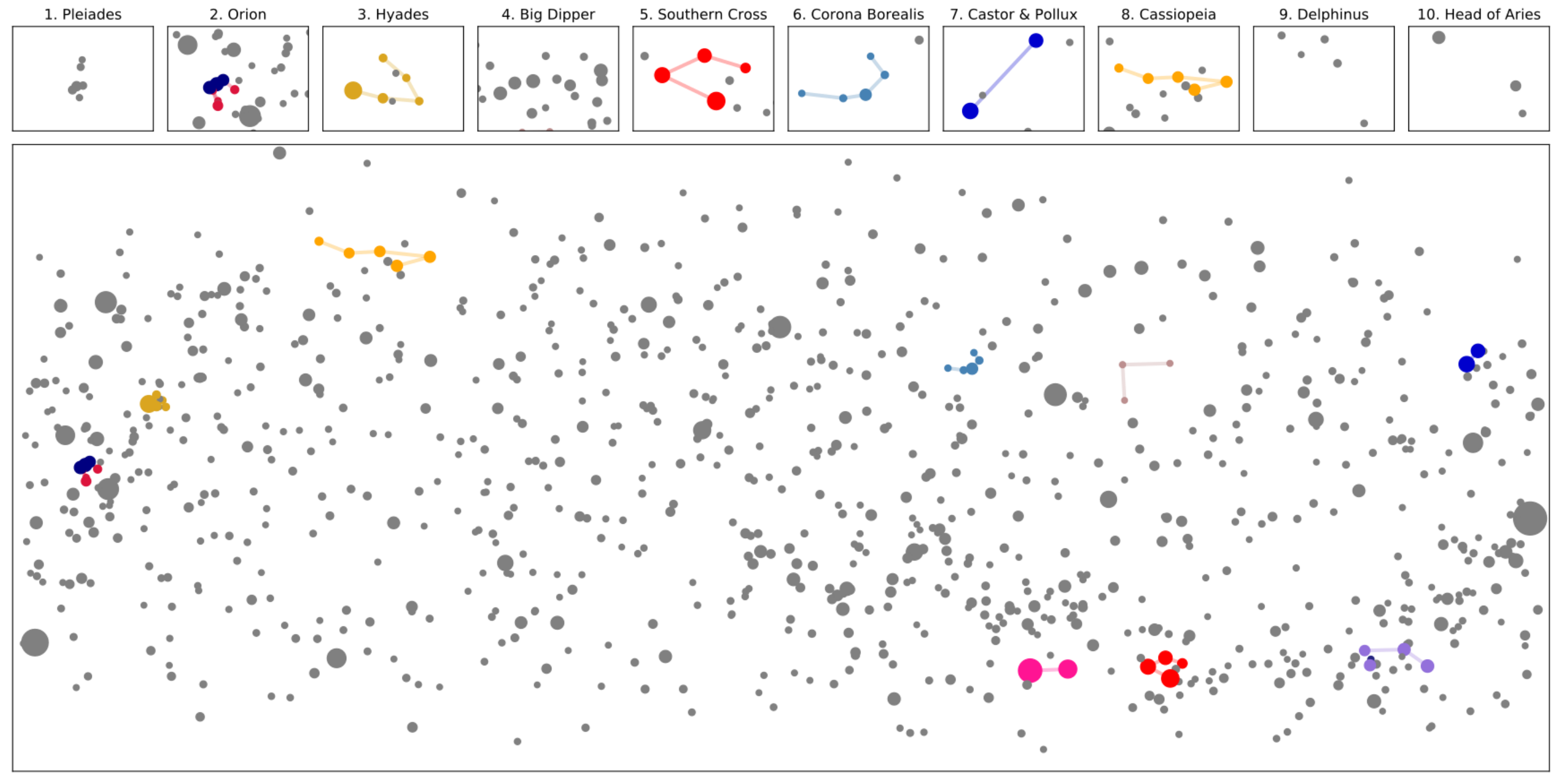}
\label{tongan}
\caption{Tongan (Stellarium).}
\end{center}
\end{figure}

\begin{figure}[h!]
\begin{center}
\includegraphics[width=\textwidth]{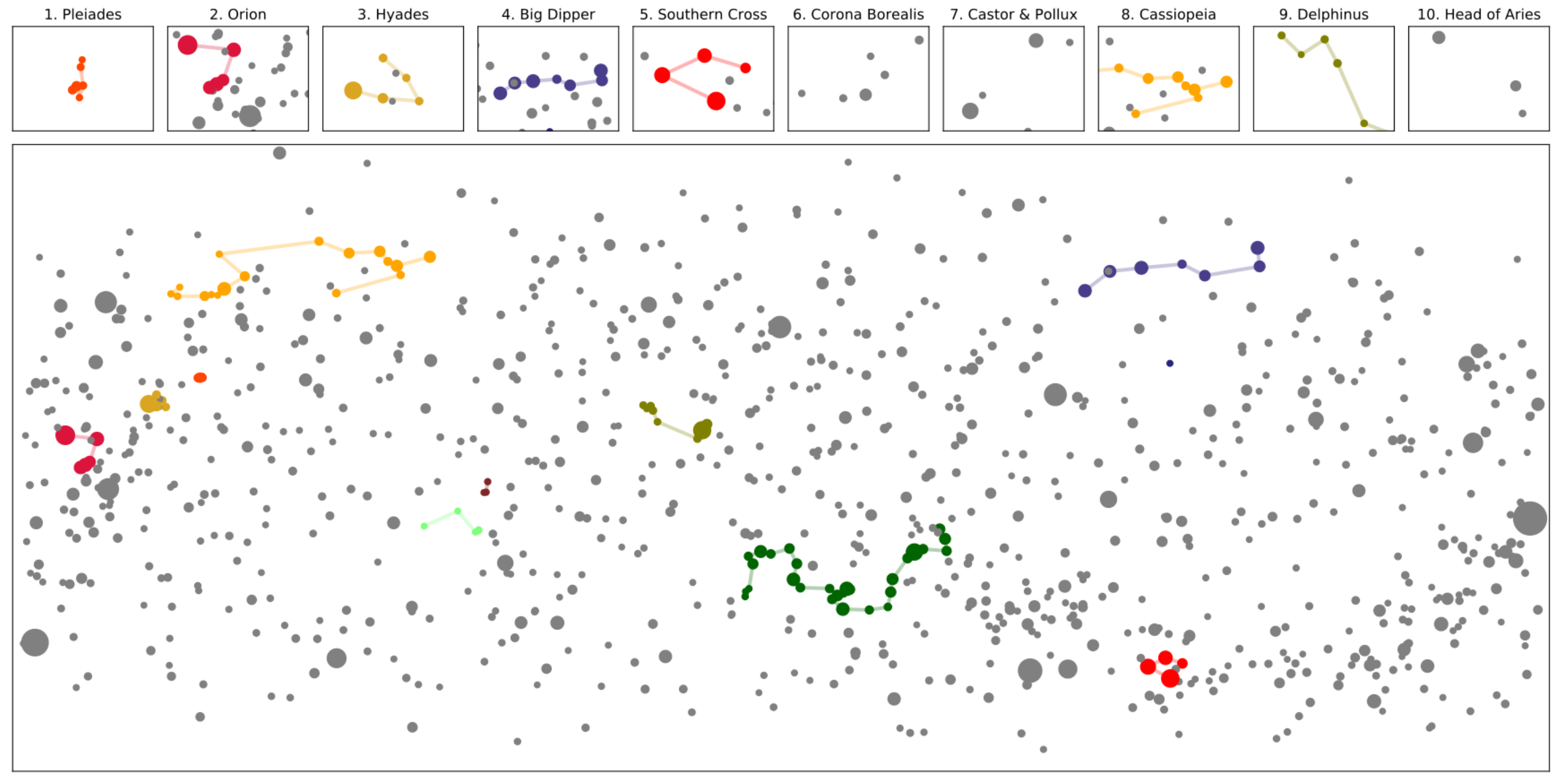}
\label{tukano}
\caption{Tukano (Stellarium).}
\end{center}
\end{figure}

\begin{figure}[h!]
\begin{center}
\includegraphics[width=\textwidth]{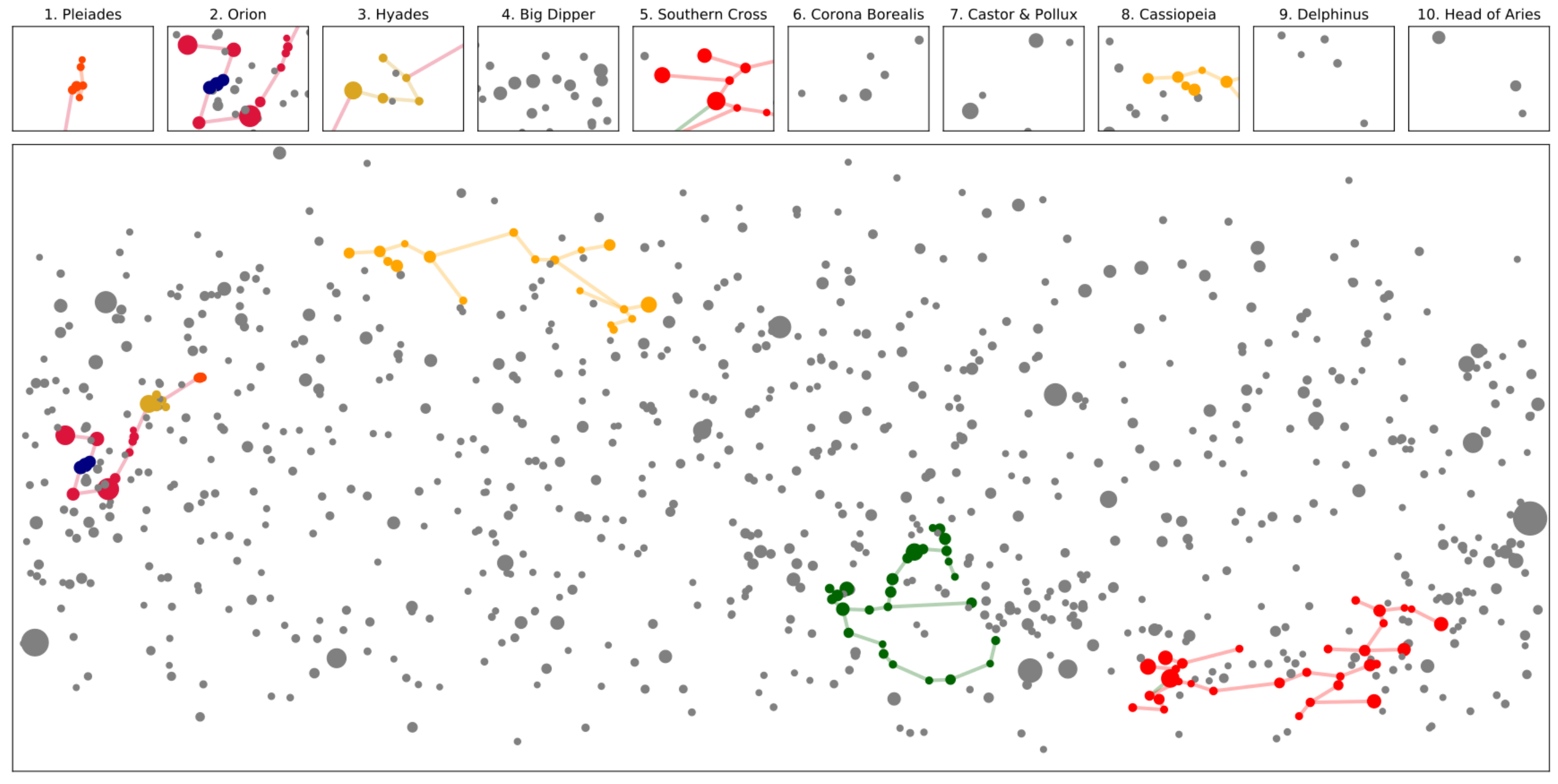}
\label{tupi}
\caption{Tupi (Stellarium).}
\end{center}
\end{figure}

\begin{figure}[h!]
\begin{center}
\includegraphics[width=\textwidth]{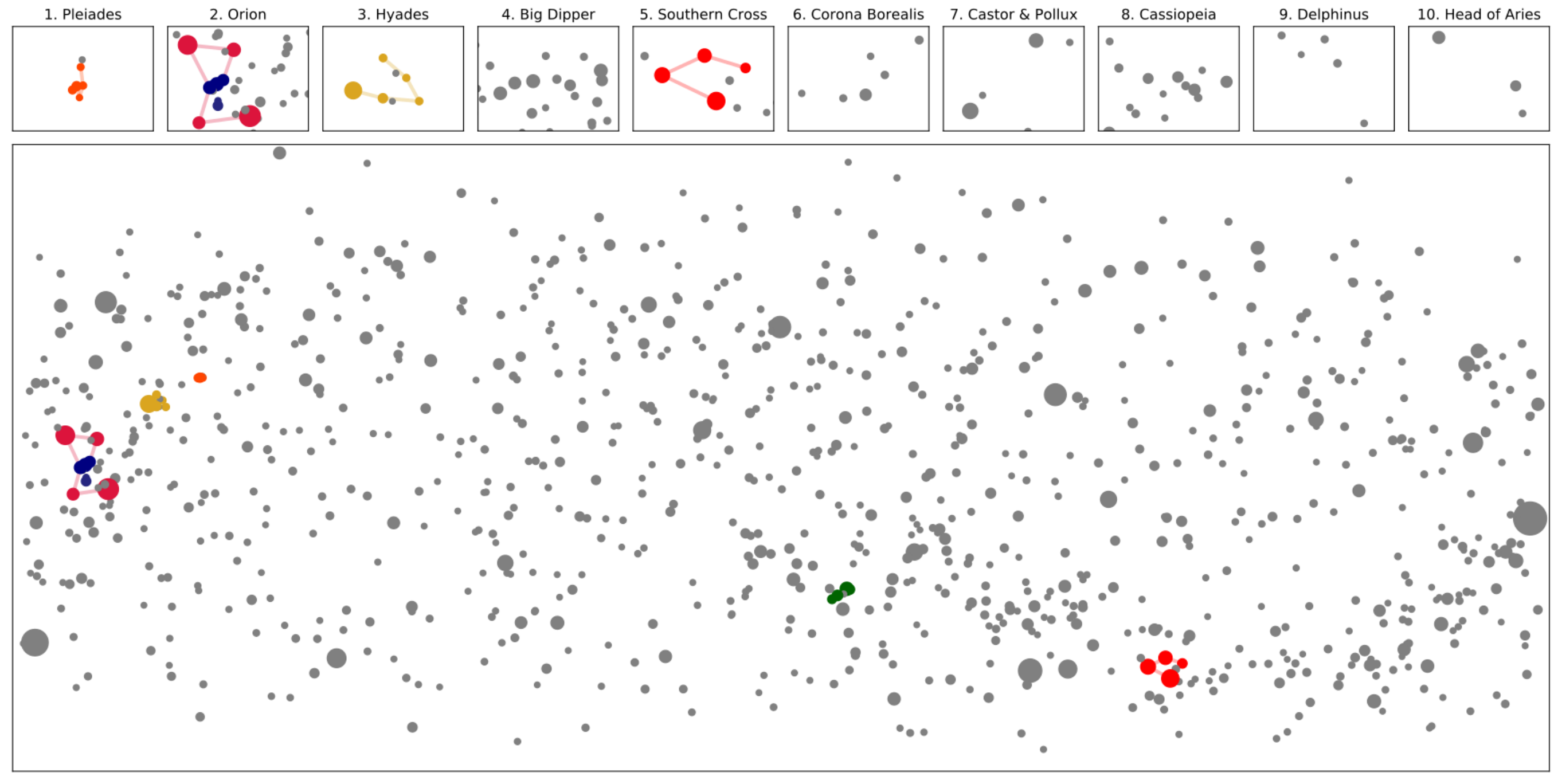}
\label{lenakel}
\caption{Lenakel (Vanuatu) (Netwar sky culture in Stellarium).}
\end{center}
\end{figure}

\begin{figure}[h!]
\begin{center}
\includegraphics[width=\textwidth]{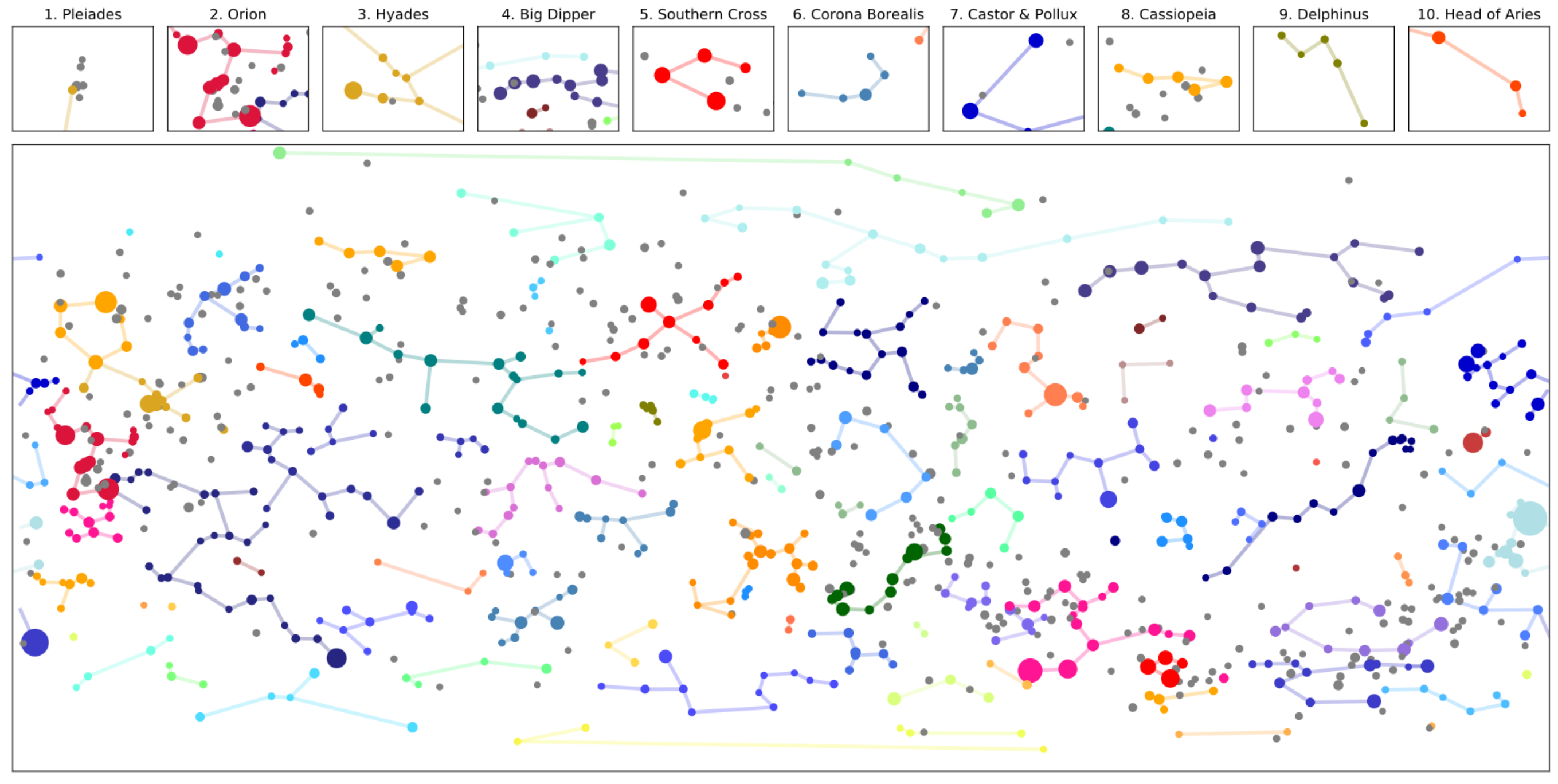}
\caption{Western (Stellarium).}
\label{western}
\end{center}
\end{figure}

\afterpage{\clearpage}

\newpage

\section{Common asterisms} 
\label{commonasterismsec}


\captionsetup{width=\textwidth}
\renewcommand\baselinestretch{1.1}\selectfont
\begin{longtable}[t]{rp{0.5in}p{0.5in}p{0.5in}p{3.4in}l}
\footnotesize
 & \textbf{Human Score} & \textbf{Weighted Human Score} & \textbf{Model Score} & \textbf{Stars} & \textbf{Description} \\

1 & 0.66 & 0.73 & 1.0 & 25EtaTau, 17Tau, 19Tau, 20Tau, 23Tau, 27Tau & Pleiades \\
2 & 0.65 & 0.64 & 1.0 & 34DelOri, 46EpsOri, 50ZetOri &  Orion's Belt \\
3 & 0.61 & 0.52 & 1.0 & 87AlpTau, 54GamTau, 61Del1Tau, 74EpsTau, 78The2Tau & Hyades \\
4 & 0.58 & 0.47 & 0.88 & 50AlpUMa, 48BetUMa, 64GamUMa, 69DelUMa, 77EpsUMa, 79ZetUMa, 85EtaUMa &  Big Dipper \\
5 & 0.43 & 0.4 & 1.0 & Alp1Cru, BetCru, GamCru, DelCru & Southern Cross \\
6 & 0.38 & 0.27 & 0.83 & 5AlpCrB, 3BetCrB, 8GamCrB, 13EpsCrB, 4TheCrB &  Corona Borealis \\
7 & 0.35 & 0.28 & 1.0 & 66AlpGem, 78BetGem &  Castor and Pollux \\
8 & 0.35 & 0.34 & 0.45 & 58AlpOri, 24GamOri, 34DelOri, 46EpsOri, 50ZetOri &  \\
9 & 0.31 & 0.35 & 0.6 & 34DelOri, 46EpsOri, 50ZetOri, 44IotOri, 42Ori &  Orion's Belt and Sword \\
10 & 0.31 & 0.25 & 0.56 & 50AlpUMa, 48BetUMa, 64GamUMa, 69DelUMa, 77EpsUMa, 79ZetUMa, 85EtaUMa, 1OmiUMa, 29UpsUMa, 63ChiUMa, 23UMa &  \\
11 & 0.3 & 0.22 & 0.71 & 18AlpCas, 11BetCas, 27GamCas, 37DelCas, 45EpsCas & Cassiopeia \\
12 & 0.3 & 0.35 & 1.0 & 9AlpDel, 6BetDel, 12Gam2Del, 11DelDel & Delphinus \\
13 & 0.28 & 0.23 & 0.38 & 11AlpDra, 50AlpUMa, 48BetUMa, 64GamUMa, 69DelUMa, 77EpsUMa, 79ZetUMa, 85EtaUMa, 23UMa, 26UMa, 12Alp2CVn &  \\
14 & 0.27 & 0.25 & 0.75 & 46EpsOri, 50ZetOri, 48SigOri &  \\
15 & 0.24 & 0.21 & 1.0 & 13AlpAri, 6BetAri, 5Gam2Ari & Head of Aries \\
16 & 0.24 & 0.22 & 1.0 & 53AlpAql, 60BetAql, 50GamAql & Shaft of Aquila \\
17 & 0.24 & 0.34 & 1.0 & Alp1Cen, BetCen & Southern Pointers \\
18 & 0.23 & 0.33 & 1.0 & AlpCrA, BetCrA, GamCrA & Corona Australis \\
19 & 0.23 & 0.15 & 0.75 & 1AlpUMi, 7BetUMi, 13GamUMi, 23DelUMi, 22EpsUMi, 16ZetUMi &  Little Dipper \\
20 & 0.22 & 0.19 & 1.0 & 50AlpUMa, 48BetUMa &  \\
21 & 0.22 & 0.22 & 1.0 & 8Bet1Sco, 7DelSco, 6PiSco & Head of Scorpius \\
22 & 0.21 & 0.21 & 0.04 & 54AlpPeg, 53BetPeg &  \\
23 & 0.21 & 0.2 & 1.0 & 35LamSco, 34UpsSco & Stinger of Scorpius \\
24 & 0.21 & 0.21 & 1.0 & Iot1Sco, KapSco, 35LamSco, 34UpsSco &  \\
25 & 0.2 & 0.17 & 0.75 & 32AlpLeo, 41Gam1Leo, 17EpsLeo, 36ZetLeo, 30EtaLeo, 24MuLeo &  Sickle \\
26 & 0.2 & 0.15 & 0.83 & 1AlpCrv, 9BetCrv, 4GamCrv, 7DelCrv, 2EpsCrv &  Corvus\\
27 & 0.19 & 0.18 & 0.56 & 21AlpSco, 8Bet1Sco, 7DelSco, 6PiSco, 20SigSco &  \\
28 & 0.19 & 0.19 & 0.0 & 21AlpAnd, 88GamPeg &  \\
29 & 0.18 & 0.16 & 0.58 & 21AlpSco, 8Bet1Sco, 7DelSco, 26EpsSco, Zet2Sco, Mu1Sco, 6PiSco, 20SigSco, 23TauSco &  \\
30 & 0.18 & 0.13 & 0.33 & 50AlpCyg, 6Bet1Cyg, 37GamCyg, 18DelCyg, 53EpsCyg, 21EtaCyg & Northern Cross \\
31 & 0.17 & 0.16 & 0.56 & 26EpsSco, Zet2Sco, EtaSco, TheSco, Iot1Sco, KapSco, 35LamSco, Mu1Sco, 34UpsSco & Tail of Scorpius \\
32 & 0.17 & 0.17 & 1.0 & 21AlpSco, 20SigSco, 23TauSco &  \\
33 & 0.17 & 0.09 & 0.83 & 3AlpLyr, 10BetLyr, 14GamLyr, 12Del2Lyr, 6Zet1Lyr &  Lyra \\
34 & 0.16 & 0.13 & 0.01 & 21AlpAnd, 54AlpPeg, 53BetPeg, 88GamPeg & Square of Pegasus \\
35 & 0.16 & 0.08 & 0.56 & 6Alp2Cap, 9BetCap, 40GamCap, 49DelCap, 34ZetCap, 23TheCap, 32IotCap, 16PsiCap, 18OmeCap & Capricornus \\
36 & 0.16 & 0.12 & 0.35 & 21AlpSco, 8Bet1Sco, 7DelSco, 26EpsSco, EtaSco, TheSco, Iot1Sco, KapSco, 35LamSco, Mu1Sco, 6PiSco, 23TauSco &  Scorpius \\
37 & 0.15 & 0.13 & 0.2 & 34DelOri, 46EpsOri, 50ZetOri, 87AlpTau, 54GamTau, 61Del1Tau, 74EpsTau, 78The2Tau, 17Tau &  \\
38 & 0.14 & 0.15 & 1.0 & 39LamOri, 37Phi1Ori, 40Phi2Ori &  \\
39 & 0.14 & 0.14 & 1.0 & 58AlpOri, 24GamOri &  \\
40 & 0.13 & 0.09 & 0.5 & 42AlpCom, 43BetCom, 15GamCom &  \\
41 & 0.12 & 0.15 & 0.4 & 53AlpAql, 60BetAql, 50GamAql, 9AlpDel, 6BetDel, 12Gam2Del, 11DelDel, 2EpsDel &  \\
42 & 0.12 & 0.1 & 1.0 & 7BetUMi, 13GamUMi, 5UMi &  \\
43 & 0.12 & 0.06 & 0.53 & 16AlpBoo, 42BetBoo, 27GamBoo, 49DelBoo, EpsBoo, 30ZetBoo, 8EtaBoo, 25RhoBoo, 5UpsBoo & Bo\"{o}tes\  \\
44 & 0.12 & 0.11 & 1.0 & 68DelLeo, 70TheLeo &  \\
45 & 0.11 & 0.09 & 0.33 & 37DelCas, 45EpsCas, 33TheCas &  \\
46 & 0.11 & 0.08 & 0.23 & 13AlpAur, 34BetAur, 37TheAur, 3IotAur, 112BetTau & Auriga \\
47 & 0.11 & 0.1 & 0.57 & EpsCar, IotCar, DelVel, KapVel & False Cross \\
48 & 0.11 & 0.08 & 1.0 & 10AlpCMi, 3BetCMi &  \\
49 & 0.11 & 0.06 & 0.09 & 11AlpDra, 23BetDra, 33GamDra, 57DelDra, 63EpsDra, 22ZetDra, 14EtaDra, 13TheDra, 12IotDra, 5KapDra, 1LamDra, 25Nu2Dra, 32XiDra, 60TauDra, 44ChiDra &  Draco \\
50 & 0.11 & 0.09 & 0.5 & 9Alp2Lib, 27BetLib, 38GamLib &  \\
51 & 0.11 & 0.1 & 1.0 & 48GamAqr, 62EtaAqr, 55Zet2Aqr &  \\
52 & 0.11 & 0.09 & 1.0 & 22ZetDra, 14EtaDra, 13TheDra &  \\
53 & 0.11 & 0.09 & 0.33 & 1AlpCrv, 9BetCrv, 4GamCrv, 7DelCrv, 2EpsCrv, BetHya, 29GamVir, 51TheVir &  \\
54 & 0.1 & 0.1 & 1.0 & 4DelHya, 11EpsHya, 16ZetHya, 7EtaHya, 22TheHya, 13RhoHya, 5SigHya &  \\
55 & 0.1 & 0.06 & 0.33 & 67AlpVir, 29GamVir, 43DelVir, 47EpsVir, 3NuVir &  \\
56 & 0.1 & 0.05 & 0.17 & 65AlpCnc, 17BetCnc, 47DelCnc, 48IotCnc &  \\
57 & 0.1 & 0.09 & 0.33 & 58AlpOri, 24GamOri, 39LamOri, 37Phi1Ori, 40Phi2Ori &  \\
58 & 0.1 & 0.06 & 0.75 & 23BetDra, 33GamDra, 57DelDra, 25Nu2Dra, 32XiDra &  \\
59 & 0.1 & 0.06 & 0.21 & 66AlpGem, 78BetGem, 24GamGem, 55DelGem, 27EpsGem, 43ZetGem, 13MuGem &  Gemini \\
60 & 0.1 & 0.07 & 0.32 & 7EtaGem, 13MuGem, 58AlpOri, 19BetOri, 24GamOri, 34DelOri, 46EpsOri, 50ZetOri, 53KapOri, 39LamOri, 61MuOri, 8Pi5Ori, 40Phi2Ori &  \\
61 & 0.09 & 0.08 & 1.0 & 10Gam2Sgr, 19DelSgr, 20EpsSgr, 38ZetSgr, EtaSgr, 22LamSgr, 34SigSgr, 40TauSgr, 27PhiSgr & Teapot \\
\caption{An extended version of Table 1 that includes 61 asterisms in total.}
\label{commonasterism}
\end{longtable}

\section{Asterisms for the GC model with $n=320$}
\label{secmodelasterism}

\renewcommand\baselinestretch{1.1}\selectfont
\begin{longtable}[t]{rrp{3.7in}l}
\footnotesize
 & \textbf{Score} & \textbf{Stars} & \textbf{Description} \\
1 & 1.0 & 25EtaTau, 17Tau, 19Tau, 20Tau, 23Tau, 27Tau & Pleiades \\
2 & 1.0 & Alp1Cen, BetCen & Southern Pointers\\
3 & 1.0 & 10Gam2Sgr, 19DelSgr, 20EpsSgr, 38ZetSgr, EtaSgr, 22LamSgr, 34SigSgr, 40TauSgr, 27PhiSgr & Teapot \\
4 & 1.0 & 53AlpAql, 60BetAql, 50GamAql & Shaft of Aquila \\
5 & 1.0 & 34AlpAqr, 48GamAqr, 55Zet2Aqr, 62EtaAqr & Water Jar (part)\\  
6 & 1.0 & 9AlpDel, 6BetDel, 12Gam2Del, 11DelDel & Delphinus\\
7 & 1.0 & AlpCrA, BetCrA, GamCrA & Corona Australis \\
8 & 1.0 & 5AlpSge, 6BetSge, 12GamSge, 7DelSge & Sagitta \\
9 & 1.0 & 13AlpAri, 6BetAri, 5Gam2Ari & Head of Aries \\
10 & 1.0 & 5GamEqu, 7DelEqu & Judge of right and wrong (Chinese) \\
11 & 1.0 & 95 Her, 102Her & Textile ruler (Chinese)\\
12 & 1.0 & 60BetOph, 62GamOph & Official for the royal clan (Chinese) \\
13 & 1.0 & 67PiHer, 75RhoHer  & Woman's bed (Chinese)\\
14 & 1.0 & 25IotOph, 27KapOph & Dipper for solids (Chinese) \\
15 & 1.0 & 13EpsAql, 17ZetAql & ar in Mejleb (Marshall Islands) \\
16 & 1.0 & 40TauLib, 39UpsLib & Celestial spokes (Chinese) \\
17 & 1.0 & 7BetUMi, 13GamUMi, 5UMi & Jemenuwe (Marshall Islands) \\
18 & 1.0 & EpsBoo, 25RhoBoo, 28SigBoo & Celestial lance (Chinese)\\
19 & 1.0 & MuCen, NuCen, PhiCen & Ujela (Marshall Islands) \\
20 & 1.0 & 42ZetPeg, 46XiPeg & Thunder and lightning (Chinese) \\
21 & 1.0 & 33LamUMa, 34MuUMa & Kam Anij (Marshall Islands)\\
22 & 1.0 & 39LamOri, 37Phi1Ori, 40Phi2Ori & Al-Hekaah (Arabic) \\
23 & 1.0 & 13TheCyg, 10Iot2Cyg, 1KapCyg & Xi Zhong (Chinese) \\
24 & 1.0 & 20EtaLyr, 21TheLyr & Nin-SAR and Erragal (Babylonian) \\
25 & 1.0 & 90PhiAqr, 91Psi1Aqr, 93Psi2Aqr & Mhua (Tukano)\\
26 & 1.0 & 57DelDra, 63EpsDra & Celestial kitchen (Chinese) \\
27 & 1.0 & 1Pi3Ori, 3Pi4Ori & Lulal and Latarak (Babylonian)\\
28 & 0.88 & 87AlpTau, 54GamTau, 61Del1Tau, 68Del3Tau, 74EpsTau, 78The2Tau, 71Tau & Hyades \\
29 & 0.86 & 50AlpUMa, 48BetUMa, 64GamUMa, 69DelUMa, 77EpsUMa, 79ZetUMa, 85EtaUMa, 80UMa & Big Dipper \\
30 & 0.83 & 18AlpCas, 11BetCas, 27GamCas, 37DelCas, 45EpsCas, 17ZetCas, 24EtaCas & Cassiopeia \\
31 & 0.8 & 23BetDra, 33GamDra, 25Nu2Dra, 32XiDra & Head of Draco \\
32 & 0.8 & 5AlpCrB, 3BetCrB, 8GamCrB, 13EpsCrB, 4TheCrB, 49DelBoo & Corona Borealis \\
33 & 0.8 & 53BetPeg, 44EtaPeg, 47LamPeg, 48MuPeg & Resting palace (Chinese) \\ 
34 & 0.75 & Alp1Cru, BetCru, GamCru, DelCru, EpsCru & Southern Cross \\
35 & 0.75 & 32AlpLeo, 41Gam1Leo, 36ZetLeo, 30EtaLeo, 31Leo & Sickle (part) \\
36 & 0.75 & 37Xi2Sgr, 39OmiSgr, 41PiSgr & Establishment (Chinese) \\
37 & 0.67 & TheSco, Iot1Sco, KapSco, 35LamSco, 34UpsSco, 6630 & Tail of Scorpius \\
38 & 0.67 & 4BetTri, 9GamTri & Triangulum (part) \\
39 & 0.67 & 3AlpLyr, 12Del2Lyr, 4Eps1Lyr, 6Zet1Lyr & Lyra (part)\\
40 & 0.67 & 66AlpGem, 78BetGem, 62RhoGem, 75SigGem & Castor \& Pollux\\
41 & 0.67 & 9IotUMa, 12KapUMa & \\ 
42 & 0.67 & 16AlpBoo, 8EtaBoo, 4TauBoo, 5UpsBoo & \\ 
43 & 0.67 & 42TheOph, 44Oph  & \\ 
44 & 0.67 & 2XiTau, 1OmiTau  & \\ 
45 & 0.67 & Bet1Sgr, Bet2Sgr  & \\ 
46 & 0.67 & 22ZetDra, 14EtaDra  & \\ 
47 & 0.67 & 67Oph, 70Oph  & \\ 
48 & 0.67 & 10BetLyr, 14GamLyr  & \\ 
49 & 0.67 & 16PsiCap, 18OmeCap  & \\ 
50 & 0.6 & TheCar, 4050, 4140  & \\ 
51 & 0.6 & 5AlpCep, 3EtaCep, 2TheCep  & \\ 
52 & 0.6 & BetAra, GamAra, ZetAra  & \\ 
53 & 0.5 & 68DelLeo, 70TheLeo, 60Leo & \\ 
54 & 0.5 & 10AlpCMi, 3BetCMi, 4GamCMi  & \\ 
55 & 0.5 & 4GamCrv, 7DelCrv, 8EtaCrv  &  Corvus (part) \\
56 & 0.5 & AlpMus, BetMus  &  Musca (part) \\
57 & 0.5 & 16LamAql, 12Aql  & \\ 
58 & 0.5 & 7AlpLac, 3BetLac  & \\ 
59 & 0.5 & 10ThePsc, 17IotPsc  & \\ 
60 & 0.5 & 24GamGem, 31XiGem, 30Gem  & \\ 
61 & 0.5 & 9AlpCMa, 2BetCMa  & \\ 
62 & 0.5 & 13GamLep, 15DelLep  & \\ 
63 & 0.5 & 11AlpLep, 9BetLep  & \\ 
64 & 0.5 & 13AlpAur, 34BetAur, 7EpsAur, 8ZetAur, 10EtaAur, 35PiAur  & \\ 
65 & 0.5 & 22TauHer, 11PhiHer  & \\ 
66 & 0.5 & 5Alp1Cap, 6Alp2Cap, 9BetCap  & \\ 
67 & 0.5 & AlpGru, BetGru, EpsGru, ZetGru  &  Grus (part) \\
68 & 0.5 & MuCep, 10NuCep  & \\ 
69 & 0.5 & AlpPhe, EpsPhe, KapPhe  & \\ 
70 & 0.44 & 50AlpCyg, 37GamCyg, 18DelCyg, 53EpsCyg, 58NuCyg, 62XiCyg, 31Cyg, 32Cyg  &  Northern Cross \\
71 & 0.43 & 24AlpPsA, 22GamPsA, 23DelPsA  & \\ 
72 & 0.43 & 21AlpSco, 8Bet1Sco, 7DelSco, 14NuSco, 6PiSco, 20SigSco, 23TauSco, 10Ome2Sco, 9Ome1Sco  &  Head of Scorpius \\
73 & 0.43 & 25DelCMa, 21EpsCMa, 31EtaCMa, 24Omi2CMa, 22SigCMa, 28OmeCMa  &  Rear of Canis Major \\
74 & 0.43 & 11EpsHya, 16ZetHya, 13RhoHya  & \\ 
75 & 0.43 & 17IotAnd, 19KapAnd, 16LamAnd  & \\ 
76 & 0.4 & 78IotLeo, 77SigLeo & \\ 
77 & 0.4 & 86Aqr, 88Aqr, 98Aqr, 99Aqr  & \\ 
78 & 0.4 & 44ZetPer, 38OmiPer  & \\ 
79 & 0.4 & 31EtaCet, 45TheCet  & \\ 
80 & 0.4 & 43BetAnd, 37MuAnd  & \\ 
81 & 0.4 & 26EpsSco, Mu1Sco  & \\ 
82 & 0.36 & 67BetEri, 69LamEri, 58AlpOri, 19BetOri, 24GamOri, 34DelOri, 46EpsOri, 50ZetOri, 28EtaOri, 43The2Ori, 44IotOri, 53KapOri, 48SigOri, 20TauOri, 29Ori, 32Ori, 42Ori, 1887  &  Orion \\
83 & 0.33 & BetPhe, GamPhe  & \\ 
84 & 0.33 & 51And, PhiPer  & \\ 
85 & 0.33 & 40GamCap, 49DelCap  & \\ 
86 & 0.33 & 26BetPer, 25RhoPer  & \\ 
87 & 0.33 & 40AlpLyn, 38Lyn  & \\ 
88 & 0.3 & 24AlpSer, 13DelSer, 37EpsSer & \\ 
89 & 0.29 & AlpCol, BetCol  & \\ 
90 & 0.29 & 51MuPer, 48Per  & \\ 
91 & 0.29 & 31DelAnd, 30EpsAnd  & \\ 
92 & 0.29 & 33AlpPer, 23GamPer, 39DelPer, 15EtaPer, IotPer, 35SigPer, 18TauPer, 37PsiPer  &  Perseus (part) \\
93 & 0.25 & 28BetSer, 41GamSer, 35KapSer  & \\ 
94 & 0.25 & 7EtaGem, 13MuGem & \\ 
95 & 0.25 & TheGru, IotGru  & \\ 
96 & 0.25 & 27BetHer, 20GamHer  & \\ 
97 & 0.25 & 27DelCep, 23EpsCep, 21ZetCep  & \\ 
98 & 0.25 & EpsCar, IotCar, DelVel, KapVel, 3447, 3659, 3803  &  False Cross \\
99 & 0.22 & 41Ups4Eri, 43Eri  & \\ 
100 & 0.21 & 17EpsLeo, 4LamLeo, 24MuLeo  & \\ 
101 & 0.2 & 1DelOph, 2EpsOph  & \\ 
102 & 0.2 & 76DelAqr, 71Tau2Aqr  & \\ 
103 & 0.2 & GamLup, DelLup  & \\ 
104 & 0.2 & EtaCen, KapCen, AlpLup, BetLup  & \\ 
105 & 0.2 & PhiEri, ChiEri  & \\ 
106 & 0.18 & 14ZetLep, 16EtaLep  & \\ 
107 & 0.18 & 23DelEri, 18EpsEri  & \\ 
108 & 0.15 & 1Lac, 8485  & \\ 
109 & 0.15 & 39Cyg, 41Cyg  & \\ 
110 & 0.13 & 86MuHer, 94NuHer, 92XiHer, 103OmiHer  & \\ 
111 & 0.12 & GamCen, DelCen, TauCen  & \\ 
112 & 0.11 & 67SigCyg, 65TauCyg  & \\ 
113 & 0.06 & 46LMi, 54NuUMa, 53XiUMa  & \\ 
114 & 0.0 & 37TheAur, 32NuAur  & \\ 
115 & 0.0 & AlpCar, TauPup  & \\ 
116 & 0.0 & 21EtaCyg, ChiCyg  & \\ 
117 & 0.0 & PiPup, 2787  & \\ 
118 & 0.0 & ZetPup, 3080  & \\ 
119 & 0.0 & AlpEri, AlpHyi  & \\ 
120 & 0.0 & 25TheUMa, 26UMa  & \\ 
121 & 0.0 & Del1Gru, Del2Gru  & \\ 
122 & 0.0 & 43PhiDra, 44ChiDra  & \\ 
123 & 0.0 & 3445, 3487  & \\ 
124 & 0.0 & 65Kap1Tau, 69UpsTau & \\ 
\caption{Asterisms picked out by the GC model with $n=320$. The scores roughly indicate how similar each asterism is to the closest asterism in the human data (1.0 indicates a perfect match). In some cases the descriptions are approximate only---for example, the asterism labeled ``Corona Borealis'' includes an extra star (49DelBoo). Labels for stars without a Bayer designation are HR identification numbers from the Yale Bright Star catalog.}
\label{modelasterism}
\end{longtable}

\section{SI Movie}

The SI movie shows asterisms identified by the GC model as the threshold $n$ is increased from 1 to 2000.

\bibliography{stars}
\bibliographystyle{naturemag}